# Gauge symmetry, chirality and parity violation in four-particle systems: Coulomb's law as a universal molecular function.

G. Van Hooydonk, Ghent University, Department of Library Sciences/Department of Physical and Inorganic Chemistry, Rozier 9, B-9000 Ghent (Belgium)
E-mail: *guido.vanhooydonk@rug.ac.be*

**Abstract.** *Following recent work in search for a universal function (Van Hooydonk, Eur. J. Inorg. Chem., 1999, 1617), we test four symmetric $\pm a_n R^n$ potentials for reproducing molecular potential energy curves (PECs). Classical gauge symmetry is broken, which results in generic left-right asymmetric PECs for $1/R$ potentials. A pair of symmetric perturbed Coulomb potentials is in accordance with the shape of observed PECs. For a bond, a four-particle system, charge inversion (parity violation, atom chirality) is the key to explain this shape generically. A parity adapted Hamiltonian reduces from ten to two terms and to a soluble Bohr-like formula, the Kratzer potential $(1-R_e/R)^2$. The result is similar to the combined action of spin and wave functional symmetry effects upon the Hamiltonian in the Heitler-London theory. The corresponding analytical perturbed Coulomb function varies simply with $(1-R_e/R)$ and scales attractive and repulsive branches of PECs for 13 bonds $H_2$, HF, LiH, KH, AuH, $Li_2$, LiF, KLi, NaCs, $Rb_2$, RbCs, $Cs_2$ and $I_2$ in a single straight line. Turning points for 13 bonds are reproduced with an absolute deviation of 0,3 % (0,007 Å) for about 400 points at both branches. For 230 points at the repulsive side, the deviation is 0,2 % (0,003 Å). Available turning points for $I_2$ are in need of revision. This universal molecular scaling function is the classical electrostatic perturbed Coulomb law, which reduces the complex four-particle system to a central force system on one nucleon. The Kratzer function relates to two central force systems, one on each nucleon. A minimum of parameters is required and even the ab initio zero molecular parameter function gives PECs of acceptable quality, just using atomic ionisation energies. The function can be used as a model potential for inverting levels and gives a first principle's comparison of short- and long-range interactions, of importance for the study of cold atoms. The theory may be tested with wave-packet dynamics: femto-chemistry applied to the crossing of covalent and ionic curves. We anticipate this scale and shape invariant scheme applies to smaller scales in nuclear and high-energy particle physics. For larger gravitational scales (Newton $1/R$ potentials), problems with super-unification are discussed. Reactions between hydrogen and anti-hydrogen, feasible in the near future, will probably produce normal $H_2$.*

## 1. Introduction

Ehrenfest's theorem (1927) states that in the limit quantum mechanical expectation values behave classically. The most classical form of physics is elementary statics (Stevin, 1605) and is directly linked to Euclidean geometry. Symmetry, statics and geometry are scale-invariant and of fundamental importance for describing particle interactions. Symmetry is independent of dynamics (Gross, 1996). The effects of symmetry are discrete, permanent or time-invariant: parity, mirror symmetry and left-right asymmetry (chirality, handedness) and show clearly in polyatomic molecules (Kellman, 1996, Dunitz, 1996). Until 1956 is was generally accepted that parity was never violated but the discovery of parity violation in the weak interactions (Lee and Yang, 1956) led to the new physics, culminating in the Standard Model and beyond. But parity is also violated in atoms and in polyatomic molecules and the significance of this observation can not be underestimated. According to Bouchiat and Bouchiat (1997) low energy physics still has a role to play in the exploration of the Standard Model. Looking for systems where symmetry is broken is an important issue in physics, in particular SUSY (Gel'fand and Likhtman, 1971, Wess and Zumino, 1974, Witten, 1981, 1982). This powerful new tool for physics between Fermi- and Planck-scales (Lopez, 1997) found applications at the Bohr-scale (Kostelecky, 1992, Cooper, 1993, Lévai, 1994, Roy and Varshni, 1991, Blado, 1996, Dutt et al., 1995, Mukherjee et al., 1995, Guerin, 1996) and even in biology (Bahsford et al., 1998). SUSY uses algebraic quadratic super-potentials in the framework of quantum mechanics. The basis for SUSY was laid with the method of factorisation (Dirac, 1935, Schrödinger, 1940). It was developed by Infeld and Hull (1951) with a major interest in oscillator- and generalised Kepler systems, important for molecular spectroscopy. Algebraic schemes apply to domains varying from molecular to particle and nuclear physics (Alhassid et al, 1983, Iachello, 1981, Cooper, 1993, Lévai, 1994). Nevertheless, SUSY might remain a mathematical artefact. Desperately breaking SUSY is the motto of today's physics (Lopez, 1997, Poppitz, 1997, Poppitz and Trivedi, 1998).

*However, symmetry effects are scale-invariant. If parity is violated in real systems (atoms and molecules at the Bohr-scale, sub-atomic particles at the Fermi-scale), other examples must exist in nature.*

Parity violating effects in atoms and in polyatomic molecules are small in terms of energy. Sophisticated experimental and computational methods are required to disclose the mechanisms (for atoms: Bouchiat and Bouchiat, 1997; for molecules: Bakasov et al., 1998). Chiral molecules consist of *four atoms* (particles) not in a linear alignment. But any *diatomic* bond can be considered as a *four-particle* system with the *four particles* not in a linear alignment either. This makes a diatomic bond a theoretical candidate for observing parity violation. The total charge of the system is zero and it is symmetric with respect to charge but not with respect to mass. Four-particle systems are of fundamental interest (Richard, 1994, Abdel-Raouf et al., 1998, Benslama et al., 1998) because of the quark-antiquark model and the prospects on hydrogen/anti-hydrogen reactions (Armour and Zeman, 1999, Russell, 1999).

The $H_2$ molecule, two electrons and two protons, remains the standard to test any theory on four-particle systems but it does not show parity violation. Partly due to its ordinary scale, the system got less attention in recent years, since Heitler and London (1927) solved the problem of chemical bonding 80 years ago. The quality of their potential energy curve (PEC) for $H_2$ was poor but James and Coolidge (1933) soon succeeded in calculating a better one. Theoretical physics has evolved drastically ever since but theoretical chemistry remained focused upon developing better computational methods (Pople, 1999). For $H_2$, an exact PEC was computed 30 years ago (Kolos and Wolniewicz, 1968). The next molecules in the Periodic Table are LiH and $Li_2$, which was also studied extensively (Hessel and Vidal, 1979). Femto-chemistry, an application of wave-packet dynamics (Garraway and Suominen, 1995) gave a new impetus to the study of PECs at the critical distance, where ionic and covalent curves cross (Rose et al, 1988). Here, quasi-classical approximations are used to describe long-range phenomena (Aquilanti et al., 1997, Garraway and Suominen, 1995, Remacle and Levine, 1999, Hutchinson et al., 1999). Long-range potentials explain the physics of ultra-cold atoms (Zemke and Stwalley, 1999, Wang et al., 1997, Marinescu et al., 1994, Hajigeorgiou and Leroy, 1999, Stwalley and Wang, 1999). Fitting PECs at long-range requires accurate potentials and is a delicate matter, given the small energy differences. In this respect, the Dunham series (Dunham, 1932) is not useful at all, since the series does not converge. The Morse-function (Morse, 1929) is only reliable when thoroughly adapted (Hajigeorgiou and Leroy, 1999) and lacks a theoretical basis. Therefore a universal first principle's potential is badly needed as a reference for inverting observed levels into PECs and for studying long-range behaviour in particular. If a universal PEC is available, long-range behaviour must be assessable from short-range behaviour (the repulsive branch of a PEC). A correct quantitative evaluation of the long-range behaviour (Côté and Dalgarno, 1999) is also of importance to test QED (Quantum Electro-Dynamics) as in trapped deuterium (Schmidt-Kaler et al., 1992).

Finding a first principle's relation for short- and long-range atomic interactions is still a challenge, although the problem of calculating PECs can theoretically be considered as solved. Nevertheless, there is the question whether or not a universal potential or a species independent PEC exists, the *Holy Grail of Molecular Spectroscopy* (Tellinghuisen et al, 1989). This function should rationalise the behaviour of spectroscopic constants, account for the shape invariance of PECs and lead to global scaling. Exactly here, *the Heitler-London theory can not give a simple straightforward answer*.

Symmetry effects in particle systems show in PECs. Typically, a two-centre two-electron bond gives rise to two PECs (fermion behaviour): one for a repulsive triplet state, another for the stable singlet state (sigma-state). The two branches of the singlet-state PEC are not symmetrical with respect to the minimum but show left-right asymmetry (Herrick and O'Connor, 1998). In terms of the algebra of 1/R-potentials and according to convention, attraction follows -1/R, repulsion +1/R but this elementary symmetry is broken. The attractive branch has a finite asymptote at R = ∞, the repulsive branch almost invariantly goes to infinity at R = 0. Therefore, many empirical asymmetrical 1/R-potentials were suggested (Varshni, 1957, Steele et al., 1968, Varshni and Shukla, 1963). Most are successful for related (ionic) molecules.

The invariant shape and the asymmetric chiral behaviour of singlet PECs point towards a universal function. Shape invariance indicates that two-dimensional scaling should be possible (Varshni, 1957, Calder and Ruedenberg, 1968, Jenc, 1990, 1996, Graves and Parr, 1985, Tellinghuisen et al., 1989, Van Hooydonk, 1999) but in practice it is not. Therefore the final solution is *doomed* to be generic, i.e. hidden in first principles, but this truly universal, perfectly scalable *ab initio* function still remains to be found.

Claims have been made that three, probably four or even more molecular parameters are needed for a universal function (Varshni 1957, Graves and Parr, 1985). The 3 standard parameters are the equilibrium inter-nuclear distance $R_e$, the dissociation energy $D_e$ and the force constant $k_e$. But we showed recently that even a two-parameter function can have universal character (Van Hooydonk, 1999). Unfortunately, the H-L theory can not help to solve this problem, as it is impossible to derive analytically a universal function from this theory. Traditionally, we have a right to expect that the complete theory contain well behaving empirical relations found previously. This is not so. Despite the many good empirical 1/R functions available, the H-L theory can not at all predict whether a function A will perform better than B and why this is so. We therefore wonder if the H-L theory is really complete.

With a universal function f(x), all observed PECs must reduce to a perfectly symmetric and linearly scalable V-shape with a slope equal to one. Algebraically, the two branches of different PECs should reduce to a single straight line with variable x.

*Despite all previous efforts, we show that molecular PECs are quantitatively dominated by the universal Coulomb 1/R-potential. Using classical gauge symmetry, we prove that the universal molecular function derives from a pair of symmetric Coulomb potentials. This symmetry has to be broken by a perturbation mechanism that must fit into the classical Hamiltonian. A well known yet overlooked form of chiral symmetry can achieve this, i.e. the handedness of atoms (their mirror symmetry).*

But a static Coulomb law seems much too old-fashioned and even inappropriate, as no dynamics is involved. Yet this law contains interesting continuous, discrete and scaling ingredients:

- continuous: the 1/R-dependence,
- discrete: 1/R is a power law: only *a positive world* is allowed*; a symmetry (parity) effect*: attraction and repulsion; a *species independent* unit of charge, and

- scaling: the asymptote of a Coulomb system is determined by $R_e$, which makes it perfectly suited for two-dimensional scaling (scale invariance). In addition, Coulomb's law is valid both on the micro- and macro-scale and explains both the short and long rang interaction of charged particles. This issue is of central importance for a quantitative assessment of short- and long-range behaviour (Côté and Dalgarno, 1999).

The present contribution deals with many different applications of Coulomb's first principle's law in chemistry and physics. We follow a classical procedure. Elementary steps show what kind of symmetry is generically broken and why Coulomb's law can indeed be a universal *molecular* function, which allows perfect scaling. Useful references are Bouchiat and Bouchiat (1997) for parity violation in atoms, Varshni (1957), Tellinghuisen et al. (1989) for PECs, Bakasov et al. (1998) for parity breaking in molecules, Garraway and Suominen (1995) for femto-chemistry. Recent work by Hajigeorgiou and Leroy (1999), Stwalley and Wang (1999), Côté and Dalgarno (1999) reviews long-range potentials.

Section 2 gives a summary of observed PECs and empirical potentials and the effects of gauge symmetry in general. Sections 3-9 contain the elementary steps to arrive at theoretical Coulomb-based PECs: gauge symmetry for discrete and continuous elements of Coulomb's law, perturbation theory, and the parity violation adaptations for the classical Hamiltonian of a four-particle system. Section 10 gives the theoretical results on the universal function, whereby a generic perturbation is identified. The results are confronted with experiment in section 11. Here we show how well a zero molecular parameter function fits experimental PECs and how 13 different PECs can be brought back into a single straight line. Section 12 discusses generic effects of charge inversion. Sections 13-14 deal with other consequences.

**2. Observed PECs: potentials and scaling**

*2.1. Potentials for a four-particle system. Asymptotes. Series expansions*

Denoting the lepton-nucleon system in atom X as (a,1) and in atom Y as (b,2) the interatomic potential V(R), deriving from the Hamiltonian $H_{XY}$, is

$$V(R) = H_{XY} - (H_X + H_Y) = -e^2/R_{1b} - e^2R_{2b} + e^2/R_{12} + e^2/R_{ab} \qquad (1a)$$

The asymptote is assumed to be the atomic dissociation limit or $D_e = -V(R_e)$. This assumption is probably not true. V(R) consists of 4 potentials, only 1 is related to the internuclear separation $R_{12}$, the standard variable for PECs. No information is available about this potential or its character, except that it is zero at infinite internuclear separation, which is trivial. There is no hint as to triplet-singlet splitting. To decide whether V(R) is basically attractive or repulsive, data for the singlet-state at equilibrium are available but the minimum must be supposed to be generic. In first order $R_{1b} = R_{2b}$, $R_{12} = R_{ab}$ and $R_{12} = R_{ab}$   $2R_{1b} = 2R_{2b}$ at the minimum $R_e$. $V(R_e)$ is then in a good approximation equal to

$$V(R_e) = -2e^2/R_{12}$$

or, due to nucleon-lepton attractions, V(R) is attractive between $R_e$ and   as expected. $V(R_e)$ is two times the asymptote of two charges in a Coulomb model, 2Ryd or 220000 cm$^{-1}$. This value points to the absolute well depth $H_{XY}$, the asymptote $IE_X + IE_Y + D_e$ ($IE_X$ is the ionisation energy of atom X), rather than to the atomic dissociation limit $D_e$. In fact, the maximum value for the dissociation energy $D_e$ of bonds between two monovalent atoms is about 50000 cm$^{-1}$. If true, the H-L Hamiltonian would be in error by 400 % in an otherwise legitimate approximation based upon available equilibrium data, which is impossible. As a result, Coulson (1959) said that *quantum mechanics weighs the captain of a ship by weighing the ship when he is and when he is not on board*. This means one has to solve the Hamiltonian first and then subtract the atomic energies. This is a cumbersome procedure, since observed PECs, the result of V(R) in (1a), are shape invariant with only $R_{12}$ as a variable.

In an ionic approximation, the potential V'(R) is

$$V'(R) = H_{XY} - (H_{X+} + H_{Y-}) = - e^2/R_{2a} - e^2/R_{2b} + e^2/R_{12} \qquad (1b)$$

As with (1a) singlet-triplet splitting is not obtained but the R-dependence is more specific. The well depth for an ionic system is $IE_X + EA_X + e^2/R_e$, if $EA_X$ is the electron affinity of X. We get

$$e^2/R_e = IE_X - EA_X + D_e \qquad (1c)$$

the classical ionic bond energy. At large R, this approximation (1b) gives $R_{2a} = R_{2b}$  $R_{12}$ or

$$V'(R>>R_e) = - e^2/R_{12} \qquad (1d)$$

which starts off as an ionic Coulomb attraction at the asymptote, although interchanging an inter-nuclear with a nucleon-lepton term is rather artificial. With the same conventions as for (1a), (1b) leads to

$$V'(R_e) = - 3e^2/R_{12}$$

about 3Ryd or 330000 cm$^{-1}$, larger than the covalent one, because a repulsive term in (1a) is suppressed. It seems that $V(R_e)$ refers to a well depth of order 2Ryd >> $D_e$. Then V'($R_e$) must be about half as large, 1Ryd or the ionic asymptote (1c), situated between 0 and the absolute well depth. Things go wrong when extrapolating this long-range ionic behaviour (1d) to the minimum. The only conclusion possible is that potential V(R) has an asymptote of order Ryd, in any case larger than $D_e$. But no information is found about the minimum, the existence of a triplet state or the shape of the singlet PEC. Bonding is secured by nucleon-electron interactions, conforming to the H-L theory, which leads to the cumbersome procedure referred to above. Just in the ionic one case, a classical picture (1d) emerged, which would lead to an acceptable asymptote of 1Ryd. Unfortunately, this is of restricted validity (just applicable at large R).

The behaviour of Coulomb PECs is illustrated in **Fig. 1**, where for comparison the semi-empirical RKR-curve (Rydberg, 1931, 1933, Klein, 1932, Rees, 1947) for $H_2$ is included (Weissman et al., 1963). We use the Coulomb asymptote 116400 cm$^{-1}$, twice of which is about the absolute well depth (order 1 hartree). The minimum derives from an algebraic Coulomb law |1-0,74144/R|, which seems like an artefact (see further below). The PEC with asymptote $D_e$ (38283 cm$^{-1}$) is computed similarly. In comparison with the RKR, the slopes of Coulomb PECs are much too large and the curvatures have the wrong sign. This is as far as one can go with Coulomb's law, the only first principle's law available for a system of four charged particles. Fig. 1 is the reference for this work. The situation is completely hopeless if one tries to explain bonding in $H_2$ with a static Coulomb law.

Nevertheless, for ionic bonds, such as alkali-halides, the middle PEC in Fig. 1 is a good first order approximation for the PEC *away from the minimum*, and remains useful for calculating ionic curves (Russon et al., 1997). The H-L theory accounts for the lower PEC for $H_2$ in Fig.1, not a Coulomb law situation but it can not account for ionic bonds obeying the Coulomb PEC in the middle of Fig. 1. Coulomb's law can account for bonds obeying this middle PEC but it can not account for the lower PEC for $H_2$. This dilemma led to an almost 100 year old compromise: there are *two* kinds of bonds, *covalent and ionic*. But exactly this compromise is in contradiction with spectroscopic evidence, the good empirical Coulomb 1/R potentials available, the invariant shape of PECs and the dependence of the harmonic frequency on 1/R for both ionic and non-ionic bonds (Van Hooydonk, 1999). This evidence is not covered by the H-L theory. In fact, at least molecular spectroscopy shows that there is no spectroscopic distinction whatsoever between covalent and ionic bonds: they behave alike when properly scaled (Van Hooydonk, 1981, 1999). A curious result is that this spectroscopic information invariantly points towards Coulomb's law and its asymptote (middle PEC in Fig. 1) as being valid for all bonds, *ionic and covalent as well*.

Therefore, molecular spectroscopy indicates that the H-L theory may not be complete indeed. Additional empirical evidence can be found in the many persistent studies on scaled or reduced potentials. The question of bonding may be solved in principle and calculating PECs may no longer be a problem, this kind of empirical research is going strong for decades, see Frost and Musulin (1954), Varshni (1957), Steele et al. (1968), Jenc (1988), Zavitsas (1992), Tellinghuisen et al. (1988), Graves and Parr (1986), Jhung et al. (1989) and Van Hooydonk (1999). One is entitled to do so: a universal potential lies probably hidden in the molecular spectra but the H-L theory is unable to identify this function.

The most universal potential imaginable is a *first principles* local, static, mass-less Coulomb 1/R potential: exactly this type of potential is among the favourites in *empirical* approximations. But a simple one term Coulomb law is much too rigid and probably not flexible enough. It is commonly generalised using a power series. This has disadvantages (Varshni, 1957), but the popular Dunham potential (1932) is of this type. Consider the three series

$V(R) = a_n R^n + a_{n+1} R^{n+1} + a_{n+3} R^{n+2} + a_{n+3} R^{n+3} + \ldots$

$V(R_e) = a_n R_e^n + a_{n+1} R_e^{n+1} + a_{n+3} R_e^{n+2} + a_{n+3} R_e^{n+3} + \ldots$

$V(R)-V(R_e) = a_n(R^n - R_e^n) + a_{n+1}(R^{n+1} - R_e^{n+1}) + a_{n+3}(R^{n+2} - R_e^{n+2}) + a_{n+3}(R^{n+3} - R_e^{n+3}) + \ldots$ (1e)

The third series gives a different picture than (1f), a series expansion in $(R-R_e)$

$V(R-R_e) = a'_n (R-R_e)^n + a'_{n+1}(R-R_e)^{n+1} + a'_{n+2}(R-R_e)^{n+2} + a'_{n+3}(R-R_e)^{n+3} + \ldots$ (1f)

Only for n = 1 the corresponding two terms in (1e) and (1f) are identical. Starting a function at any n does not lead to loss of generality. In practice, it is convenient to use n = 2 in (1f) to obtain oscillator models (Varshni, 1957, Dunham, 1932). But, the variables in (1e) and (1f) can be scaled in two mathematically equivalent ways. With the Dunham-variable

$d = (1-R/R_e)$ (1g)

and n = 2, potential (1f) starts off at

$V(R) = a'_2 R_e^2 d^2$ (1h)

and (1e) starts with $R_e^n((R/R_e)^n - 1)$. Using the Kratzer (1920) variable

$k = (1-R_e/R)$ (1i)

and n = 2 also, potential (1f) starts off at

$V(R) = a'_2 R^2 k^2$ (1j)

identical with (1h). Using k, (1e) starts with $R^n((R_e/R)^n - 1)$. Both (1h) and (1j) imply harmonic oscillator behaviour in function of $(R_e-R)^2$ in the (1f) expansion. The distinction between (1e) and (1f) and between d and k may seem subtle but it is not: it produces different short- and long-range behaviour, see below. A closed formula like Coulomb's, i.e. n = -1 in (1e) and neglecting all other terms, is by all means more challenging and interesting, since convergence problems are avoided, although both (1e) and (1f) are more flexible. We choose for analytical rigour instead of flexibility from the start and use a few single terms in (1e) as a starting point in our analysis. This choice must lead to mathematical, physical and chemical problems, as all coefficients in (1e) and (1f) are known to be species dependent.

*2.2. Empirical potentials and scaling. Short- and long-range behaviour. Observed PECs. Benchmarks*
The classical Born-Landé (1918) function

$$V(R) = -e^2/R + B/R^n \qquad (2a)$$

for a singlet PEC uses 2 non-consecutive terms in (1e) with exponents -1 and -9 (n is accessible through compressibility measurements). Function (2a) gives reasonable PECs for ionic bonds without computational difficulties, which is amazing if we recall the complexity of quantum-mechanical calculations and compare the analytical forms of (1a) and (2a). Its eigenvalue is about $0,9e^2/R_e$, close to the generic Coulomb asymptote, see section 2.1. Any PEC generated by (2a) is close to the middle PEC in Fig. 1. But this is not the end of the story. The predictions of (2a) for spectroscopic constants are reasonably accurate *for ionic and covalent bonds* (Van Hooydonk, 1982, 1999).

Kratzer (1920) introduced an even more intriguing potential, with n in (2a) equal to 2 and for which the wave equation can be solved (Fues, 1926). The Kratzer potential

$$V(R) = -Ae^2/R + B/R^2 \qquad (2b)$$

uses 2 consecutive terms in (1e) and can directly be rewritten in reduced form

$$U(R)/(Ae^2/2R_e) = V(R)/(Ae^2/2R_e) + 1 = (1-R_e/R)^2 \qquad (2c)$$

(for references see Znojil (1999) and Van Hooydonk (1999)). This strange potential (2c) has always been overshadowed by Morse's (1929). It is a generalised Kepler system (Infeld and Hull, 1951) but it also mixes atomic and molecular behaviour. With $A = 1$ and $R_e = 1r$, where r is the atomic radius, it is an *atomic potential*, a generalisation of the Bohr equation. This shows after taken the first derivative in function of R. With $A = 2$ and $R_e = 2r$, (2c) is a *molecular potential*. These *generic* aspects of Kratzer's potential are discussed further below. In practice, around $R_e$, the RHS of (2c) secures the PEC shows an-harmonic oscillator behaviour with left-right asymmetry, always better than a harmonic oscillator. The repulsive quadratic term $(R_e/R)^2$ refers to the kinetic energy of interacting particles, the Planck-Bohr quantum condition for central force systems. A generalisation of (2c) due to Varshni (1957) is consistent with the spectroscopic constants of hundreds of bonds (Van Hooydonk, 1999). Varshni introduced an exponent v for $R_e/R$

$$U(R)/(e^2/2R_e) = V(R)/(e^2/R_e) + 1 = (1-(R_e/R)^v)^2 \qquad (2d)$$

This *two* parameter ($R_e$ and v) Kratzer-Varshni-potential (2d) almost behaves like a universal function and is superior to Morse's *three* parameter potential (Van Hooydonk, 1999). Morse's and (in part) Dunham's oscillator models (1h) would be perfect if there was left-right symmetry in PECs. In Dunham's case, deviations from left-right asymmetry leads to the cumbersome series of Dunham coefficients, all needed to get only moderate agreement with observed PECs and a bad convergence. Moreover, the elementary connection with the energy consequences of 1/R-potentials *seems* to be lost (Van Hooydonk, 1999). Exactly these form the basis of interactions at the Bohr/Fermi-scale. Conversely, the invariantly observed left-right asymmetry of singlet PECs at $R_e$ gives an idea about the nature of the interactions.

The Morse potential is

$$W(R) = D_e (1-e^{-d})^2 \quad (2e)$$

with a species dependent constant. Dunham's is

$$W(R) = a_0 d^2(1 + a_1 d + a_2 d^2 + ...) \quad (2f)$$

where $a_n$ are the so-called Dunham coefficients, related to the spectroscopic constants. This function can never converge at large R.

For long-range interactions the situation gets more complex in general, since these lead to PECs described by functions like $V(R) = D_e - C_n/R^n$, with n >> 1 and $C_n$ a parameter. Many empirical fitting procedures have been presented in the literature but some lead to 'pathological' behaviour (Coxon and Hajigeorgiou, 1991). The best known are the Ogilvie-Tipping anharmonic oscillator (Ogilvie, 1988), the generalised Morse oscillator (Coxon and Hajigeorgiou, 1990, 1991), the modified Morse oscillator (Hedderich et al., 1993) and the modified Lennard-Jones oscillator (Hajigeorgiou and Le Roy, 1999), the latter being a mixture of Morse and Kratzer potential elements. Most of these fitting procedures are based upon the Morse potential, which is inferior to Kratzer's when it comes to rationalise the behaviour of the lower order spectroscopic constants (Van Hooydonk, 1999). The Morse-function is confined to the observed dissociation limit $D_e$, which guarantees it will always converge to unity for RKRs scaled with $D_e$.

**Fig. 1b** illustrates the observed situation. RKRs reduced with $D_e$ are shown for 13 bonds (details are given below) in function of the reduced distance $R/R_e$. Although both the y- and x-axis are scaled consistently, the PECs do certainly not coincide. Nevertheless, the shape invariance referred to above shows clearly and needs to be explained. The Morse function f(R) in (2e) is a better measure for the x-axis. If the Morse function is universal, as claimed by Jhung et al. (1989), all PECs of Fig. 1b should reduce to a perfectly symmetric V-shape with perpendicular legs with the Morse function at the x-axis. The actual result is shown in **Fig. 1c**. The required symmetric V-shape is only obtained in the vicinity of $R_e$. At the attractive side, the agreement seems satisfactory, although this is for the larger part due to the fact that the asymptote of Morse's function is $D_e$. Despite this 'imposed' asymptote the different PECs do not collapse into a single line, although all have a slope near unity. It would appear that Morse's function describes long-range behaviour rather well but it does not. This moderate agreement is not confirmed by the data at the repulsive branch. Here the slopes show large divergences. At the extreme short side, 'turn over' points appear. This leads to a strange almost contradictory situation: the 'simpler' repulsive branches are not well reproduced by Morse's function, whereas the description of the complex attractive side, where long-range potentials interfere, is better. Nevertheless, Fig. 1c sets the standard for other analytical potentials. Morse's function (2e) is *exponential* in R and uses *three* parameters, $R_e$, $D_e$ and $a_0$ to get only moderate agreement with experiment.

This shows even better in **Fig. 1d**, a linear plot of algebraic attractive and repulsive branches against the algebraic Morse function. The relative good V-shape in Fig. 1c is not confirmed, since a straight line is not obtained. This imposes restraints on the universal character of the 3-parameter Morse-function and sets a clear benchmark for other scaling approaches. Morse's function is more complex than Coulomb's but it certainly does not result in perfect scaling. It can not properly account for the 'simple' repulsive branches, which are more suited for scaling (Gardner and Von Szentpaly, 1999).

With this experimental background and these benchmarks in mind, we start from scratch. We eliminate all parameters and test the only parameter-less first principles potential available: Coulomb's. A few symmetric single-term $R^n$-potentials appearing in the series (1e) will illustrate the procedure. Two-dimensional scaling is essential. The consequences of the next and ultimate step i.e. perfect symmetry between attraction and repulsion, the basis of Coulomb's law, is a challenge. But the major problem with a pair of symmetric potentials is $V(R) = 0$ (annihilation) for all R in (2a) and (2b). But working with a closed formula, devoid of any parameter, may lead to the simplest analytical form possible for a universal potential and to that single symmetry element in the Hamiltonian, needed to account generically for shape invariant asymmetrical molecular PECs.

*2.3. Classical gauge symmetry and two-dimensional scaling*

Consider two particles with identical masses m interacting through a potential V(R), for which the classical Hamiltonian $H_{12}$ reads

$$H_{12} = \tfrac{1}{2}mv^2 + \tfrac{1}{2}mv^2 + V(R) \tag{3a}$$

The reduced mass would be equal to 0,5m. The particles interact through a single term in (1e)

$$V(R) = a_n R^n \tag{3b}$$

with n an integer, $a_n$ a potential dependent constant (with dimensions energy *times* length$^{-n}$) and R the particle separation. Let n be equal to - 2, -1, 1 or 2 (n = 0 leads to a one-dimensional PEC). The x-axis R is scaled by a characteristic distance for the system, $R = mR_e$, giving

$$V(R) = a_n R_e^n m^n \tag{4}$$

A pair of symmetric potentials (attractive negative, repulsive positive) can be created using a parity operator $\boldsymbol{P}^2 = 1$ or $\boldsymbol{P} = \pm 1$. With $(-1)^t$ this gives

$$V_+(R) = a_n R_e^n m^n = -V_-(R) \tag{5a}$$

$$V_\pm(R) = (-1)^t V(R) \tag{5b}$$

where the exponent t, the **t**ype of interaction, is equal to 0 (even) for repulsion and 1 (uneven) for attraction. The zero reference point is free (gauge-symmetry). Adding a constant C gives

$$W(R) = V(R) + C \tag{6}$$

and by virtue of (5a)

$$W_+(R) = -W_-(R) \tag{7}$$

independent of the value of C. Scaling C on the y-axis in terms of a typical energy for the system

$$C = a_n R_e^n = V(R_e) \tag{8a}$$

leads to two scaling operations on an equal and consistent basis (two-dimensional scaling, gauge symmetry, see Introduction). C is the asymptote and is determined by

$$C = |V(\ ) - V(R_e)| \tag{8b}$$

which shows why its sign is a matter of convention. The levels $\pm C$ are symmetrically distributed around the original x-axis, the gap being |2C|. Using a second parity operator $(-1)^g$ for C leads to four different symmetric states for (3b) potentials

$$W(R) = \pm a_n R_e^n \pm (a_n R_e^n) R^n / R_e^n$$
$$= C(-1)^g + C(-1)^t R^n / R_e^n$$

$$W(R)/C = w(m) = (-1)^g (1 + (-1)^{t-g} R^n/R_e^n)$$

$$w(m) = (-1)^g (1 + (-1)^{t-g} m^n) \qquad (9a)$$

if the exponent g for the **g**auge is equal to 0 (even) or 1 (uneven). Result (9) represents the algebraic effect of classical gauge symmetry for potentials: it is a well-known generic result, independent of the analytical form of the potential. The R- or m-dependence is extremely simple and is the same for all four states and the scheme is scale invariant. The distinction between the four states is only due to *one* algebraic symmetry operator, parity. The meaning of the four states is discussed below for a Coulomb potential. These scaling and symmetry operations artificially *create* two symmetric isospectral worlds W < 0 and W > 0 but only one can be real. There are mathematical techniques to arrive at the same results without negative worlds, see below. Classically, parity is never violated. If the two worlds remain separated and PECs do or can not cross, parity is not violated and convention is sufficient. But, two symmetric t - g = 1 states (asymptote and interaction have opposite signs), always cross at $R = R_e$ or at W(R) = 0. For mass-less particles and for systems with $m_1 = m_2$, symmetry may never be broken and no stable states are produced.

States with t - g = 0 can never cross. The two symmetric states with t = 0 and t = 1 around + C and - C will never cross. Symmetry breaking for two crossing t - g = 1 states is necessary, independent of n in (3b).

If a perturbation P is present for these two t - g = 1 states, scaling P with C gives p = P/C and applying the non-crossing rule to a symmetric pair gives in first order

$$w'(m) = ((1 - m^n)^2 + p^2)^{1/2} - p \qquad (9b)$$

as a generic perturbed function. A 'classical' definition of p is needed, as is its dependence on m.

*For a pair of algebraic potentials, crossing of symmetric states at the minimum is generic and the breaking of this symmetry must therefore also be generic. This can only be achieved by finding a generic perturbation. The physical origin of this must be found. If so, perfect scaling is theoretically obtained.*

*2.4. Symmetry of attraction and repulsion*

The symmetry of attraction $-R^n$ and repulsion $+R^n$ in (5) does not lead to simple solutions for N-particle systems. It is more practical to use different n-values or analytical expressions for attraction and repulsion, which is contrary to their algebraic symmetry. This is a characteristic of many empirical approaches. A typical one consists in choosing *two* terms in (1e), each having a different sign. For instance, (2a) and (2b) both use n = -1 for attraction but n-values different from -1 for repulsion, i.e. -9 and -2 respectively. Quantum mechanics, where a 1/R-dependence is invariantly applied for attraction and repulsion, shows that calculations can get complicated, see (1a) and Fig. 1a on account of this basic symmetry. Obtaining theoretical PECs is difficult in contrast with (2a) and (2b). Some of the difficulties in trying to find solutions for N-particle (particle-antiparticle) systems have been studied before (Richard, 1992, 1993, 1994, Abdel-Raouf, 1992, Abdel-Raouf et al, 1998). In this work, this algebraic symmetry is respected throughout, as it is a first principles element of Coulomb's law, conforming to gauge symmetry, and secures that scaling can not lead to distortions when going from the left to the right branch of PECs.

*2.5. Theoretical shape invariant PECs from a pair of symmetric $a_n R^n$ potentials and symmetry breaking*

At this stage, the shape of PECs generated by closed formula algebraic functions (9a) can only be discussed in the *ad hoc* hypothesis that crossing of t - g = 1 curves is avoided, see (9b). The five major invariant characteristics of PECs are:

(a) triplet-singlet splitting occurs at large R; (b) the triplet-state is repulsive (unstable state); (c) the singlet-state is attractive (stable state) and shows a minimum with asymmetric left and right branches at either side of the minimum; (d) the left branch is repulsive with a very large slope and does not reach a finite asymptote (it reaches infinity at R = 0 or at least becomes extremely large before eventually reaching the united atom energy at R=0 or even lower); (e) the right branch is attractive with a slope less than in the left branch and reaches a finite asymptote at R = ∞ (for other minor characteristics, see Varshni, 1957).

The general behaviour of (3b) in terms of (9) must be discussed referring to these characteristics (a)-(e). For the singlet-state, a reference PEC is generated with the reduced Kratzer potential (2c). Crossing is avoided by using a small constant *ad hoc* perturbation $p^2 = 0.1$ with C = 1 instead of the *generic* perturbation needed in (9b) but not yet identified.

Results are shown in **Fig. 2 and Fig. 3**. The morphology of PECs deriving from n > 0 potentials in Fig. 2a (n = 1) and 2b (n = 2) is not consistent with patterns (a)-(e). Although a splitting and a minimum is generated, the shape of the bonding PEC is wrong (reversed) at the tree level. Even the n = 1 case in Fig. 2a, the better of the two with respect to the Kratzer PEC, leads to an opposite picture. The situation gets worse for the n = 2 potential in Fig. 2b, although this is exactly the potential for the harmonic oscillator. This n = 2 potential is used frequently as a model in a variety of symmetry problems (see Witten, 1981, 1982, for a classical example in SUSY). The left branch instead of the right would give the finite asymptote at R = 0. The right instead of the left branch goes to infinity for large R. This apparent reversed left-right asymmetry simply calls for a switch from n > 0 to n < 0 potentials. In addition, it seems unlikely that any justifiable non-crossing scheme (perturbation) would improve the situation with respect to the Kratzer-reference PEC, see Fig. 2a and 2b. *But perturbation alone can not remedy the wrong left-right asymmetry inherent to n > 0 potentials*. Part of this so-called wrong behaviour has to do with convention: 1/R as a variable instead of R for n = 1 would give the correct result, see below. Fig. 2a is a Landau-Zener model: curves 2 and 3 are diabatic levels, curve 6 is (one of the two) adiabatic levels in (9a).

The morphology of PECs generated with n < 0 potentials in Fig. 3a (n = -1) and Fig. 3b (n = -2) is in line with (a)-(e) and conforming to the Kratzer singlet PEC. At the tree level, the n = -1 potential performs better than the n = -2 potential, if the Kratzer potential is used as a reference. As above, avoiding crossings is illustrated in Fig. 3a and 3b. In Fig. 3a, the resulting PEC is close to the Kratzer-approximation for larger R but gets automatically worse in the left branch. For the repulsive (triplet) state, the generic shape is almost correct by definition in the case of an n = -1 potential. At the tree level, a pair of symmetric 1/R-potentials leads to the correct *shape* of PECs in all aspects (a)-(e). Fig. 3a is a Demkov-Kunike model: also here curve 2 and 3 are diabatic levels whereas 6 is (one of the two) adiabatic levels in (9a).

Splitting, the shape of the triplet state, the existence of a minimum for the singlet state and the left-right asymmetry at the minimum all secure that, in first order, a pair of symmetric 1/R potentials is at work in cases for which Kratzer's potential gives the reference PEC. This general and very consistent model is one of the main reasons why people remain interested in empirical 1/R potentials (Fig. 3a)

24/01/00                                     14:26 G. Van Hooydonk                                     11

hoping one of them has universal characteristics. Parameters give the flexibility needed to apply the potential to more than one bond. We choose for rigour, by eliminating parameters, to find out what a truly generic one-term potential is capable off.

The main difficulty of scaling in molecular spectroscopy is to find a (generic) species independent variable which would allow a smooth transition from Fig. 3a (1/x-situation) to the perfect V-shape of Fig. 2a (x-situation) using the same experimental data, see above Fig. 1c-e. In addition, closed formula potentials, conforming to observations (a)-(e), as in Fig. 3a, cause problems for the H-L theory, bound to the complex potential V(R) in (1a). The internuclear 1/R potential in the present scheme is just one out of the four potentials in (1a). Exactly this important internuclear potential in (1a) is repulsive instead of attractive.

*2.6. Symmetry breaking and gauge symmetry*

The n = -1 potential is a power law, confined to a positive world. However, the convention is that two symmetric ± solutions are possible (attraction and repulsion). This can only be achieved by shifting the zero of the system, C in (9), and allowing worlds wherein attraction and repulsion are possible at the same time (classical gauge symmetry and a characteristic of potential theories). *The maximum range R for attraction is then automatically governed by the asymptote C*. Therefore, at the zero, the symmetry of attraction and repulsion must be broken in order for the conventions to remain valid. For n = -1 potentials, finite reference points have been replaced by asymptotes. The n = -2 case runs similar. The next problem with the two n < 0 worlds is scaling, since asymptotes are now the reference points, needed for scaling. Mathematically, there is no substantial difference between n = 1 and n = -1 cases, since we can always replace x with 1/x. The use of 1/x is more a tradition (Newton, Coulomb), see above. With respect to scaling, the information contained in PECs is more manageable with x as a variable rather than with 1/x, especially with respect to the asymptotes. Conflicting situations can occur for systems subject to two different laws. In classical and wave mechanics, n = 2 behaviour is needed for the kinetic energy, n = -1 behaviour for the potential. Conventions about the real world must be in line with these two behaviours, if both are allowed. If a conflict occurs, the symmetry of the behaviour that causes the conflict must be broken immediately at the critical point where the conflict occurs. *Determining what exactly this symmetry breaking effect means in terms of the physics of the interactions between particles is the next problem to be dealt with*. Gauge symmetry must also be confronted with the Hamiltonian to find out if this symmetry breaking effect can be incorporated. In the H-L theory, there is no *direct* link *within the Hamiltonian* to symmetry breaking effects leading to (a)-(e) as observed in PECs.

**3. Algebraic Coulomb 1/R-potentials. A new degree of freedom. Perfect Coulomb scaling**

Consider a system of two charges, interacting through a Coulomb 1/R-potential, see Fig. 3a. The charge symmetry is not (yet) broken by the (identical) particle masses giving

$V_{\pm}(R) = \pm e^2/R$ (10a)

in which, according to convention, the + sign refers to charges with equal sign (repulsion), the - sign to charges with opposite sign (attraction).

The problem case is

$$V(R) = -e^2/R + e^2/R = 0 \quad (10b)$$

when referring to (2a) and (2b), for any R. **Fig. 4** represents the general shape of these classical Coulomb PECs, which can not show a minimum, since they never cross (fermion-behaviour). Due to charge invariance, the four states ++, +-, -+ and - - reduce to two degenerate pairs, +-, -+ and ++, --. Generalising (10a) is possible using the procedure of section 2.3

$$W(R) = V_\pm(R) \pm C = \pm (e^2/R_e)(1 \pm R_e/R)$$

$$W(R)/(e^2/R_e) = w(m) = \pm(1 \pm R_e/R)$$

$$w(m) = \pm(1 \pm 1/m) = (-1)^g (1 + (-1)^{t-g}/m) \quad (11)$$

The Coulomb asymptote C can be *divided or multiplied* with any constant without loosing generality, since $R_e$ is *multiplied or divided* by the same constant. Result (11) is illustrated in **Fig. 5**.

*Referring to Fig. 3a, Coulomb forces imply a new degree of freedom.* The symmetry of attraction and repulsion resides in the interaction of *two* charges. Any system of *two* interacting charges obeys charge invariance. This principle secures that, within a given world, the energy of the system is not altered when the signs of two interacting charges are interchanged, say from +- to -+ for attraction or from + + to - - for repulsion. But, a transition from a positive world into a negative one as in Fig. 5 affects *all* symmetry aspects. At the minimum, attraction (+-) changes into repulsion (- -), if the original conventions hold. A symmetry breaking effect would explain the minimum, the asymmetric branches and the different slopes in observed PECs automatically. But there is no reason whatsoever for the interaction of two charges in the real world to suddenly change from attraction into repulsion. The effect of the two worlds is that, in contrast with (9), the four states in (11) are classically identified and further diversified, i.e. ++, +-, -- and -+, exactly in this order when starting from asymptote +C, see Fig. 5 and the arrow notation therein. As a result, Fig. 5 is more complex, since the sigma bonding state is now described, on account of gauge symmetry, by + - and - - states, whereas the two triplet states are ++ and - +. But the repulsive (- -) state +1/R - C, is also the *attractive* state in the negative world and would cross the (+-) *attractive* state -1/R + C of the positive world at $R_e/R = 1$. *But two states with the same basic gauge symmetry -attractive or t - g = 1- are not allowed to cross and perturbation must be invoked as in (9b).*

Looking at Fig. 5, the - - attractive state +1/R - C in the negative world is a repulsive state with a downshift with respect to the repulsive state +1/R + C in the positive world, the shift being equal to |2C|. For a system of *two* charges, all this may seem meaningless. Charge invariance allows a shift from + - to - + for attractive states without energy implications. But gauge symmetry overrules charge invariance: now there are two not degenerate attractive (repulsive) states + - and - - (- + and ++) with totally different energies away from $R_e$ (Fig. 5).

*But we know that switching fermion chiralities simply corresponds with switching a sign in the Hamiltonian, which is the result of charge conjugation combined with the particle-hole transformation* (Neuberger, 1999). *Which sign(s) must be switched will be demonstrated below. It is important to realise that PECs generated by gauge symmetry (Fig. 3a) are observed for systems consisting of four particles, such as chemical bonds. Therefore, we must find out how the gauge symmetry dictated scheme can be translated into a physical law in these four-particle systems.*

As in SUSY, the problem is to find a real N-particle system, where this mathematical artefact applies. For a system of two charges, it is impossible, in the absence of external effects, to imagine any perturbation, if self-perturbation is excluded. Positronium and protonium systems are extreme but straightforward examples. The 1/R attraction is used in full and can not, at the same time, perturb itself. The origin of the minimum remains a mystery and must be considered as generic, since only an electrostatic 1/R potential is used to describe the system (dynamics is not -yet- involved) but we know $R_e$ derives from classical equilibrium conditions, involving dynamics.

Up till now only a generic *ad hoc* $V(R_e)$ is available as a Coulomb asymptote $C = e^2/R_e$, also present in (2a) and (2b) and this derives naturally from gauge symmetry (parity). Classical equilibrium conditions in *two* particle central force systems respect the virial theorem and lead to an asymptote, $(1/2) e^2/R_e = C/2$, although multiplying $e^2/R_e$ with any constant (as the virial ½) does not alter Coulomb scaling.

Returning to two-dimensional scaling, (11) applies to any asymptote $C_b = C/b$, where b is a constant, see Introduction. The R-dependent part of (11) will still vary as $R_e/R = 1/m$ for any b-value. As a result, w (m) may be multiplied with any constant without loosing generality or universality. The b-values will affect the shape of unscaled PECs, as illustrated in **Fig. 6a, b** and **c**. Fig. 6a shows that, with the same Coulomb 1/R potential, a smaller asymptote reduces the slopes of the branches and increases $R_e$-values. Scaling R with $R_e$ (Fig. 6b) draws the attention exclusively to asymptote differences. Perfect two-dimensional scaling results in a reduced universal PEC as in Fig. 6c for any b-value. If b is very small, physics at the Fermi- and, in the limit, even at the Planck-scale is obtained. If b is very large, we would eventually get physics at the Newton-scale, see below. *Fig. 6c results from perfect two-dimensional scaling, the key being the universal character of the 1/R potential, its closed analytical form and gauge-symmetry.* These three Figures represent the major issue of scaling, symmetry and the n-dependence for R in (1e) or (3b). Even for seemingly weak interactions as in Fig. 6a (large $R_e$-values, small slopes) or as in Fig. 6b (small asymptotes), it is tempting but not always necessary to invoke a different n-dependence on R in the potential (1e) to account for these effects. But if gauge symmetry is applicable, a unique universal Coulomb 1/R law applies to all cases, see Fig. 6c.

The dissociation products in a Coulomb system must be charged particles. This is not trivial: in practice, observed molecular PECs are usually scaled with the atomic dissociation limit $D_e$, where the dissociation products are two *neutral* atoms. Theoretically, these can never represent the natural eigen-value of a true Coulomb system. Exactly the dissociation limit of four-particle systems (chemical bonds and the long-range behaviour of atomic interactions) cause severe problems when interpreting RKRs. As a matter of fact, Fig. 1b clearly shows that $D_e$ is certainly not a Coulomb scaling asymptote in the strict sense as the 1/R behaviour is not reproduced. The long-range behaviour is usually described by terms in (1e) of the form $C_n/R^n$, where n is large, see above. The range can be estimated by means of the Le Roy radius (Le Roy, 1973). Then it is of primordial importance to describe as exactly as possible the normal R-dependence of particles in this region, about which wave-packet dynamics (femto-chemistry) could provide us with new information. With a universal function it may be possible to scale long-range behaviour too, which, to the best of our knowledge, is not yet possible.

We applied this generic Coulomb scheme to see whether the exclusive use of R and $R_e$ leads to a generally consistent physical picture. **Fig. 6d** shows the behaviour of 13 observed RKRs in an algebraic Coulomb scheme with gauge symmetry (with a small perturbation added at the minimum). It is readily verified that this extremely simple 1/R and 2C-gauge system reproduces the general shape of observed molecular PECs. At the repulsive side, Coulomb's law is closely followed in the whole range. At the attractive side, the branches do not follow symmetrically Coulomb's law, probably due to long range forces. A perturbation is needed to shift the minima of the PECs upwards by a small amount, as in Fig. 3a. This gives a better correspondence with Coulomb 1/R behaviour at both branches. The underlying perturbation mechanism is very important, but must be identified. Fig. 6d reveals that PECs may indeed validly subject to a Coulomb scaling process (Van Hooydonk, 1999). The only exception seems to be the RKR for $I_2$, as it does not follow the general trend, especially when compared with $Li_2$, since bot have approximately the same $R_e$-values of 2,67 Å. It appears that, if we scale locally all observed PECs with $D_e$, all RKR-data-points will be contained between 0 and 1, irrespective of the value of $R_e$. To get a scaled result, all PECs reduced with $D_e$ must somehow also be rescaled using relative $R_e$ values. For instance, bringing up (scaling) the $Rb_2$ RKR to the position of that of $H_2$ in Fig. 6d does not depend on the value of $D_e$ in the first place but on $R_e$. How to achieve this is illustrated below.

**4. Problems with generic shape invariant PECs and a four-particle Hamiltonian. The switch**

Reconstructing *qualitatively* the shape invariance of observed PECs is no problem, as long as symmetry is broken: gauge symmetry and a pair of symmetric 1/R-potentials seems sufficient (see Fig. 3a and 6d) and guarantees Coulomb scaling for PECs. With respect to *quantitative* aspects (slopes and curvatures), problems are met if we confront this simple picture with the complex four-particle Hamiltonian
$H = ½ m_1 v^2 + ½ m_2 v^2 + ½ m_a v^2 - e^2/R_{1a} + ½ m_b v^2 - e^2/R_{2b} - e^2/R_{1b} - e^2/R_{2a} + e^2/R_{ab} + e^2/R_{12}$
Not less than 10 terms appear, 4 of which are kinetic energies and 6 are Coulomb potentials. We must select the signs to switch (see above) in the 6 charge conjugated terms. This complex Hamiltonian generates problems for a 1/R scheme dictated by elementary gauge-symmetry:

1. Prove that, like in (5), introducing a *parity operator* (a switch) in this Hamiltonian suffices to reproduce *quantitatively* shape invariant PECs. In the H-L theory, both splitting and the asymmetric minimum result from the combined (external) symmetry effects of wave functions and electron spin. *The symmetry breaking effect we need must be a generic consequence of a physical 1/R process.*

2. Prove that the resulting PEC is in a good first order approximation, obtained by a pair of *Coulomb 1/R potentials*. The H-L Hamiltonian contains 6 potential energy terms, only one describes the internuclear separation R, needed to construct PECs and is repulsive. At least this term needs a switch, a parity operator. *The asymptote must be identified.*

3. Find a suited, realistic and even generic *perturbation* to break the symmetry and to apply the non-crossing rule for the intersecting pair of t - g = 1 states in Fig. 3a, a problem in its own right. A perturbation does not leave the asymptote unaffected. Alternatively, we can look for lower states with the same symmetry, intersecting the generic state away from the minimum, which avoids crossing at the minimum.

4. Not the least, find *a real and observable N-particle system*, where these theoretical deductions apply. It seems impossible to avoid composite particle systems, be it alone to bring more complexity into the rigid analytical form of a simple Coulomb potential for a two-particle system.

## 5. A theoretical four-particle system. Coulomb-potentials and charge-distributions. Perturbation. The Born-Oppenheimer (B-O) approximation

*5.1. Symmetric perturbed Coulomb 1/R-potentials for a theoretical four- particle system*

A system with only two charges (*two fermions*) with particles of equal mass must be ruled out, see above. The next *neutral* system possible consists of at least *four* charges and must be subdivided in such a way that a Coulomb interaction between two point-like sub-particles emerges. The sub-particles must have a global mass asymmetry and carry opposite charges. Partitioning the four-particle system into two *neutral* subsystems, consisting of *two neutral* particles each, can never produce the Coulomb interaction $e^2/R$ needed. Therefore, the only possible and *ideal* partitioning leads to two asymmetric charged subsystems: a charged *composite* particle $X^{\pm}$ interacting with an oppositely charged *not composite* particle $Y^{\pm}$. The total charge of the composite $X^{\pm}$ particle being $\pm e$ implies that the number of charged particles in X is 3. These are confined to the X domain and their mutual separation is close to $R_e/2$, negligible at large R. The masses of $X^{\pm}$ and $Y^{\pm}$ must be nearly equal to secure that the charge symmetry will not be broken by the X-Y mass difference.

As in section 2.1, the 3 charges in X are labelled as (1,a,b) and the one in Y as (2). Since the total charge is zero, two pairs of fermions with opposite charges (a,b) and (1, 2) are needed. One pair resides within X, the other is distributed between X and Y. We can distinguish two charge distributions in the fermion pairs forming the neutral four-particle system:

(a) fermion pairs (2 leptons, 2 nucleons) have equal charges: $+e^2$ (++ and --, or vice versa);

(b) fermion pairs have opposite charges: $-e^2$ (+- and -+, or vice versa).

For the classical case (a) in this theoretical four-particle system, partitioned in this asymmetric way, the long rang interaction V(R) is equal to (1b), the standard ionic model discussed in Section 2.1.

For case (b) we get instead

$$V(R) = - e^2/R_{12} + e^2/R_{2a} - e^2/R_{2b} \qquad (12)$$

When compared with (1b), the three terms in (12) are the same but two of them have switched signs. The last two terms are equal in magnitude, the fermion pair (a,b) being confined to X. A small perturbation $V_{12}$ at large R results from (12)

$$V(R) = - e^2/R_{12} + (e^2/R_{2a} - e^2/R_{2b}) = - e^2/R_{12} + V_{12} \qquad (12a)$$

Apart from two signs, potential (12a) has the same structure as (1b). Therefore, only this theoretical four-particle system, case (b), would give rise to the situations already depicted in Fig. 3a and 5: a pair of two perturbed Coulomb potentials, if we only could justify *the switch* in sign for this part of the Hamiltonian. The standard non-crossing rule would simply lead to

$$W(R) = (C^2k^2 + V^2_{12})^{1/2} - V_{12} \qquad (12b)$$

in agreement with (9b) and used for obtaining Fig. 3a in the *ad hoc* hypothesis that $V_{12}$ is a small *constant*. **Fig. 7** gives the comparison of the results obtained with the generic potentials (12b) for $V_{12} = 0,1$ and 0,35 respectively and with the Dunham, Born-Landé and original Kratzer potentials of section 2.2. Reminding Fig. 1, the PEC for a theoretical four-particle system, (12b) may be close to reality, if the Born-Landé/Kratzer PECs are useful references. The famous oscillator PEC (n = 2 potential), the leading term in the Dunham expansion, is also shown in Fig. 7 (curve 1) but is completely out of range, at both short and long ranges.

*5.2. Can the real four-particle system consist of ions interacting through Coulomb's law? The switch*

Although the PEC generated by the present scheme seems to have prospects, see Fig. 7, composite asymmetric systems like those described in Section 5.1 may seem exotic. Fortunately, case (a) resembles systems very common in chemistry: two ions interacting through Coulomb's law. An anion is like the composite particle, say $F^-$ and a cation the non-composite particle, say $H^+$. Anions consist of three particles: one positive nucleon and two negative valence electrons. Strictly spoken a division into a molecular ion $XY^{\pm}$ and a lepton $e^{\pm}$ is also possible. But in that case the charge symmetry is spontaneously broken by the large mass difference, which must be avoided. In fact, this essentially atom-like interaction will generate an atom-like spectrum.

The masses of the partners, anion and cation, can not be identical. The mass difference is at maximum, when the nucleon masses are identical and small. For H it is 2/1836 or about $10^{-3}$. This difference leads to a small perturbation and to symmetry breaking as required.

Selecting the case of ionic bonding would secure that the non-crossing rule applies on account of (12), if only the signs of the terms in (12) were in agreement with those in (1b). But if all this can be rationalised, many PECs would be available for testing. The interactions confined to the X-domain do not vary with R. Hence $e^2/R_{1a}$, $e^2/R_{1b}$ and $e^2/R_{ab}$ are *frozen* in first approximation, an IIM (Ions In Molecules) approach (Van Hooydonk, 1999).

But since the theoretical system (12a) gives different signs for the potential than that for the classical ion-pair (1b), an ionic approximation seems to be forbidden on account of symmetry effects connected with charges, which reduces to a switch problem. These were discussed in Section 3 and we will now show how to remedy generically this symmetry- or parity-anomaly.

*5.3. Confrontation with the Hamiltonian*

The B-O approximation freezes nucleons and implies central force systems. Reduced masses appear. The eight terms in an order related to a standard AIM (Atoms In Molecules) approach are

$$H_{12} = (½\ _av^2 - e^2/R_{1a} + ½\ _bv^2 - e^2/R_{2b}) - e^2/R_{1b} - e^2/R_{2a} + e^2/R_{ab} + e^2/R_{12} \quad (13a)$$

The first 4 are the (intra-) atomic terms, the last 4 form a perturbation of the atomic states leading to V(R) in (1a). The inter-nuclear term is not dominant and is *repulsive* instead of *attractive*. With an IIM (Ions In Molecules) approach the eight terms in (13a) are rearranged as

$$H_{12} = (½\ _av^2 + ½\ _bv^2 - e^2/R_{1a} - e^2/R_{1b} + e^2/R_{ab}) + e^2/R_{12} - e^2/R_{2a} - e^2/R_{2b} \quad (13b)$$

The first 5 terms belong to the domain of the anion $X^{\pm}$, the latter 3 to the cation-anion interaction V'(R) in (1d). Although the relative influence of the internuclear term in (13b) is more pronounced than in (13a), it remains repulsive, not in agreement PECs either. To get an ionic interaction, dominated by $-e^2/R_{12}$ (case b. of the Section 5.1) the following identity could be artificially imposed

$$- e^2/R_{12} = e^2/R_{12} - e^2/R_{2a} - e^2/R_{2b}$$

leading to

$$e^2/R_{12} = (1/2)( e^2/R_{2a} + e^2/R_{2b}) \quad (13c)$$

The consequence of trying to reproduce an ionic attraction - $e^2/R_{12}$ in this way automatically rules out the possibility of finding a small perturbation term, which vanishes due to (13c). A way out is considering one of the nucleon-electron interactions in (13b), say - $e^2/R_{2a}$, to act as a *pseudo-ionic* attraction or

- $e^2/R_{2a}$ + ($e^2/R_{12}$ - $e^2/R_{2b}$)                                                                 (13d)

Rewriting the last 3 terms in (13b) accordingly gives the *almost* correct picture needed for the theoretical scheme (12a) and the correct perturbation result (12b). In fact the two terms between brackets in (13d) can be considered as a perturbation $V_{12}$ of the pseudo-ionic interaction - $e^2/R_{2a}$ but *it will only be small at large R*. This solution is artificial: we interchanged *nucleon-nucleon repulsion* and *nucleon-electron attraction* to get *pseudo-ionic* attraction. This artificial mechanism for going from cas (a) to case (b) in Section 5.1 does not give the correct switch. The charge symmetry of the system is broken, which is not allowed. Nevertheless, this solution would allow to apply the non-crossing rule as in (9b) or (12b), if only the internal consistency of the model was not broken so drastically as in (13d). But all evidence of observed molecular PECs points out that attraction depends naturally on $1/R_{12}$ and not on a nucleon-electron potential as in (13d). In addition, the scheme must be valid for all R, not only for R>> $R_e$.

## 6. Intra-atomic charge inversion. A generic switch (parity operator) to apply gauge-symmetry to a chemical bond

There is a relatively simple way to realise the transition from case (a) to case (b) in Section 5.1. This makes the sign in the *ionic potential* (1b) correspond with that of *theoretical model* (12), restores the broken internal symmetry in (13d) and secures at the same time that $V_{12}$ will remain *small at all R-values*. By definition, the main interaction must be attractive and must depend on $R_{12}$, not on $R_{2a}$ or $R_{2b}$. For the perturbation to be small and in order to retain the internal symmetry of all six interactions in (13a) and (13b), $V_{12}$ should be related to a difference $e^2/R_{2a}$ - $e^2/R_{2b}$ instead of to an asymmetrical solution (13d). In this hypothesis

$|V_{12}|$        ~ $|e^2/R_{2a}$ - $e^2/R_{2b}|$   << $|e^2/R_{12}|$                                                   (13e)

the non-crossing rule could be applied correctly as in (12b) and the internal charge-symmetry of the four-particle system is restored.

Although the geometry of the four-particle system and the reference axis are unknown, it is difficult to describe exactly which interactions could switch sign generically. Yet, making the signs of (1b) *generically* congruent with those in (12) has an unexpected consequence in connection with the principle of charge invariance, discussed in Section 3. The difference in (13e) can only be obtained in any geometrical arrangement of the four particles, *when the two electrons (leptons) in the anion have opposite charges, which is contrary to convention and especially to the standard H-L bonding theory.* The next consequence of charge invariance is *that this charge inversion also applies to the nucleons*. This provides exactly the key for going from case (a) to case (b) charge distributions, discussed in Section 5.1. We remind that the conventional case (a) has dominated theoretical chemistry in the post H-L era (Pople, 1999). The possibility that case (b) might be involved in theoretical chemistry has never been considered.

But due to the principle of charge invariance, case (b) would never have changed the energy of an atom, since the strength of the intra-atomic interaction does not change when going from +- to -+. *Intra-atomic charge* *inversion can only show when two ions interact according to Coulomb's law*: then a ++ interaction can change into a +- interaction of the same character, as discussed in Section 3. This intra-atomic charge inversion provides us exactly with the generic switch, the single parity operator we wanted above. The effect of a charge inversion in one of the two atoms is different for a parallel and an anti-parallel alignment of the charges *within the two atoms*. A simple example of this effect is the interaction of two magnets (dipoles). Attraction and repulsion are immediately felt when turning one of the magnets upside down when bringing them together. *The effect of this generic switch, not available in the H-L theory, is not dependent on the shape or geometry of the four-particle system: it is a generic chiral symmetry effect implied in gauge symmetry and only one algebraic operator is needed to describe it, a parity operator, as expected. The mathematical symmetry breaking effect, inherent to t - g = 1 Coulomb states has now become a real physical even generic effect.*
(Footnote remark: The effect of this chiral symmetry has been brilliantly visualised by the Belgian surrealist Magritte in a painting showing a man seeing his back when looking in a mirror).

The only way to make the Hamiltonian of the four-particle system in a chemical bond congruent with the recipe provided with gauge-symmetry (see Fig. 3a and 5), is to adapt it for parity violation: a repulsive nucleon-nucleon or lepton-lepton H-L state ++ (+C + $1/R_{12}$) becomes attractive + - (+C - $1/R_{12}$). The opposite applies to two of the four nucleon-lepton interactions. This is needed to get charged dissociation products at large R (fermion behaviour), in agreement with the generic gauge-symmetry based Coulomb scheme, see conclusion of Section 3.

All these preliminary considerations led us to very specific and clear benchmarks for an alternative way to explain chemical bonding or the stability of 4-particle systems in general.

The first criterion to justify this rather drastic measure will be the agreement with experimental PECs (published RKRs): the theory should result in a really universal function, whereby the ultimate challenge is to reduce all 13 PECs in Fig. 1b to a single straight line.

The second is the character of this theoretical result, which depends upon the number of parameters required or allowed to fit theory with experiment (Pople, 1999). If this number is zero, the *ab initio* status applies.

The third is an objective quantitative criterion for the acceptance of *ab initio* methods and was clearly set by Pople (1999) in his recent review about the evolution in quantum chemistry since 1927: heats of formation should be reproduced with an accuracy of 1 kcal/mol or about 400 cm$^{-1}$. In terms of molecular spectroscopy this is a relatively wide margin when level energies are under discussion but it is a clear quantitative benchmark for theoretical approaches.

**7. Intra-atomic charge inversion in the Hamiltonian leads to symmetric 1/R potentials and shape invariant PECs**

Permutational symmetry requirements and their effect upon the wave function in the classical H-L theory (antisymmetry for fermions and symmetry for bosons) are now taken over by charges and charge distributions. Mirror symmetry for the two charges in a boson produces a switch which transforms an attraction into a repulsion or vice versa. This gives fermion behaviour for a pair of bosons (see Fig. 5). As a result, 4 out of 6 potential energies in Hamiltonian (13a) change sign (are switched), if charge inversion, atom handedness or atomic chirality in *one* atom or boson is introduced (Van Hooydonk, 1985). We get

$$H_{XY} = (½ _av^2 - e^2/R_{1a} + ½ _bv^2 - e^2/R_{2b}) + (-1)^t (- e^2/R_{1b} - e^2/R_{2a} + e^2/R_{ab} + e^2/R_{12})$$
$$= H_X + H_Y + (-1)^t V(R) = H_X + H_Y \pm V(R) \quad (14a)$$

with the switch t, as in (5), the type of interaction determined by charge inversion in X or Y. But (14a) can not simply be generalised as in (9a) or (11). For writing down a system like (9a) or (11) the real asymptote C = V( ) - V($R_e$) is needed, see (8). This asymptote derives from the internal mechanics of the system. The dissociation products of systems obeying Coulomb's law must be *charged* particles like ions, not *neutral* species like atoms. The asymptote of a Coulomb system can therefore never be equal to $D_e$, unless, during the long-range interaction, the 'ionic' interaction transforms smoothly into an atomic interaction. If so, the shape of observed PECs and the curvatures must contain a (hidden) message that this conversion has taken place. Evidence points out that the true asymptote can be more than 5 times as large than $D_e$, see Section 2.1. This rather extra-ordinary scaling hypothesis, stating that only an ionic Coulomb asymptote $e^2/R_e$ is effective for scaling has been detected recently (Van Hooydonk, 1999). Only

$$V(R_e) = H_{XY}(R_e) - H_{XY}( ) \quad (14b)$$

can give the real value of the asymptote. For the *attractive* branch of a molecular PEC, the conventional value for w (m) would be

$$W(R)/D_e = w(m) = 1 - R_e/R = 1 - 1/m \quad (14c)$$

which proves to be fallacious (see above and also Fig. 1b). But an intermediate (virial) asymptote, situated at about half the well depth (in any case larger than $D_e$), gives

$$W(R)/IE_X = w(m) = 1 - R_e/R = 1 - 1/m \quad (14d)$$

This fits in a generic Coulomb scheme and leads to different prospects (see Section 2.1, middle PEC in Fig. 1). Unfortunately, it is difficult to imagine, given the complexity of the H-L theory, that this extremely simple static Coulomb picture (14d) could ever be applicable to the internal dynamics of real molecules and would be the basis of chemical bonding. Nevertheless, it gives us a hint for a universal potential fitting in a Coulomb model with gauge symmetry, in a form ± (1-1/m) we wanted and roughly in accordance with observation, see Fig. 6d.

When t = 1, the Hamiltonian (14a) can directly be rewritten, if the small perturbation is neglected, as

$$H_{XY} = H_{X^-} + H_{Y^+} - e^2/R_{12} \quad - (IE_X + EA_X) + 0 - e^2/R_{12} \quad (14e)$$

the simple one-term *generic* Coulomb solution, nearly conforming to (9a) or (11). We first discuss some implications of (14e) and of intra atomic charge inversion in the Hamiltonian.

## 8. Reducing the 10 term four-particle Hamiltonian to a Kratzer potential, a Bohr-like equation and a Coulomb potential

Formally in both AIM and IIM approaches, the singlet Hamiltonian with charge inversion t = 1 (14a) has a generic advantage over (13a), since the 4 nucleon-lepton interactions pair-wise get opposite signs instead of being attractive all four (Van Hooydonk, 1985). Unfortunately in the *covalent* (AIM) *approach*, the 2 nucleon-lepton pairs refer to different domains. For large R, $R_{1a} \ll R_{1b}$ and $R_{2b} \ll R_{2a}$, so they can not cancel for all R. Cancelling these terms is only allowed at $R = R_e$.

In the *ionic approximation* (IIM), the switch $t = 1$ has a different effect

$$H = H_{X+Y-} = (\tfrac{1}{2}\,_av^2 + \tfrac{1}{2}\,_bv^2 - e^2/R_{1a} + e^2/R_{1b} - e^2/R_{ab}) - e^2/R_{2a} + e^2/R_{2b} - e^2/R_{12}$$
$$= H_{X+} + H_{Y-} - V'(R) \qquad (15a)$$

the basis of (14e). The advantage of using an ionic instead of a covalent model now shows, since $R_{1a} = R_{1b}$ and $R_{2b} = R_{2a}$ *for all R*. The 4 nucleon-lepton interactions cancel pair-wise *for any R* in an ionic model. This leads to a first simplification of Hamiltonian (15a) to *four* terms, *applicable for all R*

$$H = (\tfrac{1}{2}\,_av^2 + \tfrac{1}{2}\,_bv^2 - e^2/R_{ab}) - e^2/R_{12} = -IE_X - EA_X - e^2/R_{12} \quad H_{pos} - e^2/R_{12} \qquad (15b)$$

The 3 terms between brackets correspond with the anion, a positronium-like system with Hamiltonian $H_{pos}$. The last term is the nucleon attraction, a protonium-like system. This is the ionic Coulomb interaction, we needed to construct shape invariant observed PECs. We get

$$H - H_{pos} = -e^2/R_{12} = V(R)$$

exactly of the Coulomb form required in our generic model, see section 3, if the small $V_{12}$ is neglected.

In general, four-particle systems are supposed to be stable (Richard, 1992, 1993, 1994, Abdel-Raouf et al, 1998), although trying to solve these systems requires a number of hypotheses, one being the dissociation products (the asymptote) of the system. A B-O approximation in $H_2$ with charge inversion is simply a *positronium-protonium system* (15b) or a *hydrogen-antihydrogen* system. If our derivations are valid, we can easily compute a PEC for this system, see below.

Depending on the (unknown) equilibrium geometry of the 4-particle system, different *static* cases can be distinguished one of which was already discussed by Luck (1957) without using charge inversion.

If $R_{ab}$ is perpendicular to $R_{12}$, (15b) is in first order equal to $-0{,}5IE_H - e^2/R_{12}$, the 1st term being the Bohr-solution for a positronium-like system $H_{pos}$ (the reduced mass for positronium is $m_e/2$). In any case, this *is a species independent (universal) molecular 1/R potential*, having $R_e = 2r_B$ if $r_B$ is the Bohr-length, the asymptote being 0.5Ryd, about 54800 cm$^{-1}$. This is of the correct order of magnitude for bonds between monovalent atoms, see Section 2.1. In this model, the two leptons in the positronium part would act as a classical Watt-regulator on the protonium system.

If $R_{ab}$ is parallel or antiparallel to $R_{12}$, we leave the B-O-approximation and get, under the conditions above, that

$$H = (\tfrac{1}{2}m_av^2 + \tfrac{1}{2}m_bv^2 - e^2/R_{ab}) + (\tfrac{1}{2}m_1v^2 + \tfrac{1}{2}m_2v^2 - e^2/R_{12})$$

an explicit *positronium-protonium* system in the case of hydrogen.

If $R_{12} \gg R_{ab}$, we can substitute this 2-component system with two *hydrogen-like* systems and get

$$H \approx 2(\tfrac{1}{2}m_ev^2 + \tfrac{1}{2}m_pv^2 - e^2/R) \qquad (15c)$$

if $m_e$ is the electron and $m_p$ the proton mass. Deviations occur from the value of 2 in (15c), if the static alignment of $R_{ab}$ and $R_{12}$ is parallel or anti-parallel. It is also different, for intermediate cases. Nevertheless, $R_{12} \gg R_{ab}$ is a reasonable assumption and both will be equal to about $2r_B$. The *sequence* of the particles -alternating or not- is important in a not perpendicular arrangement. These are all classical situations (Luck, 1957) but we can never solve this 4-particle system exactly. For a positronium and a protonium (Kong and Ravndal, 1998) system, the reduced mass equals m/2. This charge-mass symmetry will be broken in the two hydrogen-like systems, where a B-O approximation reappears. The leptons in (15c) are to be redistributed between the nucleons. If applicable, (15c) reduces to only 2 terms

$$H \approx 2(½ v^2 - e^2/2r) \qquad (15d)$$

which, as in Bohr's theory, can be solved, if these would be central force systems. Since this is not certain at all, hydrogen-like systems can be expected to obey the Planck-Bohr quantum condition, meaning that $½ v^2$ in (15d) varies as $B/R^2$. Instead of (15d) we now get (Van Hooydonk, 1985)

$$H \approx 2(-e^2/R + B/R^2) \qquad (15e)$$

nothing else than the *empirical* Kratzer potential (2b) in Section 2.2. The first derivative of (15e) at $R = R_e$ yields $B = e^2 R_e/2$. *This leads to the reduced molecular form (2c) of the Kratzer potential for A = 2*, an extremely useful molecular potential (Van Hooydonk, 1999). The result is the Kratzer PEC

$$W(R)/e^2/R_e = (1 - R_e/R)^2 = k^2 \qquad (15f)$$

These derivations show why the Kratzer potential can be useful for both *atomic and molecular spectroscopy*, see Section 2.2. In an ionic charge inverted model, the 2-term Hamiltonian (15d) is a good theoretical approximation for the original classical 10-term four-particle Hamiltonian. The Kratzer potential (15f) can safely be applied in the whole range $0 < R < \infty$. Its use is not restricted to $R_e$, where an an-harmonic oscillator results, see Fig. 3a. The strange thing is that a near Bohr solution (15d) for *atoms* produces an ionic Coulomb solution with asymptote $e^2/R_e$ for *molecules* in (15f). We do not need perturbation for the Kratzer potential which produces the required shape for a PEC, see above. These results were presented some time ago in a different version (Van Hooydonk, 1985).

Reminding that $R_e \approx 2r_B$ (see above), the universal asymptote in this case obeys

$$C = -H \approx -2(-0,5 e^2/2r_B) = e^2/2r_B = IE_H = 13,595 \text{ eV} = 109600 \text{ cm}^{-1} \qquad (15g)$$

or 1Ryd, see (1d), which, as in the perpendicular case above, can never be equal to $D_e$. For $1\mu$, this is 116400 cm$^{-1}$. If this universal asymptote has any meaning, it must appear when scaling PECs.

Nevertheless, the m-dependence in a Kratzer potential is quadratic and different from that in a gauge-symmetry based generic scheme, where it is linear. Using (14e) and (15g), the generic solution from the total well depth of bond XX is

$$H_{XX}(R_e) = -(IE_X + EA_X + e^2/2r_X) = -(2IE_X + D_e) = -(IE_X + EA_X + e^2/R_e)$$

and, *without perturbation*,

$$H_{XX}(R) - H_{XX}(R_e) = e^2/R_e((1-R_e/R)^2)^{1/2} = IE_X((1- 1/m)^2)^{1/2} = k \qquad (15h)$$

independent of $D_e$, as required and consistent with guess (14d) but different from (15f). Moreover

$$V_{12} = e^2/R_{2a} - e^2/R_{2b} \quad (<< e^2/R_{12}) \qquad (16)$$

is the important perturbation term, exactly as required, see (13e) and this is in line with the ionic charge inverted Hamiltonian. All conditions to apply (12b) are now satisfied. But it appears we found not one but two generic solutions: (15h) a Coulomb solution varying as $k = (1-1/m)$ for which perturbation is required, and (15f) a Kratzer solution varying as $k^2 = (1-1/m)^2$ for which no perturbation is needed, except at the crossing with the atomic dissociation limit.

The difference in the physics of the two model potentials has to do with the 'central force' character of a 4-particle system. With $k^2$ two central force subsystems appear centred on each of the nucleons. With k, a central force system is assigned to only one nucleon, i.e. in the anion. The $k^2$-potential is closer to the H-L theory than the k-potential, which relates to simple hyper-classical ionic bonding.

Charge inversion (a generic switch in the H-L Hamiltonian) leads to a straightforward introduction of polarisation effects as the major contribution to $V_{12}$, since the leptons have opposite charges. Without charge inversion polarisation occurs only between nuclei and leptons, since the leptons have equal charges.

This section shows that even a static Coulomb based generic model (15b) or (15e) can theoretically be relevant for chemical bonding. It justifies part of the criticism given above on the H-L theory. This theory does not contain any well behaving empirical 1/R potential out of the many available in the literature. The relatively small but fundamental adaptation (charge inversion, chiral behaviour of a two-fermion atomic system) opens the way for really getting first hand theoretical, generic, approximations from parity-adapted Hamiltonians in perfect agreement with these earlier empirical findings about 1/R-potentials. This result is consistent with spectroscopic evidence offered by the generalised Kratzer-Varshni potential (2c) and by 1/R potentials (Van Hooydonk, 1999). Here the H-L theory does not give an explanation, on the contrary: it immediately led to the total rejection of ionic bonding models, the very basis of the Coulomb k-potential. The last attempt to explain chemical bonding with electrostatics was given by Luck (1957), more than 40 years ago.

But if the present theoretical derivations are confirmed by experimental RKR-curves, it seems that even an hyper-classical static Coulomb law is theoretically capable of coping with PECs for the so-called dynamic process of bond formation and if so, the scaling problem is automatically solved (see Fig. 6a-c).

Both solutions (15f) and (15h) even indicate that a zero molecular parameter function can exist: this means that, theoretically, molecular PECs can be constructed just using atomic data. Then the PEC of $H_2$ must have a connection with the Rydberg, the connection being of universal form $k = (1-1/m)$ or $k^2 = (1-1/m)^2$. When confronting these derivations with experiment, the number of parameters required for fitting the data will determine the *ab initio* character of the final result.

### 9. Applying the non-crossing rule at the minimum for states governed by 1/R potentials. Perturbation. Consequences for the asymptote

In order to perform the calculations, we first need to identify the perturbation for the k-potential. For Kratzer's solution (15f), no perturbation is needed, unless for the crossing with the atomic dissociation limit. The generic 1/R solutions, (14e) or (15h) can now be used in combination with the generic perturbation (13e) or (16). Reminding (1e), this is still a one-term solution for V(R). To avoid crossing, we need the sum and the difference of two diabatic potentials

$$W_1 + W_2 = -e^2/R_{12} + C + e^2/R_{12} - C = 0 \qquad (17a)$$

$$W_1 - W_2 = 2C(1 - R_e/R_{12}) = 2C(1 - R_e/R) = 2C(1-1/m) \qquad (17b)$$

with $R_{12} = R$. The classical result for the non-crossing rule gives two adiabatic potentials

$$W(R)/C = w(m) = \pm((1 - 1/m)^2 + (V_{12}/C)^2)^{1/2} \qquad (17c)$$

with $V_{12}$ a small perturbation (13e) or (16) and used already above, see (9a), (12b) and Fig. 3a. The perturbation in (17c) affects the asymptote. But the most important effect of perturbation around the minimum is that it affects curvature at the minimum (the force constant) and the curvature of the branches, see Fig. 2 and

3. In this respect, only perturbation would solve some of the difficulties met above, see Fig. 1 and Fig. 6a-c. In any case, with $V_{12}$ small and constant, symmetry will always be broken. The perturbation must be confined to the region where the two attractive t - g = 1 curves intersect, the minimum.

*9.1. Symmetric 1/R potentials at the minimum*

Around the minimum, a generic Coulomb solution 1-1/m can be expanded as

$$W(R_e)/C = |k| = |1 - 1/m| = |1-1/(1 + x^p)|  \quad (18a)$$

if $m = 1 + x^p$ with x > or < 0. This secures that the PEC remains shape invariant and (quasi-) harmonic around the minimum depending on the analytical form of x. By definition, $x^p$ must be small, when it remains coupled to W(R)/C as a scaled unit of energy for the system. It must also be related to a scaled difference in a distance away from the minimum $|R-R_e|$, see (1f) in Section 2.1. For scaling this distance, the variables (1g) or (1i) are available, related as

$$k = d/(1+d) \text{ or } d = k/(1-k)  \quad (18b)$$

since $R_e/R = 1/m = R_e/(R - R_e + R_e) = 1/(1 + d)$. The transition from variable k to d causes algebraic effects not visible when using $1/m = R_e/R$ but we do not discuss these here. In general

$$|R - R_e| = R_e|d| = R|k|  \quad (18c)$$

irrespective of the functional dependence of $x^p$ on exponent p. Using k or d for x, expansion gives

$$W(R_e)/C = |x^p /(1 + x^p)| \quad | \quad x^p (1 - x^p \ldots)| = |x^p - x^{2p} +...|  \quad (18d)$$

This leads to the Dunham expansion if |x| = |d|, see also the general discussion in Section 2.1 and (1f). But the |x| = |k| solution is equally valid and probably even better, since it is asymmetric again and will respect the left-right asymmetry at the minimum more explicitly. If (18a) is expanded in function of k, this leads to self-replication (continued fractions) and to all consequences thereof (Fibonacci-like series, chaos/fractal behaviour, Freeman et al., 1997). Continued fractions have been discussed recently by Molski (1999). Since an expansion of $W(R_e)/C$ in (18d) must give positive values by definition, the minimum value of exponent p should be 2 in a non algebraic context. This is always a quasi-harmonic dependence on $(m-1)^2$ or an an-harmonic dependence on $(1-1/m)^2$, in both cases an oscillator presentation confined to m = 1, the only case where |x|= |d| = |k|. A solution with k is valid for all R. A perturbation, confined to the minimum, can be supposed to vary in an oscillator mode. With the exponent p = 2 and using (18d), we can write

$$V_{12}(R) = V_{12}(R_e) (1 - x^2/(1 + x^2)) = V_{12}(R_e)/(1 + x^2)  \quad (18e)$$

$$x = k = (R-R_e)/R = 1 - 1/m  \quad (18f)$$

This relation has the disadvantage of not converging to zero for large m. It gives 0,5 $V_{12}(R_e)$ at infinite separation, and normalisation is required. Here, we use a simplified expression

$$V_{12}(R) = V_{12}(R_e)(1 - x^2) = V_{12}(R_e)(1 - k^2)  \quad (18g)$$

deriving from (18e) in first order, which vanishes at k = 1. The value of the exponent p in (18c) remains somewhat arbitrary: maybe it is 2 but it could be any other integral.

We remark that both (18e) and (18g) for the perturbation are still closed form analytical functions.

*In an algebraic context*, as with a pair of generic Coulomb potentials, the p = 2 case is not the only one possible. In the Landau-Zener formalism (Landau, 1932, Zener, 1932), the non-crossing rule leads to avoided crossings away from the minimum. Applying it at the minimum, where the intersecting states W(R) and asymptote are almost perpendicular is allowed, see Fujikawa and Suzuki (1997).

The behaviour of k, $k^2$ and $d^2$ for the leading terms is illustrated in **Fig. 8**. The $H_2$-RKR data points are included (level energies scaled with the Dunham asymptote, 78580 cm$^{-1}$). The leading term in the Dunham variable $d^2$ is completely wrong (see also Fig. 7) although this is exactly the oscillator representation used so frequently in theoretical physics (Witten, 1981, 1982). The *unperturbed* Coulomb potential using |k| is not performing well either (see Fig. 1 and 7). The Kratzer potential leads to an acceptable result. The perturbed Coulomb potential is calculated with a perturbation of 0,35C (see below). In comparison with Fig. 1a, the progress is considerable, *qualitatively and even quantitatively*.

*9.2. Intersecting states. The asymptote problem. Varshni's fifth potential.*

The *attractive* branch of W(R) operates between C and 0. A singlet Kratzer state $C(1-1/m)^2 = Ck^2$ can never cross a generic singlet state $C(1-1/m) = Ck$ starting from the same asymptote. Their 'crossing' points are m = and m = 1. The m-values for intersection points $m_b$ with a fixed *asymptote* below C, say $C_b = C/b$, can easily be calculated. With b > 1, i.e. lower lying asymptotes, such as $D_e$, are generated. Kratzer's potential gives $Ck^2 = C/b$, with an intersection point $m_b = $ b/( b -1), a generic potential gives $m'_b = b/(b-1)$. Their ratio $m'_b/m_b = $ b/( b+1) shows that the generic potential will intersect a lower asymptote $C_b$ closer to the minimum than $m'_b < m_b$ for all b. Theoretically, it is possible that the generic asymptote is not identical with Kratzer's. Then, crossing between the two states can occur away from the minimum. But this only applies when the generic asymptote is *lower* than Kratzer's. In this case, the intersection points are obtained using $Ck^2 = (C/b)k$, which is equal to

$$Ck = C/b = C_b \quad (19a)$$

and gives the same solution $m'_b$ as for the generic unperturbed potential crossing an asymptote $C_b$. The other case where intersection can occur is with a perturbed generic potential.

Let both generic and Kratzer potentials start at any asymptote $C_b$. Then

$$W(R) = C_b|k^q| \quad (19b)$$

The generic potential, q = 1, starting from the absolute well depth (see section 2.1) is generated with b = 1/2. The original Kratzer potential (2c), q = 2, starts from the ionic dissociation limit would with b very close if not identical to 1.

Next, applying a *constant perturbation* C/b for a generic potential starting from an asymptote C gives, using (17c)

$$W_q(R) = C((k^2 + 1/b^2)^{½} - 1/b) \quad (20a)$$

If b is large, 1/b can be subtracted to obtain zero at m = 1, to account for the asymptote shift $C - V_{12} = C(1-1/b)$. *Without perturbation*, the *attractive* branch obeys

$$W(R) = C(1 - 1/m) - C/b = C(1 - 1/m - 1/b)$$
$$= C(1 - (m+b)/mb) \quad (20b)$$

a simplification of (20a), i.e. neglecting powers of b higher than 1, but this does not lead to the smooth curvatures at the minimum needed (see Fig. 6a-c).

The Kratzer result (15f) can be factorised as

W(R)/C = (1-1/m)(1-1/m)                                               (20c)

Let a generic potential start at a reduced asymptote $C(1 - 1/b) = C – V_{12}$ leading, without perturbation, to

W(R)    = C(1 - 1/b)(1 - 1/m)

        = C(1 – (m+b)/mb + 1/mb)                                      (20d)

as illustrated in Fig. 6a-c. If the reduced asymptote $C(1 - 1/b)$ is a *running* asymptote, i.e. if b is a variable just like m, the result is a potential bearing a striking resemblance with Kratzer's reduced potential (20c), whereby the starting value of asymptote C remains unchanged. This is a similar picture as that obtained in (18d). Nevertheless, (20b) and (20d) can both be of help in interpreting the Varshni exponent v in (2d). In the present model, a running asymptote means it is not necessary to apply a perturbation due to lower asymptotes. The result is

W(R) = C(((1-1/b)(1-1/m))$^2$)$^{½}$                                  (21a)

No normalisation is required and this potential is a pseudo generalised Kratzer-Varshni potential. However, Fig. 6a-b show this does not lead to the correct curvatures around the minimum.

Allowing for *an m-dependent* perturbation $V_{12}$ (18g), means (20a) must be replaced by

W(R) = C((k$^2$ + ((1/b)(1 - k$^2$))$^2$)$^{½}$-1/b)                  (21b)

with the exponent in (18a) p = 2.

**Fig. 9** shows how *perturbed generic potentials* (21b) behave with respect to the *generalised* Kratzer-Varshni potential (2d), known to be consistent with the spectroscopic constants (Van Hooydonk, 1999). Three v-values, equal to 0.7, 1.0 and 2.0 are used with an asymptote C/2 and these potentials are compared with a perturbed generic scheme based upon the asymptote C, using (21b) with b-values: 1,5, 2, 3, 6, 10 and infinity. The case v = 0.7 typically applies to homonuclear bonds, whereas v = 2.0 and larger applies to heteropolar bonds. For v = 0.7 (covalent bonds), the generalised Kratzer is almost identical with the generic state with b = 3/2. *The original Kratzer state is exactly the same as the generic perturbed by its virial asymptote with b=2* (see also below). For v = 2, the attractive side coincides with the generic b = 4 state up till m = 2. Large b-values will give the generic potential almost the shape of a Born-Landé potential or, in the limit, the Coulomb potential again. The greatest deviations are found at the repulsive branch, but these are difficult to show graphically. The efficient generalisation of the Kratzer potential, suggested a long time ago by Varshni (1957) turns out to be an elegant artefact to reproduce perturbed generic 1/R potentials *if these would occur in practice*, which remains to be determined. Conversely, extrapolating these results to observed PECs, Fig. 6-7 and the closeness of the two kinds of potentials show that generic 1/R Coulomb potentials can most probably cope with observed molecular PECs. In fact, Varshni's potential (2d) leads to very consistent results for hundreds of bonds (Van Hooydonk, 1999).

The intersection of generic and Kratzer states with a lower *fixed* asymptote, such as the atomic dissociation limit $D_e$ is also of interest, given the developments in femto-chemistry in particular. The perturbation around the critical distance being small and confined to that region, we get

$W_{PEC}(R)$ = ½((W(R) + $D_e$) – ((W(R) – $D_e$)$^2$ + V'$^2$(R))$^{1/2}$)     (22)

with W(R) given by (21b) for the generic function or by (20c) for Kratzer's.

Even a crude approximation V'(R) = 0 for (22) will not influence the result drastically, provided W(R) is accurate enough, especially at long range. If valid, this justifies our previous conclusion that the usual constraint for a molecular function W($\infty$) = $D_e$ can even be disregarded (Van Hooydonk, 1999). Femto-chemistry could provide us with information about that specific region where an ionic potential crosses the covalent asymptote. We consider this asymptote as fixed in a good first approximation since the Coulomb interactions between neutral atoms are not based upon a strong $e^2/R$ law but refer to smoother polarisation terms (Aquilanti et al, 1997). Covalent interactions can be described as a function of atomic polarisabilities, having a stabilising effect only between the long-range contours (higher turning points) of an ionic Coulomb PEC (see for instance Aquilanti et al, 1997, especially their Fig. 1). We therefore expect femto-chemistry applied to the critical distance of covalent bonds (crossing of ionic $X^+X^-$ and non-ionic XX curves) can provide conclusive evidence for a generic static bonding scheme, see also below. The situation is not that simple however: finding a universal potential is essential for understanding the behaviour of interacting particles at short- and long-range. Accurate long-range potential energy curves are extremely important for studying ultra-cold atoms and related phenomena (Zemke and Stwalley, 1999, Wang et al., 1997, Marinescu et al., 1994, Stwalley and Wang, 1999). In this respect, the phenomena at the 'critical' distance are themselves also critical. We will show below that this critical region is almost perfectly scalable in the relatively small portion of the total PEC, i.e. around 10 to 20 % from the crossing point.

**10. Generic low parameter universal Coulomb function.**

To finalise quantitatively the Coulomb scheme and to conserve its *ab initio* character, non-empirical generic perturbations or b-values in (21b) are required. Two such solutions are easily found.
(a) First (21b) may be rewritten as
W(R)b/C + 1 = (1 + $k^2(b^2-2)$ + $k^4$)$^{1/2}$ (22c)
With b = 2 or $V_{12}(R_e)$ = C/2, the virial asymptote is the generic perturbation. The first generic result is therefore, as expected, the Kratzer potential
2W(R)/C = $k^2$ (22d)
see (15f). This is consistent with the Hamiltonian, see section 8 and already illustrated in Fig. 9. Applying a perturbation equal to the virial asymptote C/2 to a Coulomb 1/R potential leads to the (quadratic) Kratzer potential. This is not really a surprise when looking at Fig. 3a and 5. The gap is equal to 2C and covers the two worlds by definition. The Kratzer potential is confined to half the total gap or C, i.e. the positive world. For an attractive Coulomb potential to stop at the zero energy in this picture, it is necessary that it be confined to C. All this derives from classical gauge-symmetry.
Replacing W(R) in (22d) by $Kk^2$, where K is the Kratzer asymptote, leads to
K = C/2 (22e)
suggesting that in this case C, the generic asymptote, is the absolute well depth of a bond, the total gap. This is in agreement with an earlier guess (14d). Also, this result makes it quite acceptable that the most probable value of the exponent p in (18c) is 2, see section 9.1.

(b) A second generic b-value is obtained by equating the perturbed generic and Kratzer potentials. This unusual procedure corresponds with a redistribution of the central force character within the different subsystems of a four-particle system (see above). If a non-trivial solution can be found, this leads to generic b-values in terms of K and C. If P is the perturbation, we get

$$Kk^2 = (C^2k^2 + (P(1 - k^p))^2)^{1/2} - P \qquad (22f)$$

This leads to P = C/b = |C - K| for k = 1 and to a trivial P = P for k = 0, irrespective of the value of the exponent p in (18b). Working (22f) out analytically gives a relation between P and the asymptotes C and K. This implies a generic value for the exponent p also. We rewrite (22f) as a quadratic relation in P

$$P^2k^p(k^p - 2) - 2PKk^2 - K^2k^4 + C^2k^2 = 0 \qquad (22g)$$

After dividing by non-zero $k^2$, the solutions

$$P = (K/(k^{p-2}(k^p-2)))(1 \pm (1 - (k^{p-2}(k^p-2)((C/K)^2 - k^2))^{1/2}) \qquad (22h)$$

are obtained. For k = 1, one solution is like C – K above. But a generic non trivial solution also results *when the exponent p = 2*, since then $k^{p-2} = k^0 = 1$ for any k or

$$P = (K/(k^2-2))(1 \pm (1 - ((k^2-2)((C/K)^2 - k^2))^{1/2}) \qquad (22i)$$

When C = K, this gives the non trivial P (or b) solution

$$P = (K/2)(-1 + (1 + 2(C/K)^2)^{1/2}$$
$$= 0,5K(\sqrt{3} - 1) = 0,366025 \text{ K} \qquad (22j)$$

and solves the problem about truly generic perturbations or generic b-values, which are found to be equal to 2 and to 2,7321 (close to e). For k = 1, (22i) leads to

$$P = -K(1 \pm C/K) \qquad (22k)$$

or P = -K - C or P = C - K as above. The second generic, i.e. Coulomb, solution is now found to be

$$w(k) = W(R)/C = (k^2 + (0,366025(1-k^2))^2)^{1/2} - 0,366025 \qquad = k_{gen} \qquad (22l)$$

if both the Kratzer and generic potential refer to the same asymptote C. The only variable in (22l) is a number $1/m = R_e/R$ and therefore the RHS of (22l) can be called a generic species independent universal variable $k_{gen}$. But W(R) in (22l) is always below the original Kratzer result and leads to a more stable system. With P(R) = $P/(1+k^2)$ instead of $P(1-k^2)$, section 9.1, the same value 0.366025 is obtained.

Reminding (1e) and (1f) and our choice for a single term in (1e), a simple Coulomb potential produces, unlike the Dunham expansion, closed formulae for PECs: the Kratzer potential $k^2$ and a more complicated one $k_{gen}$ (22l), still a closed formula. For larger b-values (small perturbations) no generic solutions are available. For these ionic cases, empirical approximations will have to be used to account for the PECs, if they do not obey a generic w (m) solution.

If the virial asymptote acts as a constant perturbation for W(R), a normalised asymptote 5/2 appears, indicating a possible interference of Euclid's golden number (Van Hooydonk, 1987) and, eventually chaos/fractal behaviour, suggested by Freeman et al. (1997), see also above.

Finally, taking into account the -constant- atomic dissociation limit $D_e$ a theoretical PEC obeys

$$PEC_{theo} = 0,5(W(R) + D_e) - 0,5((W(R)-D_e)^2 + (V''(R))^2)^{1/2} \qquad (22m)$$

like (22), with W(R) either equal to the Kratzer (22d) or generic potential (22l). In this report we test (22m) mainly in the hypothesis V''(R) = 0, see Section 9, a reasonable working hypothesis. Only experiment will decide if these two generic functions, both depending solely on R in a closed analytical formula, correspond with observed PECs within reasonable limits. The constraints on PECs generated by functions (22d) and (22l) are stringent, since their analytical form is extremely simple. Both are, at this stage, one parameter (R) functions. Just one molecular parameter $R_e$ is needed if this can not be computed from atomic data. Smooth transitions from a two central force systems ($k^2$) to a one central force system ($k_{gen}$) approach in function of k and (1-k) are easily made and can be written down using the equations above. Nevertheless, at this stage we leave out computations based upon hybrid potentials, which will always give results very close to the starting potentials. The benchmark, Morse's, is in Fig. 1c-d. Apart from reducing all 13 PECs to a single straight line, the ultimate challenge is to calculate reasonable PECs with a zero molecular parameter function.

## 11. Results and discussion

The 13 PECs (RKR or IPA) used are for the bonds $H_2$ and $I_2$ (RKR, Weismann et al., 1963), HF (RKR, Fallon et al., 1960, Di Leonardo and Douglas, 1973, IPA by Coxon and Hajigeorgiou, 1990), LiH (IPA, Chan et al., 1986), KH (Hussein et al., 1986), AuH (PEC data by Le Roy's method, Seto et al., 1999), $Li_2$ (RKR, Kusch and Hessel, 1977; IPA, Hessel and Vidal, 1979), KLi (Bednarska et al., 1998), NaCs (Diemer et al., 1984), $Rb_2$ (Amiot et al., 1985), RbCs (Fellows et al., 1999), $Cs_2$ (Weickenmeier et al., 1985) and a theoretical PEC for LiF (Padé approximant, Jordan et al., 1974, Kratzer-Varshni-type, Van Hooydonk, 1982). We will use here the Morse PEC for LiF instead. Some of these data were used above for Fig. 1b-d. We did not always use all published IPA data available, which are sometimes very detailed. The number of data-points is over 500, including the 13 minimum values, or about 40 per bond. These date have been used in constructing (part of) the Fig. 1c-d and 6d.

*11.1. The asymptote problem*

Seven asymptotes are available to interpret PECs using a universal function f(R). For a bond XX these are: 1. the absolute asymptote (absolute well depth), *covalent approximation* (AIM): $C_C(abs) = 2IE_X + D_e$; 2. the absolute asymptote, *ionic approximation* (IIM): $C_I(abs) = IE_X + EA_X + e^2/R_e$; 3. the generic asymptote $G = e^2/R_e$ (C in this text). For homonuclear bonds, $R_e$ is equal to $2r_X$, where $r_X$ is the covalent radius of atom X (see above). If true, asymptotes 2 and 3 are available from *atomic data* and would not be molecular parameters; 4. the ionic asymptote $I = e^2/R_e = IE_X - EA_X + D_e$, nearly equal to G in a reasonable first approximation, although this is not a priori certain (Van Hooydonk, 1999). The main difference between G and I is the dependence on $D_e$, which is to be avoided for a generic solution. Asymptote 2 is equal to asymptote 3 only if $EA_X = D_e$ (Van Hooydonk, 1982, 1999). In this case, all *molecular asymptotes* 1-4 for a bond XX would be available from *atomic data* X, $IE_X = e^2/R_e$ and $EA_X = D_e$; 5. the atom (species independent) Coulomb asymptote in terms of the internuclear separation (in Å) and invariantly equal to 116431 cm$^{-1}$, the basis of 4 and 5 and corresponds with 1 Ryd; 6. the covalent asymptote, the atomic dissociation limit $D_e$ (apparently the first real molecular parameter met unless $EA_X = D_e$) and finally 7. the Dunham-asymptote or the first Dunham-coefficient: $A = a_0 = \omega_e^2/4B_e = 0,5k_eR^2_e$ deriving mathematically from the zero$^{th}$ order spectroscopic constants $\omega_e$ and $B_e$.

For the molecules $H_2$, HF, $Li_2$, LiH and LiF data collected earlier for asymptotes 3, 4 and 6 (Van Hooydonk, 1982) are now completed with asymptotes 1, 2, 5 and 7 and are given in Table 1. Only the Dunham asymptote is confined to the minimum in a specific mathematical way. Species dependent asymptotes show large divergences. Dunham's asymptote values are strange: they are either below ($H_2$), almost equal to ($Li_2$, LiH) or above the generic/ionic one (HF, LiF). There is no regularity in asymptote ratio's either. This species dependence shows why it is so difficult to find a universal function f(R).

Table 1. *Six species dependent asymptotes (in $cm^{-1}$) for five bonds*

| Bond | R(Å) | $C_c = IE_X + IE_X + D_e$ | $C_I = IE_X + EA_X + e^2/R_e$ | $G = C = e^2/R_e$ | $I = IE_X - EA_X + D_e$ | Covalent $D_e$ | Dunham A |
|---|---|---|---|---|---|---|---|
| $H_2$ | 0,7414 | 257584 | 272264 | 156453 | 141878 | 38283 | 79580 |
| HF | 0,9168 | 300041 | 295090 | 126526 | 131633 | 49406 | 204308 |
| $Li_2$ | 2,6729 | 95558 | 94991 | 43399 | 46909 | 8612 | 45902 |
| LiF | 1,5639 | 232580 | 242674 | 74175 | 64173 | 48122 | 154008 |
| LiH | 1,5957 | 173415 | 188404 | 72695 | 57681 | 20292 | 65747 |

The 7 asymptotes, all candidates for scaling RKRs, each have their own pros and cons. The data in Table 1 illustrate why the asymptote problem had to be discussed in detail, especially in connection with scaling semi-empirically constructed RKRs. If $D_e$ would be the unique reference asymptote, w(m) should vary between 0 and 1 invariantly for all bonds. Using larger asymptotes X compresses the maximum value of 1 to $D_e/X$, leading to different slopes and curvatures, as outlined above.

*11.2. PECs for $H_2$, $Li_2$ and $Cs_2$ from a zero molecular parameter function*

If the generic solution that the ionisation energy of an atom X generically determines the molecular PEC of the $X_2$ molecule is true (see Section 8), it must be possible to calculate $X_2$-PECs with *a zero molecular parameter function* just using atomic X-data. If moreover $D_e$ is equal to $EA_X$, the complete PEC becomes available, including the dissociation limit. If the present scheme is really universal and of first principle's nature, it must apply to the simplest bonds $H_2$, LiH and $Li_2$ in the first place. This sounds impossible, given the complex procedure to get solutions for the 4-particle Hamiltonian in the H-L theory. The case of $H_2$ is a standard example, see above. The $Li_2$ molecule has been studied extensively in the past as it is the lightest molecule in the Periodic Table after $H_2$. A review on theoretical and experimental studies on $Li_2$ is given by Hessel and Vidal (1978) of interest also because of the convergence problems with the Dunham expansion. The earliest attempts for understanding bonding in this molecule go back to Delbrück (1930) using the H-L method. Bond LiH is treated in the next Section 11.3.

The atomic data are $IE_H$ 13.595 eV and $IE_{Li}$ 5.3917 eV, giving theoretically $R_e(H_2)$ = 1.0572 Å and $R_e(Li_2)$ = 2.6731 Å. With these atomic values, the Kratzer potential predicts PECs obeying $13.595(1-1.0572/R)^2$ and $5.392(1-2.6731/R)^2$. The same asymptotes are used for the generic function $k_{gen}$. The theoretical PECs deriving from these atomic data are presented in **Fig. 10a** for $H_2$ and **Fig. 10b** for $Li_2$ and $Cs_2$ with $R/R_e$ on the x-axis to make the minima of theoretical PECs and RKRs coincide. The curve for $Cs_2$ ($IE_{Cs}$ 3,8939 eV) is included in Fig. 10b, since, unlike the RKR for $Li_2$, the turning points go near $D_e$. The mean % deviation for all 108 turning points is 11.4 %. In these very important 'simple' cases, the agreement between the zero molecular parameter and observed PECs is rather astonishing.

For $H_2$, Fig. 10a, the largest deviations are found at the repulsive branch, where experiment shows that the asymptote generated by the present method is close to the ionisation potential of H (see below, Table 3). We wonder if these RKR turning points are not in need of revision. For the attractive branch, the agreement is astonishingly good also at long-range close to the dissociation limit, which intersects the theoretical curve. This simple first principles 'atomic' PEC for $H_2$, deriving from Kratzer's generalisation of Bohr's formula, is much closer to the observed one than the PEC originally calculated by Heitler and London (1927). Exactly this poor H-L PEC for $H_2$ is at the origin of quantum chemistry, as we know it today, see also Introduction and Pople (1999).

For $Li_2$ and $Cs_2$, Fig. 10b, the agreement is better at both branches. Here the 'atomic' PECs are available up to infinity as alkali-metals have electron affinities very close to their $D_e$-values. Quantitative details for $Li_2$ are given below.

Before discussing the intermediate case of LiH, results of the same quality are obtained for all bonds between elements of the first Column in the Periodic Table if we use $(IE_X+IE_Y)/2$ as a first order approximation for the asymptote of an XY bond. In **Fig. 10c**, observed level energies are plotted versus those obtained in this zero molecular parameter approximation for 8 bonds $Li_2$, LiH, KH, KLi, NaCs, $Rb_2$, RbCs and $Cs_2$. The differences in $cm^{-1}$ are also shown. Only the repulsive branch of KH deviates from the general trend. The slope is close to unity and the goodness of fit is relatively high as indicated in Fig. 10c. The average deviation for the 310 turning points of 8 bonds is 9,42 % (for $Cs_2$ 11,3, RbCs 10,8, $Rb_2$ 9,0, NaCs 12,2, KLi 9,3, KH 7,5 and LiH 6,3, including long-range situations where applicable).

Table 2 gives the details of the results for $Li_2$. The Kratzer results are collected in Columns 2-4. The accuracy for the 30 turning points of the Kuch and Hessel (1977) RKR, calculated *in this zero molecular parameter approximation* is 2.5 %, impressive be it not of spectroscopic accuracy. The clearly visible inflection points are consistent with an expected crossing with the atomic dissociation limit at larger R, a feature inherent to the Kratzer potential (see above). Columns 2-4 represent molecular *ab initio* results acquired with atomic data only (zero molecular parameter function). In this case, the $EA_{Li}$ value is very close to $D_e$ as it is also for several other alkali-metals. This assures the complete PEC is available from atomic data only using (22m) with $EA_X$ as a substitute for $D_e$.

We multiplied $IE_{Li}$ with $k_{gen}$ for the ionic k-model and plotted the calculated levels versus the RKR. Fitting gives a scaling constant of 0,935596. This leads to the results given in Columns 5-7 in Table 2. Deviations for the zero parameter Coulomb k-function (0,6 %) are smaller than for the $k^2$-potential (2,5 %). The corresponding G(v) curve (not shown) is within 1 % of the experimental value for the 14 levels. The balance between the levels varies between 98,9 and 100,1 %, usually a problem for Morse's potential. This analysis is consistent with the results of Hessel and Vidal (1978), who extended the range to v = 18 using the Inverted Perturbation Approach. This IPA PEC is claimed to be more accurate than a RKR (differences vary from 0,1 $cm^{-1}$ to 5 $cm^{-1}$ for inner- and outermost turning points). The dependence of the IPA for $Li_2$ on the Kratzer and generic variables leads to a goodness of fit close almost equal to 1. The asymptote obtained by fitting the data by a linear equation through the origin is 45246,2 $cm^{-1}$, only 1,42 % lower than the Dunham asymptote 45902 $cm^{-1}$ in Table 1. Up till R = 4 Å, the correlation between generic PEC and IPA data-points is y = 0,97993x with $R^2$ = 0,9999156, which makes our approach reliable.

Despite the fact the agreement is not exactly 100 %, the zero molecular parameter function and its *ab initio* character suggest that conventional inversion procedures do not imply absolute certainty about the real PEC. In the extreme case, we might even claim various available turning points are in error with the same absolute deviations as given in Table 2. This is certainly a matter of further research.

The first hurdle for this theory and the Coulomb based function $k_{gen}$, deriving from gauge symmetry has been taken: it is possible to calculate realistic PECs from atomic data only. Or, PECs show definite Coulomb behaviour, dictated by atomic parameters, in agreement with our derivations above.

But this also implies that Coulomb based scaling (Fig. 6a-d) **must** apply to RKRs. If the present generic function indeed leads to an acceptable scaling scheme for RKRs of different bonds into a single one, preferably a single straight line, the Coulomb approach must be universally valid. This may ultimately lead to more reliable PECs obtained by inverting observed level energies.

Table 2. *$Li_2$ PEC calculated ($cm^{-1}$) with a zero molecular parameter Kratzer potential (columns 2-4) and a generic Coulomb potential (columns 5-7).*

| RKR | Kratzer PEC | Diff | Abs % | Generic PEC | Diff | Abs % |
|---|---|---|---|---|---|---|
| 4525 | 4694.8 | 169.8 | 3.75 | 4514,81 | -9,97 | 0,22 |
| 4252 | 4390.7 | 138.7 | 3.26 | 4246,58 | -5,15 | 0,12 |
| 3972 | 4082.4 | 110.4 | 2.78 | 3971,81 | -0,65 | 0,02 |
| 3687 | 3770.0 | 83.0 | 2.25 | 3690,43 | 3,32 | 0,09 |
| 3396 | 3454.3 | 58.3 | 1.72 | 3402,80 | 7,00 | 0,21 |
| 3099 | 3135.3 | 36.3 | 1.17 | 3108,87 | 10,22 | 0,33 |
| 2796 | 2813.4 | 17.4 | 0.62 | 2808,53 | 12,79 | 0,46 |
| 2487 | 2488.8 | 1.8 | 0.07 | 2501,86 | 14,67 | 0,59 |
| 2173 | 2161.8 | -11.2 | 0.52 | 2188,84 | 15,77 | 0,73 |
| 1853 | 1832.7 | -20.3 | 1.10 | 1869,46 | 16,00 | 0,86 |
| 1528 | 1501.8 | -26.2 | 1.72 | 1543,69 | 15,28 | 1,00 |
| 1198 | 1169.4 | -28.6 | 2.39 | 1211,54 | 13,54 | 1,13 |
| 862 | 835.7 | -26.3 | 3.05 | 872,85 | 10,59 | 1,23 |
| 521 | 501.3 | -19.7 | 3.78 | 527,91 | 6,65 | 1,27 |
| 175 | 166.7 | -8.3 | 4.75 | 176,91 | 1,88 | 1,07 |
| 0 | 0.0 | 0.0 | 0.00 | 0,00 | 0,00 | 0,00 |
| 175 | 165.7 | -9.3 | 5.32 | 176,70 | 1,66 | 0,95 |
| 521 | 494.2 | -26.8 | 5.14 | 522,00 | 0,74 | 0,14 |
| 862 | 820.7 | -41.3 | 4.79 | 859,37 | -2,89 | 0,34 |
| 1198 | 1145.9 | -52.1 | 4.35 | 1189,99 | -8,00 | 0,67 |
| 1528 | 1470.2 | -57.8 | 3.78 | 1514,78 | -13,64 | 0,89 |
| 1853 | 1794.1 | -58.9 | 3.18 | 1834,29 | -19,17 | 1,03 |
| 2173 | 2117.9 | -55.1 | 2.54 | 2149,25 | -23,82 | 1,10 |
| 2487 | 2441.9 | -45.1 | 1.81 | 2460,17 | -27,02 | 1,09 |
| 2796 | 2766.6 | -29.4 | 1.05 | 2767,66 | -28,08 | 1,00 |
| 3099 | 3092.3 | -6.7 | 0.22 | 3072,15 | -26,49 | 0,85 |
| 3396 | 3419.5 | 23.5 | 0.69 | 3374,22 | -21,58 | 0,64 |
| 3687 | 3748.5 | 61.5 | 1.67 | 3674,34 | -12,77 | 0,35 |
| 3972 | 4079.9 | 107.9 | 2.72 | 3973,07 | 0,61 | 0,02 |
| 4252 | 4414.0 | 162.0 | 3.81 | 4270,87 | 19,13 | 0,45 |
| 4525 | 4751.6 | 226.6 | 5.01 | 4568,35 | 43,57 | 0,96 |
| | | **Mean %** | **2.549** | | | **0,639** |

Although all the above results are unprecedented, the zero molecular parameter solution delivers a continuous PEC, not aware of turning points, related to vibrational levels, obeying quantum mechanics. However, Fues (1926) pointed out a long time ago how to solve the wave equation exactly for the Kratzer potential, which closes the circle. At this stage, it seems unavoidable that the nature of bonding in covalent molecules is basically ionic, conforming to gauge-symmetry and that PECs obey Coulomb's law, with an asymptote deriving simply from atomic characteristics.

It seems all three criteria given at the end of Section 6 are met.

*11.3. The procedure: LiH and the performance of the variables $d$, $d^2$, $k$, $k^2$ and $k_{gen}$*

If the zero molecular parameter function works for $H_2$ and $Li_2$, it must also apply to LiH, the third member of the critical series for an *ab initio* approach. The global data for LiH were already incorporated in Fig. 10c but we now use this bond as a test case for the general scaling procedure with various variables. The first 4 correspond roughly with the four potentials $a_nR^n$ in (3b) discussed above, the fifth is the generic variable (22l).

**Fig. 11a** gives an algebraic plot of all five variables versus the LiH-RKR, known almost to 100% of the atomic dissociation limit. We use 23 levels (46 turning points). It is obvious from Fig. 11a that a linear fit through the origin can never produce a smooth relation for d, k and even $d^2$, the basis of conventional oscillator models (Morse, Dunham) and so widely used in theoretical models (Witten, 1981, 1982). For the variables of the Coulomb approach $k^2$ and $k_{gen}$ linear relations are detected, with slopes producing asymptote values close to the Dunham, ionic and generic asymptote values. The goodness of fit is over 0,98 for Kratzer's variable and over 0,99 for the generic Coulomb variable. This proves the present procedure extends to this important bond also (see Section 11.2).

In addition, we observe diverging points at long-range (about 5) probably symmetrically followed by 5 turn over points at the short range but which are less visible. Turn over points are more clearly visible with the Kratzer than with the generic Coulomb variable. Despite this, all 46 turning points are within 1,85 % (or 0,068  ) of the RKR for the generic variable fit, which is close to but not of spectroscopic accuracy provided this RKR is correct.

**Fig. 11b** gives the results for 18 levels and their 36 turning points leaving out extreme turning points. The asymptote for the generic variable is within 0,086 % of the generic asymptote $e^2/R_e$. In this case, the accuracy of the generic Coulomb potential model is simply impressive. Turning points are within 0,54 % of R (corresponding with an average deviation of 0,012  ) the largest deviations still being found at the long-range attractive branch. For the repulsive branch the deviation reduces to 0,0047  , even closer to spectroscopic accuracy.

The quantitative possibilities of this simple non-empirical *ab initio* model potential (22l) are obvious. We can easily calculate the turning points from observed level energies. Due to the nature of the generic function, this will always produce a perfect balance. Since no parameters are involved, this may form the basis of an alternative inversion process. This example for LiH clearly shows that spectroscopic accuracy can be obtained with Coulomb behaviour. Apart from the theoretical consequences these results have, they also have practical implications for the determination of turning points by an inversion process.

In fact, we are now at a stage where we could even suggest to revise published turning points. Mainly for practical reasons we will adhere to these published turning points and derive 'operational' asymptotes in the framework of a Coulomb based scaling scheme by using a similar analysis as the one illustrated in Fig. 11a and 11b. This means that any asymptote obeying Coulomb's law and its scaling power (see Fig. 6a-c) may be used. Asymptotes not obeying 1/R behaviour must lead to deviations in the scaling procedure.

*11.4 Determining the asymptote for PECs: the universal function, 1$^{st}$ scaling procedure, classical view*

The most direct way to find the asymptote for a universal PEC is to plot the observed energy of the levels against the variable or the function and fit the data with a linear equation as exemplified in section 11.3. This procedure is generalised in **Fig. 12a** for 7 bonds with smaller $R_e$-values and **Fig. 12b** for 6 bonds with large $R_e$-values. Here all RKR- or IPA-data are plotted against $k_{gen}$. The 99 long-range points (at > 50% of $D_e$) are shown but have not been included in determining the fit. The remaining 409 turning points cover about 75% of the total PEC. The asymptote values obtained are shown with the goodness of fit (typically 0.99 or better). In general these operational asymptotes are close to but not equal to the Dunham asymptotes. This can be interpreted in two ways as indicated above: either the $R_e$ values must be (slightly) adapted or the published turning points would need revision. This is a general problem when discussing PECs constructed with inverting model potentials. In the case of scaling, slight deviations are allowed provided the general trend of inversion technique used is obeyed and physically or chemically meaningful. As indicated above, we choose, mainly for practical reasons, to calculate operational asymptotes instead of recalculating all the turning points, which is mathematically equivalent.

Table 3. *Average operational asymptote values computed from RKR/$k_{gen}$ and RKR/$k^2$ for 407 points for 13 bonds in cm$^{-1}$ (the 13 minima are not taken into account). Deviations (%) of theoretical level energies calculated with $k_{gen}$ and asymptotes (from graphical fit or pivot table) from observed ones.*

| Branch | Data | AuH* | Cs$_2$ | H$_2$ | HF | I$_2$ | KH | KLi | Li$_2$ | LiH | NaCs | Rb$_2$ | RbCs | LiF | Total |
|---|---|---|---|---|---|---|---|---|---|---|---|---|---|---|---|
| attr | # | 58 | 10 | 5 | 2 | 9 | 6 | 15 | 7 | 8 | 17 | 12 | 11 | 2 | 162 |
| | Abs% level energy (fit) | 10,31 | 6,45 | 13,86 | 25,01 | 55,01 | 16,93 | 1,33 | 0,57 | 3,40 | 5,51 | 3,00 | 5,13 | 1,42 | 10,0 |
| | RKR/$k^2$ | 156898 | 35997 | 88920 | 186973 | 250894 | 65543 | 41121 | 44881 | 65662 | 39064 | 34967 | 35482 | 156433 | 99833 |
| | StdDev of RKR/$k^2$ | 9684 | 4530 | 2811 | 5521 | 22632 | 3070 | 1590 | 1226 | 649 | 2081 | 2004 | 2956 | 379 | |
| | RKR/$k_{gen}$ | 159633 | 37010 | 98872 | 199185 | 176613 | 69352 | 42769 | 46723 | 70995 | 40276 | 35789 | 36378 | 162090 | 98162 |
| | StdDev of RKR/$k_{gen}$ | 11295 | 3965 | 8261 | 2151 | 132794 | 432 | 684 | 272 | 2332 | 1257 | 1320 | 2255 | 3226 | |
| | Abs% level energy (pivot) | 5,86 | 6,94 | 6,61 | 0,76 | 29,85 | 0,47 | 1,34 | 0,45 | 2,64 | 2,65 | 3,08 | 5,00 | 1,41 | 5,55 |
| rep | # | 36 | 28 | 15 | 5 | 20 | 14 | 15 | 10 | 23 | 26 | 19 | 19 | 2 | 232 |
| | Abs% level energy (fit) | 14,73 | 1,03 | 3,42 | 5,49 | 6,71 | 3,72 | 1,89 | 0,97 | 3,34 | 1,56 | 1,59 | 1,49 | 7,13 | 4,53 |
| | RKR/$k^2$ | 204221 | 33542 | 63727 | 215346 | 382163 | 71635 | 42142 | 43328 | 60441 | 40997 | 35548 | 35100 | 144335 | 103976 |
| | StdDev of RKR/$k^2$ | 17234 | 1127 | 4636 | 1192 | 23639 | 1379 | 904 | 1564 | 2130 | 1266 | 1297 | 1297 | 6891 | |
| | RKR/$k_{gen}$ | 206479 | 35888 | 82117 | 243113 | 389978 | 80054 | 43794 | 46138 | 71436 | 43112 | 37123 | 37014 | 152224 | 109487 |
| | StdDev of RKR/$k_{gen}$ | 15942 | 1080 | 2832 | 15868 | 28935 | 3647 | 26 | 369 | 2773 | 440 | 139 | 593 | 915 | |
| | Abs% level energy (pivot) | 5,67 | 1,34 | 2,88 | 5,32 | 6,70 | 3,83 | 0,05 | 0,69 | 3,27 | 0,87 | 0,33 | 1,14 | 0,43 | 2,73 |
| | # | 94 | 38 | 20 | 7 | 29 | 20 | 30 | 17 | 31 | 43 | 31 | 30 | 4 | 394 |
| Asymptote from graphical fit | | 175155 | 36038 | 84673 | 248983 | 390257 | 81088 | 42965 | 46552 | 73207 | 42457 | 36534 | 36736 | 163075 | 116580 |
| Average RKR/$k^2$ | | 175021 | 34188 | 70025 | 207240 | 341424 | 69807 | 41632 | 43968 | 61788 | 40233 | 35323 | 35240 | 150384 | 102273 |
| Average RKR/$k_{gen}$ | | 177574 | 36183 | 86306 | 230562 | 323761 | 76843 | 43281 | 46379 | 71322 | 41991 | 36607 | 36781 | 157157 | 104831 |
| Abs% level energies (fit) | | 12,0 | 2,5 | 6,0 | 11,1 | 21,7 | 7,7 | 1,6 | 0,8 | 3,4 | 3,1 | 2,1 | 2,8 | 4,3 | 6,8 |
| Abs% level energies (pivot) | | 5,8 | 2,8 | 3,8 | 4,0 | 13,9 | 2,8 | 0,7 | 0,6 | 3,1 | 1,6 | 1,4 | 2,6 | 0,9 | 3,9 |
| Ratio abs % energy levels | | 2,1 | 0,9 | 1,6 | 2,8 | 1,6 | 2,7 | 2,3 | 1,4 | 1,1 | 2,0 | 1,5 | 1,1 | 4,7 | 1,7 |

* PEC computed by Le Roy's potential. Although this 'empirical' PEC is reproduced qualitatively in an acceptable way, the picture is slightly distorted. Further work is needed. Without these AuH-data, the absolute % deviations at the attractive side are 5,38 %, at the repulsive side 2,19 %. For the global results on all remaining data points the errors are respectively 5,14 and 3,29 %.

To check the generic procedure further we calculated average RKR/$k^2$- and RKR/$k_{gen}$-values for the same 395 data-points used in Fig. 12a and 12b. The results are given in Table 3 together with standard deviations for the asymptote values and the resulting deviations for the level energies, calculated with asymptotes obtained from the graphical fitting procedure and the pivot table results for RKR/$k_{gen}$. The average asymptote values are all close to Dunham's (see above). The major difference resides between asymptotes for repulsive and attractive branches. Pertinent examples are AuH, HF, $H_2$ and $I_2$, the last three are older RKR-curves. Absolute deviations for level energies returned by this method are given for each bond and for each branch. Various published or computed turning points quite exactly match those predicted by our model potential (KH, KLi, $Li_2$, NaCs and RbCs) with an error of about 1 %, which justifies our decision not to recalculate the turning points. If the published tutning points are exact, the only molecular parameter that could account for the deviations in asymptote values is $R_e$, since  $_e$ and reduced masses are well known. The $R_e$-value is decisive for obtaining turning points. Even a very small shift in $R_e$ (1 % or about 0,01  ) can have a drastic influence upon the turning points calculated by semi-empirical approaches (RKR, IPA). This problem is discussed further below. Overall deviations are collected in the last rows of Table 3, starting with the headers *Abs% level energies (fit)* and *Abs% level energies (pivot)*. For five bonds, the average deviations are equal to or smaller than 1,6 %. The worst results are obtained for $I_2$ (as to be expected from Fig. 6d above) and AuH (see also footnote of Table3). The ratio of absolute deviations by the two methods is given in the last row. On average (last column), the pivot approximation (Table 3) returns level energies with an accuracy about 70 % greater than with the graphical fitting procedure. The accuracy for the attractive side (deviation 5,55 %) is less than for the repulsive side (deviation 2,73%) leading to an overall deviation of 3,9 % for the level energies in 75 % of the complete PEC for 13 bonds.

In general, the differences between the averaged asymptotes in Table 3 are minor and are not of that order to influence the basic scaling results we want. The final 'averaged' value for the asymptote at the repulsive side is almost equal to the exact Coulomb asymptote (or 1 Ryd), see Table 3, last column. This is strange but conforms to the general idea about the importance of Coulomb's law in molecular spectroscopy. We refer to Fig. 6d, where this is the unique reference asymptote for all bonds at the repulsive side. For the attractive side, this standard value is reached for about 80 %, meaning either that more turn over points are present or that the turning points at this side are too large in general. These results seem to point out that, given the definition and the correctness of the Dunham asymptote values, the published turning points of some RKRs may indeed be in need of revision.

Finer details for the short range behaviour are observed, i.e. turn over points, in agreement with the analysis by Zemke and Stwalley (1999) in the case of NaK at v=60 near the dissociation limit. Having at our disposal a 'generic' function, it is straightforward to detect turn over levels. The empirical function of Gordon and Von Szentpaly (1999) claims a high accuracy for repulsive branches of PECs but it does not detect these important turn over points at short-range. This is certainly a topic for further investigation as similar effects must also be visible at the attractive branch, where long-range behaviour appears (see Fig. 11a for LiH). For the long-range, deviations set in at a particular R-value, maybe around the critical distance or at the Le Roy radius, which we could now only empirically estimate with reference to the goodness of fit in the asymptote finding process.

**Fig. 13a** gives a graphical illustration of all the data points in Table 3. Quantitative details about the level energy accuracy are given in Table 3. At the attractive side and in cases where the RKR/IPA reaches $D_e$, it appears that atomic asymptotes must be considered as 'deviations' from Coulomb scaling. In fact, at the repulsive side Coulomb behaviour is respected throughout in a species independent way. We can not but conclude that $D_e$ is indeed not an asymptote conforming to Coulomb scaling, as argued above.

This is clearly visible in **Fig. 13b** where 506 theoretical and observed level energies are plotted against $k = 1-R/R_e$. The atomic asymptotes appear as side branches of the general reduced PECs. We easily verify that anharmonic oscillator behaviour is nicely obeyed, i.e. a near quadratic dependence on the Kratzer variable k. Of all 13 bonds, only $H_2$ seems to obey Kratzer $k^2$-behaviour at long range with the asymptote deriving from the graphical procedure. Using a higher value (closer to the atomic ionisation energy) will give $k_{gen}$ behaviour at this branch, see also Fig. 10a. Fig. 13b must be compared with the general scaled result in Fig. 1b to notice the effect of our procedure.

The data-points for AuH (Seto et al., 1999) are hardly visible. The three coinage metal PECs AuH, AgH and CuH (Hajigeorgiou and Le Roy, 1999, Seto et al., 1999) show a very similar behaviour in comparison with the generic Coulomb variable, but are slightly distorted, especially at the minimum, where the largest % deviations are found. Since these are unusually large deviations, we left out the 100 coinage metal PEC-data for AuH, which reduces the absolute deviation to almost half of the total or 3,3 % for the 300 turning points for the remaining 12 largely different bonds. The relative large number of turning points for one bond AuH in the data range (about 100) would otherwise have distorted the whole picture for the remaining bonds (quantitative results are given in footnote of Table 3).

Whether or not our theoretical conclusions given above that the atomic dissociation limit is not necessary to reveal the nature of the universal function, must become more apparent when we scale PECs with $D_e$, i.e. test $D_e$ for Coulomb scaling as in Fig. 6a-c. In essence, the results of this section are in support of the general theory outlined above, in particular with the generic and completely *ab initio* $k_{gen}$ variable in (22l) leading to acceptable results in the zero molecular parameter approach (see 11.2) even without using (or even knowing) $a_0$. These results are in support of the intra-atomic charge inversion technique.

*11.5. Effects of the first 1st step in scaling: V-shaped or linear reduced PECs coinciding at $R_e$*

Up till now, scaling effects of RKRs are usually presented in the classical form RKR(R) or RKR(R/$R_e$) as in Fig. 1b above. The algebraic linear approach is more illustrative and, to our knowledge, unprecedented. Using the generic *ab initio* Coulomb function, it is a simple matter to present V-shaped PECs with perpendicular legs or even linear PECs (see the Fig. 2a version of Fig. 3a). We reduce the levels with $a_{(piv)}$ at the y-axis and use the generic Coulomb variable $k_{gen}$ on the x-axis. For all 506 points for 13 bonds these V-shaped PECs are given in **Fig. 14a**. The agreement at the repulsive branch is again impressive, since all data collapse into a single line (left branch of the V). At the attractive side, the same effect is noticed but the various atomic dissociation limits are clearly visible, when the RKR extends to that region, as in Fig. 13b. This result must be compared with the Morse equivalent in Fig. 1c above.

The consistency of our generic Coulomb function shows even better when, as in Fig. 2a, the results are converted algebraically into a linear form (**Fig. 14b**). It is not difficult (now) to ascertain that the attractive branch is simply an algebraic continuation of the repulsive branch, the essence of a Coulomb approach (see above). It illustrates the perfect mathematical symmetry between attraction and repulsion, which proves our point about Coulomb scaling for PECs and charge inversion in full detail. The reference point here is the equilibrium distance, the origin of all V's. When compared with Fig. 1d, Fig. 14b is almost an astonishing result as so many people during so many years have been trying to find a universal function. Atomic dissociation limits, looked from the Coulomb physics of a 4-particle system are misleading, a *trompe l'oeil* effect (Van Hooydonk, 1999). We must now try to find out if the complete curves are scalable, i.e. from origin to asymptote (identical V's with scaled legs).

*11.6. The 2$^{nd}$ step: scaling from asymptotes*

The second step of the scaling process can not but deal with the 'chemical' asymptotes $D_e$. Normally, scaling simply with $D_e$ gives rise to reduced PECs varying from 0 to 1 (all legs of the V's would then be equal). We already noticed, see Fig. 6d, that for this process $R_e$ values (generic asymptotes) are unavoidable. Unfortunately, the transition from ionic to covalent behaviour falls outside the range covered by our generic approach (see Fig. 13b) and we must use an indirect method to zoom in on this part of PECs. A universal function f(R) can be rescaled from any viewpoint, if the (Coulomb) scaling mechanism is not distorted. The disadvantage of the classical view using (a) $D_e$ for scaling RKRs and (b) $R_e$ for scaling R, as in Fig. 1b, is that it is probably distorted. There is no simple smooth or scaling relation between $D_e$ and $R_e$ (determining the Coulomb asymptote), the main conclusion of this report. If we want to check the long-range forces for Coulomb behaviour, we must rescale the PECs accordingly, i.e. use the generic Coulomb potential $e^2/R_e$ instead of the atomic asymptote as a starting point (see Fig. 6d on the general Coulomb behaviour of bonds). In practice, this would correspond with setting the atomic limit equal to 1 and shifting the data for true Coulomb behaviour, which we are now able to do.

In the scaled picture Fig. 14a, all V's coinciding at the origin, the picture is completely Coulombic, atomic dissociation limits appear as disturbances. Therefore, it is of interest to see how the bonds behave when their asymptotes are aligned. The asymptote values in Fig. 14 are equal to $D_e/a_0$.

Shifting the asymptotes by their differences (the maximum $D_e/a_{0(piv)}$ value is 0,46620 for $H_2$, for the 15 turning points at the attractive branch of $H_2$ we used the ionisation energy as asymptote) gives the results presented in **Fig. 15a**. The global picture hardly shows diverging behaviour near the atomic dissociation limit. This means that, when local scaling at individual $R_e$-values is retained, no great anomalies seem to appear. However, the insert in Fig. 15a shows that this is not so in reality. The upper clustering in this insert shows that $I_2$ is behaving abnormally, the two other bonds clustering here are $H_2$ and LiH. The lower clustering set encompasses the remaining bonds, for which the PEC extends to $D_e$. The clustering of MM bonds (M alkali) seems real, KH is slightly above the MM clustering line, HF and LiF are both slightly below. Nevertheless, the global picture in Fig. 15a would suggest that the mechanism governing the transition from the ionic PEC to the atomic dissociation limit is probably very similar, even scalable, for all bonds.

In this critical region, the only mechanism applicable is obvious and is the same for all bonds: it consists of charge transfers from two ionic bonding partners to two (weakly interacting) atomic systems. However, the good agreement in the global picture Fig. 15a is in part due to the fact that the legs of the V-shaped PECs are not scaled yet for Coulomb behaviour, which we must also try to remedy.

Scaling RKRs directly *with* $D_e$ makes all legs of the V-s, regardless of the value of $R_e$, equal to 1 if the RKR extends to $D_e$. The question is whether or not this conforms with Coulomb behaviour as in Fig. 6a-d were the effects of Coulomb scaling on PECs are illustrated. This is the most intriguing case, since scaling RKRs with $D_e$ is the standard procedure (see the Sutherland parameter) and the basis of Morse's function, see Fig. 1d. No linear shifts have to be applied. The question whether or not Coulomb behaviour is respected when using $D_e$ for scaling can now be answered in more detail. The result is shown in **Fig. 15b** and this must be compared with the benchmark Morse solution in Fig. 1d. We remark that in principle a perfect scaling procedure can not be destroyed by the choice of any Coulomb asymptote (see above: any asymptote obeying the same $1/R$ law will reproduce invariantly the same scaled results, see Fig. 6c, Fig. 14 and Fig. 15). However, the use of $D_e$ as a scaling asymptote does not obey this principle. In fact, scaling with $D_e$ shows an expected duality: Coulomb scaling is retained at short range, even with $D_e$, which proves a unique Coulomb law is active at the repulsive side, but at the attractive branch the situation gets distorted more than in Fig. 15a. The atomic dissociation limit is not suited for scaling this branch of a PEC when speaking in terms of a Coulomb process. This may seem trivial (as pointed out above, since $D_e$ refers to two neutral particles) but exactly this asymptote has -up till now- invariantly been used for scaling RKRs in molecular spectroscopy. $D_e$ can not act as a scaling asymptote for a process of interacting charged particles dissociating into *neutral* particles, unless it would be itself a scaled form (a projection) of the asymptote consisting of charged particles. This latter possibility can not be excluded *a priori* but remains to be proven. Fig. 15b is nevertheless more consistent than the Morse benchmark Fig. 1d. At long range, we observe roughly three clusters: HH and LiH; HF, LiF and KH; all MM (M=alkali metal); $I_2$ is (again) a clear exception in all respects.

The only thing that is clear now is that, at long range, there is no simple $1/R$ dependence, governed by $D_e$ as an asymptote. Only not too far from equilibrium, Coulomb behaviour towards the asymptote is retained with $D_e$. It may be possible of course that $D_e(R)$ may be a scaling function in some cases at attractive branches (Morse behaviour), but this relation has not yet been found. In fact, we are convinced long range behaviour will have to be explained by the interplay between long range forces of type $C_n/R^n$ (see above) and Coulomb $1/R$ behaviour and with the charge transfer process in the critical region.

We remind the basis of our approach is essentially

$RKR = s(e^2/R_e)k_{gen} = a_0 k_{gen}$

where s is a species dependent parameter.

After multiplying with $R_e$ we get

$(RKR)R_e = R_e a_0 k_{gen}$

which should reflect more appropriately the Coulomb scaling process illustrated in Fig. 6d: it is an attempt to scale the length of the V-legs according to the position of the bond in the global $W(R)$ Coulomb field. The asymptote is now $D_e R_e$, having a maximum for LiF at 75258 cm$^{-1}$. **Fig. 15c** gives the corresponding results. As in Fig. 15a, the first clustering of lines refers to $H_2$ and LiH, the second to all other bonds except $I_2$ (an exception also at the repulsive branch) and the Morse RKR for LiF. Leaving out these latter two cases, it appears that scaling RKRs with a Coulomb asymptote $e^2/R_e$ indeed leads to more consistent results for the complete PEC than scaling with $D_e$, also the essential conclusion of our previous analysis of the constants (Van Hooydonk, 1999). This leads to the interesting prospect that the long-range situation in PECs may be derived in an elementary way from their short-range behaviour, which is an unprecedented result also. Finally, **Fig. 15d** gives observed level energies in function of the theoretical ones, both shifted by (49406 - $D_e$) where 49406 is the largest $D_e$-value in the set for HF. The anomalous $I_2$ set has been left out. Clustering now extends throughout the complete data range, except near the asymptote. The few diverging points here belong to only two molecules HF and LiF. This scaling result from the asymptote is by far the best we could obtain up till now and it does not use $D_e$ as a Coulomb scaling factor (Fig. 15b). Diverging points are now confined to the relatively small triangle anchored to the intersection of ionic and covalent curves. It strengthens our idea that scaling long-range behaviour must be possible, as it relies upon the same charge-transfer process, needed in all cases to convert ionic states to atomic states.

If this is valid, we can 'safely' put V"(R) = 0 in equation (22m) in first order. For all data points, we applied this simple working hypothesis for the complete PECs. This effortless procedure gives an absolute deviation of about 4% for all data points with the function (22m) in a zero perturbation approximation, which is almost the same deviation as that found for all points below 50 % of $D_e$ mentioned in Table 3. The corresponding graph is shown in **Fig. 16**, whereby the values obtained with (22m) are easily retraced. We remark that these data refer to 100% of the observed RKRs for 13 different bonds. On the whole, this absolute deviation for energy levels is well within the criteria for a function to be universal when it comes to reproduce PECs (Van Hooydonk, 1999, Varshni, 1998, private communication). The slope is very close to unity and the goodness of fit is almost 0,999. When looking at previous scaling attempts, also this is an unprecedented result.

*11.7. Turning points in RKRs and in PECs deriving from a Coulomb based scheme*

The accuracy of the turning points depends solely on the accuracy of the model potential. This delicate matter was recognised a long time ago. Using an RKR as an *observed PEC* -the working hypothesis of our present analysis- remains a matter of belief, faith or trust, as no real benchmark solution is available. In other words, there is no alternative for checking the present theory. The dissatisfaction with conventional RKRs and the Dunham expansion ultimately led to alternate inversion methods (IPA) to arrive at PECs.

Table 4. *Agreement between 394 RKR and Coulomb turning points below 50% of $D_e$.*

| Side | Data | Bond | | | | | | | | | | | | | |
|---|---|---|---|---|---|---|---|---|---|---|---|---|---|---|---|
| | | AuH | CsCs | HH | HF | II | KH | KLi | LiLi | LiH | LiF | NaCs | RbRb | RbCs | Total |
| attr | # | 58 | 10 | 5 | 2 | 9 | 6 | 15 | 7 | 8 | 2 | 17 | 12 | 11 | 162 |
| | abs dev in % | 0,346 | 0,583 | 1,899 | 0,133 | 1,716 | 0,0813 | 0,16478 | *0,0464* | 0,514 | 0,1571 | 0,281 | 0,29 | 0,431 | 0,4432 |
| | in Angstrom | 0,006 | 0,034 | 0,02 | 0,0017 | 0,0507 | 0,0026 | 0,00706 | *0,0016* | 0,012 | 0,003 | 0,014 | 0,0153 | 0,024 | 0,0134 |
| rep | # | 36 | 28 | 15 | 5 | 20 | 14 | 15 | 10 | 23 | 2 | 26 | 19 | 19 | 232 |
| | abs dev in % | 0,218 | 0,082 | 0,565 | 0,7009 | 0,3319 | 0,4472 | *0,00310* | 0,06716 | 0,497 | 0,0381 | 0,076 | 0,0244 | 0,079 | 0,2205 |
| | in Angstrom | 0,003 | 0,003 | 0,003 | 0,0047 | 0,0078 | 0,0075 | *0,00009* | 0,00141 | 0,005 | 0,0005 | 0,002 | 0,0008 | 0,003 | 0,0034 |
| | # | 94 | 38 | 20 | 7 | 29 | 20 | 30 | 17 | 31 | 4 | 43 | 31 | 30 | 394 |
| Average abs dev in % | | 0,297 | 0,214 | 0,898 | 0,5386 | 0,7615 | 0,3375 | 0,08394 | *0,0586* | 0,501 | 0,0976 | 0,157 | 0,1272 | 0,208 | 0,3121 |
| Idem in Angstrom | | 0,005 | 0,011 | 0,007 | 0,0038 | 0,0211 | 0,006 | 0,00357 | *0,0015* | 0,007 | 0,0018 | 0,007 | 0,0064 | 0,011 | 0,0075 |

Especially here, the H-L theory is not very useful either, as the number of approximations to be made in sophisticated quantum mechanical calculations to obtain a theoretical PEC is innumerable (Pople, 1999). We remind the techniques underlying the computations for four-particle systems, studied by Richard and Abdel-Raouf (cited above), where about 300 parameters are needed (Hylleraas type approximation) to get theoretical results with a reliable CL.

For all these reasons, we adopted the same pragmatic procedure as above to cope with this important issue. We calculated the turning points by our generic scheme for all data points below 50% of $D_e$ using the pivot $a_0$- values and the observed level energies in cm$^{-1}$ (see Table 3). The results are shown in **Fig. 17** and in Table 4, where deviations in italics are minima and those underlined are maxima in their category. The average 'deviation' in the turning points is 0,31 % and shows the over-all agreement is acceptable, although the present Coulomb model potential is extremely simple. For the attractive branches, deviations are in general 'large', i.e. 0,44 %. For the repulsive branches the global deviations are 2 times smaller in % (0,22 %) and much smaller in in comparison with the attractive branches. This latter agreement brings us closer to spectroscopic accuracy, as the 0,001  barrier or even lower can be reached.

*11.8 Specific bonds and systematics*

- AuH is one of the three coinage metal hydrides studies by Seto et al., 1999 using the Le Roy fitting technique to extract the complete PEC from a number of observed levels. It is not difficult to obtain a complete PEC using the generic model potential. The results are shown in **Fig. 18**. The agreement near $R_e$ only seems satisfactory from the plot: in reality very large % deviations are found. At the short range a slight divergence (in %) is noticed. At long range, the differences are more pronounced and are similar for all three coinage metal PECs (not shown). We used (22m) to compute the long range PEC with the generic function only.

- HF is an intriguing molecule and its PEC has been studied extensively (see Coxon and Hajigeorgiou, 1990). With its companions DF and TF, it is suited to study BOB (breaking of the Born Oppenheimer approximation). As an illustration of our procedure, we used an hybrid potential $k^2_{(rep)} + k_{gen(att)}$ and used a graphical fitting technique to determine the asymptotes in each case. We get 216756,46 cm$^{-1}$ for the repulsive Kratzer branch and 197861,72 cm$^{-1}$ for the attractive generic branch, giving an average value of 202309,1 cm$^{-1}$ within 1 % of the Dunham asymptote 204308 cm$^{-1}$ (see Table 1). The resulting PEC, using (22m) is given in **Fig. 19**. The agreement is acceptable. The average deviation for all 40 turning points, including those at long range (the triangle), is a modest 1,8 %.

- $I_2$ is a clear exception in this series of 13 bonds. The $R_e$ value compares with that of $Li_2$ but the two PECs are completely different, as easily verified in Fig. 6d. The PEC for $Li_2$ is consistent with a Coulomb model but that for $I_2$ is not (the Dunham asymptote for $I_2$ is exceptionally large). The atomic data are largely different, $IE_I$ = 10,45 eV whereas $IE_{Li}$ is only 5,4 eV. This suggests the $R_e$ value for $I_2$ is long by about 1 Å. In Fig. 6d, the PEC for $I_2$ should correspondingly be shifted towards lower R-values, where the shape of the reported RKR would be in line the shape predicted by a Coulomb model. As a matter of fact, $I_2$ is the only case out of 13 where the RKR does not fit into the gauge-symmetry based bonding model that led us to Fig. 6d. Analysing the PECs for the other halogens can be of help.

- For LiF we used a Morse-curve (Van Hooydonk, 1982), since the two alternatives available, a Padé approximant (Jordan et al., 1974) and a Kratzer PEC (Van Hooydonk, 1982) do not show asymptote behaviour near $D_e$. The Kratzer PEC would be too close to the generic scheme presented here.

*11.9 Intermediate conclusion*

At this stage (the usefulness of the zero molecular parameter function and the scaling results), the pertinent question is about the agreement between our model potential and the available inversion techniques (depending on iterative processes and higher order WKB terms). This is certainly a matter of further investigation. A correct scaling procedure can not but lead to the scaling of different PECs into a single straight line. To the best of our knowledge, only the present Coulomb scheme can achieve this with reasonable success.

These two kinds of results provide us with ample evidence that our earlier conclusions (Van Hooydonk, 1982, 1983, 1985, 1999) are essentially correct. The hyper-classical Coulomb law and its inherent scaling capacity are important for rationalising the abundant data in molecular spectroscopy. But the main conclusion of this section with quantitative results about PECs from atomic data and about scaling is that intra-atomic charge inversion (chirality, parity violation) at the Bohr-scale is a reality. A previous report on the same issue (Van Hooydonk, 1985) lacked the generic Coulomb function for calculating PECs and the experimental confirmation presented here. In fact, the only reaction on this first attempt to include the generic effect of intra-atomic charge inversion in a theory for the chemical bond came from the late Pauling (1985), who wrote that 'charge-symmetry is broken by the electron-proton mass difference' (see further below).

## 12. Generic effects of intra-atomic charge inversion

Table 5a. *Matrix of 16 states for two interacting atoms (absolute charge distributions)*
Intra-atomic charge distributions in atoms $X_1$ and $X_2$ and the sign of interatomic interactions (nucleon charge between brackets) different states for any molecule $X_1X_2$ (homonuclear bonds only)

|  | World > 0 $X_1$ | | World < 0 $X_1$ | |
|---|---|---|---|---|
|  | (+)+ | (+)− | (−)+ | (−)− |
| World > 0 $X_2$ | | | | |
| (+)+ | (+)+\|(+)+ | (+)−\|(+)+ | (−)+\|(+)+ | (−)−\|(+)+ |
| Total charge | +4 | +2 | +2 | 0 |
| (+)− | (+)+\|(+)− | (+)−\|(+)− | (−)+\|(+)− | (−)−\|(+)− |
| Total charge | +2 | **0** | **0** | −2 |
| World < 0 $X_2$ | | | | |
| (−)+ | (+)+\|(−)+ | (+)−\|(−)+ | (−)+\|(−)+ | (−)−\|(−)+ |
| Total charge | +2 | **0** | 0 | −2 |
| (−)− | (+)+\|(−)− | (+)−\|(−)− | (−)+\|(−)− | (−)−\|(−)− |
| Total charge | 0 | −2 | −2 | −4 |

Heitler and London (1927) could not foresee the power of a direct internal (algebraic) correlation between fermion-boson symmetry and potential energy in a four-particle Hamiltonian. This link shows at the Bohr-scale, operates when going from atoms to molecules by means of a generalised Bohr-formula and is measurable in eV. The scale invariance has interesting prospects. The energetic effects of a bonding process are easily described by a static Coulomb law (with charge inversion in one atom or boson) and may provide us with a simple alternative to the H-L theory, all other things remaining equal. The two approximations use the same Pauli-matrices. Since the energetic effect of spin is small, not of order eV, spin symmetry effects must operate on wave functions (H-L theory).

It is not difficult to apply the charge inversion technique to four-particle systems in general, giving 16 theoretically possible states, equally distributed over two symmetrical worlds, a chemical eight-fold way. Allowing for the principle of charge inversion, the four particles, arranged in a pair of two atoms, lead to a number of forbidden or allowed Hamiltonians, mainly on account of the *total charge of the four-particle system*. These are not discussed here in full, since their characteristics can be derived from the formulae above. Table 5a gives the matrix for atom combinations. Only the neutral states in the centre correspond with allowed states. Neutral states with **0** are H-L states. For these states to become bonding, the combined symmetry effects of electron-spin and of electron wave functions must be invoked. Neutral states with **0** lead to a PEC deriving from parity-violation adapted Hamiltonians, i.e. without needing spin and wave function symmetry effects but using charge inversion instead.

Table 5b gives the matrix for ionic interactions in the generic Coulomb scheme for bonding. Singlet-states in bold result from charge inversion. Doublets in Table 5b do not arise from charge inversion but from lepton-rotation. At the real asymptote, a four-particle Coulomb system must not be divided into two interacting boson pairs (two neutral atoms) but into two interacting fermion pairs. This generates splitting within the Hamiltonian, which is then soluble classically, see Section 8. Each world has one H-L state and two degenerate states with intra-atomic charge inversion (parity violating states in the conventional way). These are shown with a total zero charge in bold in Table 5b.

Table 5b. *Matrix of 16 states for two interacting ions (absolute charge distributions)*
Intra-ionic charge distributions in anion and cation and the signs in homonuclear bonds (nucleon charges between brackets)
Anion states +(+)- and -(+)+ do not imply charge inversion, only rotation is involved.

|  | World <0 anion | | | | World >0 anion | | | |
|---|---|---|---|---|---|---|---|---|
|  | +(+)+ | +(+)- inversion | -(+)+ =rotation | +(-)+ | -(+)- | +(-)- Inversion | -(-)+ =rotation | -(-)-- |
| Total; mean charge* | +3; +1 | +1; +1/3 | +1; +1/3 | +1; +1/3 | -1; -1/3 | -1; -1/3 | -1; -1/3 | -3, -1 |
| World > 0 Cation (+) Bond | +(+)+(+) | +(+)-(+) | -(+)+(+) | +(-)+(+) | -(+)-(+) | **+(-)-(+)** | **-(-)+(+)** | -(-)-(-) |
| Total charge | +4 | +2 | +2 | +2 | 0 | **0** | **0** | -2 |
| Remark |  |  |  |  | HL state |  |  |  |
| World < 0 Cation (-) Bond | +(+)+(-) | **+(+)-(-)** | **-(+)+(-)** | +(-)+(-) | -(+)-(-) | +(-)-(-) | -(-)+(-) | -(-)-(-) |
| Total charge | +2 | **0** | **0** | 0 HL state | -2 | -2 | -2 | -4 |

* mean charges are discussed below

We have restricted the present analysis to bonds between mono-valent atoms but it is straightforward to apply the principle of charge inversion to polyvalent atoms. Consider a bond between two divalent atoms in the 2nd or 6th Column of the Periodic Table. The classical atom has a charge distribution -(++)-, where the charges between brackets relate to the nucleon. Inverting charges in only one lepton-nucleon pair gives -(+-)+ and in both gives +(--)+. Interactions between two divalent atoms can proceed in two ways either as: a. -(+-)+/-(+-)+ in which one pair is inverted in each atom or as: b. -(++)-/+(--)+, whereby two pairs are inverted in one atom. In terms of Coulomb's law, type b. bonding will lead to strong nucleon-nucleon bonding, whereas type a. bonding will be (very) weak, since only lepton-lepton interactions can generate a bond. Atoms of the 6th Column ($O_2, S_2,...$) form strong bonds indeed, whereas those of the 2nd Column ($Be_2, Mg_2, Ca_2,...$) form weak bonds.

## 13. Born-Oppenheimer (B-O) approximations

BOB, breaking of the B-O approximation, shows when applying a Dunham type expansion to isotopomers. The B-O-approximation freezes nucleons and considers two central force systems (reduced masses). Temporarily freezing leptons instead leads to an inverted B-O and gives

$$H_{12} = (½m_1v^2 + ½m_2v^2 - e^2/R_{12}) - e^2/R_{ab} \qquad (23a)$$

using the charge inverted, parity violating Hamiltonian. The further treatment runs exactly as in section 8 in the case of the B-O approximation. The minimal solution ($R_{ab}$ perpendicular to $R_{12}$) for vibrating nucleons leads to the generic force constant and the harmonic frequency

$${}^2R = m_1 \quad {}^2R/2 = e^2/R_e^2$$

taking into account that the reduced mass equals $m_1/2$. With $m_1{}^2 = k_e$, the force constant, we get

$$k_e = 2e^2/R_e^3 \qquad (23b)$$

This expression has been used successfully for scaling the Dunham asymptote in the parameter t, proposed by Varshni and Shuckla some time ago (1963) for ionic bonds and which leads to acceptable results for all kinds of bonds (Van Hooydonk, 1982). Depending on the (unknown) alignment of the particles, i.e. $R_{ab}$ relative to $R_{12}$, different values are obtained for $k_e$, conforming to the analysis in Section 8.

Instead of freezing nucleons or leptons, one can freeze nucleon and lepton motion temporarily at the same time, which corresponds with mass annihilation and gives a static mass-less four-particle system.

The charge inversion technique here leaves only two electrostatic terms in the Hamiltonian

$$H = - e^2/R_{12} - e^2/R_{ab} \qquad (23c)$$

which brings us directly at the top of the absolute well depth. Exactly as above, and depending whether or not the 2 two-particle subsystems are perpendicular, starting at the top of the well depth, the Coulomb potential (23c) is $-2e^2/R_{12}$. Starting half way, with one subsystem fixed at an intermediate asymptote, gives

$$H_{X-X+} = -e^2/R_{ab}$$

Exactly as for the chemical bonds studied here, it can be expected that these mass-less systems are stable. This is also the essential and general conclusion of Richard (1994) and Abdel-Raouf et al. (1998), although the charge inversion technique is not used in their works. But their conclusions are similar: insoluble four-particle systems appear to be stable, *given a particular charge-distribution*.

The experimental evidence collected above about charge inversion in molecules can be used in atoms too, although it will not be detectable in two-particle systems. But, theoretically, an atomic Hamiltonian $H_X$ can now be generalised as in (9a) but, due to the large mass difference between an electron and a proton, charge symmetry is broken almost by definition (Pauling's remark, cited above). Its effect can only be observed through the interactions between two atoms (bonds). The atomic Hamiltonian

$$H = T + V = (1/2)mv^2 - e^2/r$$

would be, referring to (9a) or (11), a t - g = 1 state (interaction and asymptote have opposite signs), just one of the four states allowed by gauge symmetry. The 4 atomic Coulomb states are theoretically

$$W(r) = (-1)^g C(1 + (-1)^{t-g} r_e/r) \qquad (23d)$$

of which two belong to a real, the other two to an imaginary world. But (23d) does not lead to an *atomic* PEC, it results in a *molecular* Coulomb PEC $|IE_X(1-R_e/R)|$ with $R_e = 2r_e$. The classical equilibrium condition gives the absolute gap, $2C = e^2/r_e = mv_e^2$ and in the algebraic scheme used here, equilibrium is obtained half way this gap (the virial theorem) or at $C = e^2/2r_e = mv_e^2/2$ (see Fig. 3a and 5).

Using the charge inversion technique, boson-boson interactions can be rewritten as fermion-fermion interactions. Due to the appearance of intra-atomic charge inversion, the analysis of PECs at the Bohr-scale is ideally suited for examining the behaviour of (composite) particle-antiparticle systems. Then, the interaction of a composite particle (atom, positronium, protonium) with its antiparticle (anti-atom, i.e. a charge inverted atom, anti-positronium, anti-protonium), follows exactly the generic scheme proposed here. Reference systems are the positronium molecule (Richard, 1994) and, of course, the much underestimated hydrogen molecule itself, see Introduction. *At the molecular level*, the mystery surrounding the absence or presence of anti-matter may be solved, if our derivations are valid. Both kinds of matter are present in exactly the same amount in bonds since, for example, $H_2$ = (H, AH) where AH is an anti-hydrogen atom (a charge inverted hydrogen atom, Van Hooydonk, 1985). This is the more intriguing since decisive experiments are planned to reveal the mechanism of anti-hydrogen reactions, as relatively large amounts of anti-hydrogen can now be produced (Armour and Zeman, 1999 and references therein). For a variety of reasons, the outcome of these future experiments is of crucial interest.

If applying the charge inversion technique to atoms from the start is valid in chemistry and leads to Coulomb scaling (as demonstrated here) it should apply at all scales in physics since it is scale-invariant. Compactifying $H_2$ to smaller R-values must lead to a similar composite but smaller system, the atom D. Then, the *ionic version* of a deuterium atom would have as constituent particles a proton and an anti-proton, surrounded by two oppositely charged leptons. This entity must be related to $H_2$. In the above, the ionic variant $(XY)^+$ and $e^-$ was avoided above, since it would lead to an atomic spectrum.

Schemes deriving from classical gauge symmetry like those shown in Fig. 3a for a gap of 2C will have repercussions when similar systems with larger or smaller gaps are considered. For large differences between the gaps, fine structures could be generated generically, leading to double well PECs. Extrapolating these results to nuclear physics and neglecting this possible generic fine structure, the so-called Z-instability of nuclei can now be avoided. Only two types of stable nuclear *ground states* would remain: Z-even (singlets) and Z-uneven (triplets). Singlets are stable with packings similar to those in crystals (solid-state physics). Triplets are stable too whatever the value of Z. These consequences are as observed (Bohr and Mottelson, 1969). There would be no need to consider a separate class of nuclear forces to hold nuclei together since intra-nuclear *repulsions* roughly of order ($Z^2/4$) are replaced by *attractions* of the same order of magnitude. This makes nuclei in their ground state naturally *stable* instead of theoretically unstable. Here, the transition from repulsive states in nuclei on account of positive Z-values to attractive states is similar to the situation met above for bonds. Intra-atomic charge inversion secures that the intra-nuclear repulsion changes into an attraction (see Sections 3 and 6). On the chemical level, uneven Z-nuclei would act as monovalent atoms, whereas even Z-nuclei would behave as noble gases. This is confirmed by experiment: molecules like $Be_2$ and $Ca_2$ are very unstable and have small dissociation energies (Van Hooydonk, 1999) just like noble gases (see also Section 12).

To understand classically the stability of composite stable neutral systems in nature, they must be partitioned in oppositely charged mass-asymmetric subsystems (or particles) if Coulomb's law applies universally. Neutral mass-asymmetrical molecules in a dielectric medium such as water (reducing the power of Coulomb's law by a factor of 80) dissociate in a *charge-symmetric* but *mass-asymmetric* ion pair. The only concurrent 1/R law available, Newton's, uses particle masses instead of charges in the coefficient $a_n$ of R-dependent potentials such as (3b).

**14. Further consequences**

*14.1. Decisive experimental evidence*

The results in Section 11.2 show that also covalent $X_2$ bonds have a critical region where there is interplay between a bond $X_2$, atoms X and ions $X^+$ and $X^-$ since $D_e$ is not a Coulomb asymptote. To determine the intersection point of the ionic potential and a nearly *fixed* $D_e$, the classical result for the critical distance

$IE_X + EA_X + e^2/R_{crit} = 2IE_X$

leads to

$e^2/R_{crit} = IE_X - EA_X$

In the case of $Li_2$, this gives a critical distance of about 3,3 Å. We can calculate the perturbation in the right branch between W(R) and the lower lying asymptote at $2IE_{Li}$ (i.e. $D_e$) provided $EA_{Li} = D_e$. The complete PEC can then be computed using (22m). Using our generic function we find different intersection points ($R_{crit}$). For $Li_2$, the intersection at the left branch occurs at 1,815 and at 5,069 for the attractive branch. On account of atomic long-range interactions varying with $C_n/R^n$, this critical distance at the attractive side can be different. But these crossing points open the way *for a possible experimental proof of the generic scheme based upon electrostatics*. Femto-chemistry can help to find out (i) whether or not the atomic dissociation limit is a scaled form of the ionic asymptote, which seems unlikely but can not be excluded or (ii) if crossing of ionic and atomic states is really avoided. As in the case of NaI (Rose et al, 1988), femto-chemistry can be of assistance to verify which species are present in this critical region of covalent molecules $X_2$ like $Li_2$. If ions $Li^-$ and $Li^+$ are found experimentally, this would be conclusive evidence in favour of an electrostatic approximation to chemical bonding.

*14.2. Universal Coulomb function as an ab initio model potential. Scaling the Dunham coefficients. Chaos and fractal behaviour: an open question?*

Coulomb's law in a zero$^{th}$ order approximation with $V_{12} = 0$ can never account for the (lower order) spectroscopic constants as it leads to meaningless F and G Varshni-parameters, based upon Dunham's analysis. This is not so for the $V_{12} \neq 0$ solution, the perturbed Coulomb function. The corresponding derivations were not made. Maybe this is not even necessary, since the generalised Kratzer-Varshni potential can rationalise the lower order spectroscopic constants quite easily (Van Hooydonk, 1999). Other generalisations are possible. The shift from the generic variable R to R + r, accompanied by a similar shift for $R_e$, where r is relatively small and constant, can not be excluded (Van Hooydonk, 1983, 1999). The introduction of a small r might lead to a consistent extension to a united atom model and even to high-energy (nuclear) physics for R close to zero (Van

Hooydonk, 1983). On the other hand, Varshni-like generalisations (v 1) can not be excluded, such as $(1/m)^x$, with x not too different from 1 (see below).

Dunham's expansion is a poor approximation to the reality of Coulomb based PECs. Ionic asymptotes must be used to scale Dunham-coefficients. The relation between the Dunham variable d and the Coulomb/Kratzer variable $k = d/(1+d)$ is at the origin of this process. An expansion of the Kratzer potential in function of the Dunham variable leads to ideal Dunham coefficients $a_n = (-1)^n n$ (Van Hooydonk, 1983), in agreement with observation in a relatively large number of cases (Van Hooydonk, 1999). The fluctuations in the Dunham series increase linearly with n and illustrate the convergence problems associated with Dunham's series. The Dunham expansion must be abandoned as many people realised (Molski, 1999, Lemoine et al., 1988, Bahnmaier et al., 1989, Urban et al., 1989, 1990, 1991, Maki et al., 1989, 1990).

Gauge symmetry leads to a very simple generic and effective model potential. It derives from first principles and does not contain parameters. It can be expected that fitting observed levels with this scheme automatically leads to a better inversion procedure to construct PECs, transferable from bond to bond, a vacuum in the now available methods. Further work must reveal the universal character of this process.

If so, it can not be excluded that we would end with chaos and fractal behaviour. Given the results obtained here and the possibility of functions varying as $(1/m)^x$, it seems interesting to analyse the behaviour of power laws with fractional exponents like n/3, with n an integer, especially in connection with the non Coulomb $D_e(R)$ dependence. We already constructed Kratzer-Varshni type potentials of this type, some of which are very close to the generic potential used here. Above, we remarked the self-replication of expansions in terms of k. Molski (1999) recently reported on the interest of potentials with continued fractions and March et al. (1997) suggested chaos or fractal behaviour of the lower order spectroscopic constants. They observed that G varies with $F^{4/3}$, a formula we already tested on a large scale (Van Hooydonk, 1999). Further research along these lines remains interesting, reminding we suggested a long time ago that Euclid's golden number might interfere in the physics of interacting charges (Van Hooydonk, 1987). Above, we mentioned an asymptote wherein this strange number also appears. The generic perturbation depends on the square root of 3, which may point in the same direction. Its (geometrical?) origin must be understood.

**Fig. 20** gives the Coulomb asymptote and $D_e$ in function of $R_e$ for the 13 bonds with a power law fit. A fractional exponent appears for $D_e$ almost as expected. But our data about 400 bonds (Van Hooydonk, 1999) do not allow a generalisation of the power law for $D_e$ in Fig. 20.

A new attempt to smoothly generalise Kratzer's potential was recently proposed by Hall and Saad (1998). We did not yet investigate the possibility that Coulomb's law may be adapted (generalised) with a slowly varying maybe exponential R-dependence, as in Yukawa's potential.

*14.3. Charge inversion: a generic consequence of Coulomb's law. Holes. Cooper pairs. Fractional charges*
If the present scheme is scale invariant and if the charge inversion Coulomb mechanism is generic, there must be examples in other domains where analogous effects might have been observed already.
- In solid state physics, the concept of *holes*, bearing a positive unit of charge, has long been accepted and is now an essential part of the theory. With the charge inversion technique, holes can be considered as positive electrons, *positrons*. Also, ionic bonding models have an excellent reputation in solid state physics (we refer to the Born-Landé potential very close to a Coulomb potential). The hole-concept gives circumstantial evidence for this scheme. A similar situation applies to Cooper pairs and to the absence of isotope effects in high-$T_C$ superconductors (Van Hooydonk, 1989).

- As suggested above, the present scheme must also be applicable to reactions between hydrogen and anti-hydrogen, to be observed in the years to come. The outcome of this reaction, according to the present scheme, should be a normal hydrogen molecule.

- Part of the mechanism leading to the quark concept and the quark-anti-quark interaction scheme to explain the stability of (composite) elementary particles, in the first instance baryons, bears an analogy with the present scheme. Quarks are elementary particles with ±1/3 of the unit of charge. For the present approach to make sense in the simplest possible case two singly charged ions of comparable masses are required, one of which *must* be a composite particle, consisting of *three* charged sub-units. A unit of charge ±e for 3 particles leads to an average of ±e/3, see Table 5b. At the Bohr- scale, the mass difference between the constituent fermions is of the order of $10^{-3}$. Nevertheless, fractional charges equal to ±e/3 in the present scheme do not exist in reality as they are *averaged* values.

*14.4. Charge inversion and theoretical chemistry. Reactivity. Charge alternation*

- Molecular wave functions consist of a linear combination of atomic orbitals (LCAO) of Slater-type (ST)- or of (quadratic) Gauss-type (GT) in a SCF-HF procedure to describe bonding with a classical *not parity adapted H-L Hamiltonian*. Using ST-AO's with an exponential 1/R dependence typical for Coulomb's law (implying a central force system) is equivalent with the *a priori* assumption that the atomic dissociation limit is the natural asymptote for the Hamiltonian. This convention is now proved to be incapable to get a scaling procedure for PECs, see above. Coulomb's law and the generic or ionic asymptote are the necessary elements in achieving universal scaling. Unfortunately, ionic wave functions and a parity adapted Hamiltonian, the better of choices in the present analysis, have thus far never been used to our knowledge to construct molecular PECs below and beyond the atomic dissociation limit. Ionic contributions are commonly classified only as (very) small perturbations of the H-L scheme (Weinbaum, 1933). But the present interpretation of symmetry effects in bonding is a generic consequence of using a pair of symmetric Coulomb potentials and gauge symmetry. Conversely, the H-L theory even *seems unnecessarily complicated when it comes to understand the mechanism leading to bond formation*. The complexity of standard quantum-chemical calculations (SCF, HF using atomic STO or GTOs in a LCAO) is a natural consequence of the H-L theory, its Hamiltonian and the conventions about fermion charges (Pople, 1999). The H-L scheme can only produce theoretical PECs through a rather cumbersome procedure. On the basis of the present results, we predict that very accurate theoretical PECs can be obtained without too much effort in the quantum-chemical way by using a charge inversion adapted Hamiltonian and ionic instead of atomic wave functions (one central force system approximation). The atomic dissociation limit can be described by higher order Coulomb interactions in function of atom polarisability, a standard practice in modern physics (Aquilanti et al., 1997, Côté and Dalgarno, 1999). These covalent potentials, varying with $C_n/R^n$ with large n, were approximated in this work even by means of a *fixed* lower asymptote, $D_e$.

The apparatus of wave mechanics remains necessary in the present scheme. If the bonding procedure were really electrostatic, homonuclear bonds in an ionic approximation would show a dipole moment, which is not so. Hence, a homonuclear bond $X_1X_2$ has the wave mechanical hybrid ionic counterpart ($X_1^+X_2^- + X_1^-X_2^+$), a straightforward effect of permutational symmetry. For bonds of intermediate polarity, the weights of the two possible ionic structures are different from 1/2 (Van Hooydonk, 1999).

- Ionic bonding implies a smooth picture for chemical reactivity. For common displacement reactions

AB + CD ↔ AD + CB

a simple electrostatic model has advantages. Electrostatic approximations in the field of chemical reactivity are numerous (see for instance Van Hooydonk, 1975, 1976).

- A challenging example is aromatic reactivity with its many empirical rules, believed to be a purely covalent problem, soluble only in an approximation of the H-L model such as Hückel's MO theory (1930). These rules are easily accounted for by Coulomb electrostatics and *charge alternation* (Van Hooydonk and Dekeukeleire, 1983).

- If the present Coulomb approach is valid, an ionic approximation implicitly contains the important Coulomb principle of *charge alternation*, an additional consequence of Coulomb's law, not yet discussed. Charge alternation can account for a number of details in organic reactivity (Van Hooydonk and Dekeukeleire, 1991, Klein, 1983, 1989).

*14.5. Kratzer-Bohr: a consistency or a dilemma*

The peculiarities of Kratzer's potential with respect to Bohr's were discussed above (see also Van Hooydonk, 1983, 1984). Kratzer's function is more general than Bohr's as it opens the way to the molecular level. Some of the *continuous* $R/R_e$-, m-, or l- dependences (with l = m - 1) in Kratzer's potential, rewritten in function of m and d, are remarkably similar mathematically with the dependence on *discrete* quantum numbers in Bohr's theory (Van Hooydonk, 1984). This can hardly be a coincidence.

*14.6. Possible consequences for the other 1/R potential: Newton's*

The scale and shape invariance of the present scheme implies that it stands irrespective of the nature of $a_n$ in (1e) or (3b). The most typical consequence of algebraic 1/R potentials is that they remain valid whatever the scale (Newton, Bohr, Fermi or Planck, see Fig. 6a-b). If gauge symmetry is universal, gravitational forces can be considered as attractive and repulsive too. This can not yet be proven with gravitational PECs, because these are not available. Compactifying systems towards *smaller* (Planck-) or blowing them up to *larger* (Newton-) scales should have no impact whatsoever on the general features of 1/R potential governed systems and their PECs (see Fig. 6a-c). This aspect of shape- and scale invariance is a challenge in physics and in SUSY, especially for *super-unification*. If a gravitational PEC can be found with a shape like the one generated by a Coulomb 1/R potential, systems with positive and negative masses will have to be allowed in exactly the same way as for positive and negative charges. Newton's law has the right structure already. Yet, $\pm m_1m_2$ unlike $\pm e_1e_2$ states seem to be forbidden. But, if atoms can be annihilated in a molecule, this can not be done properly, if negative masses are not allowed for algebraically. What a negative mass really means is a different question. It could be a mathematical artefact related to another intrinsic property of matter (*not mass*), similar to the mathematical distinction between worlds, between variables like x and 1/x or between gaps, as discussed above.

Qualifying gravitation as weak by a factor of $10^{40}$ led to the exploration of physics at the Plank-scale, where the effect of interacting masses would compare with that of interacting charges (the definition of the Planck-mass). At comparable distances, a *unit of charge* has a mass about $10^{20}$ times a *unit of mass*. The charge/mass ratio for the electron is about $10^{20}$. The centre of a system of particles with $m_1$ and $m_2$, each with a unit of charge e and in equilibrium, must be governed by charges, as these have the larger mass. The centre must be in the middle of the distance, separating the two charges, since $|e_1e_2/(e_1 + e_2)| = e/2$. This corresponds with $e_1/r_1 = e_2/r_2$ if $r_1 + r_2 = r$. If masses determine the centre, $m_1/r_1 = m_2/r_2$ applies. If charges are equal but masses differ by a factor of 2000, the centre should be displaced in favour of the heavier particle. The total mass of a *charged* particle with mass m is $m(10^{20} + 1) = e(1 + 1/10^{20})$, with mass 2000m it is $m(10^{20} + 2000) = e(1+2000/10^{20})$. The particles' distances $r_1$ and $r_2$ from the centre then obey $e_1(1+1/10^{20})r_2 = e_2(1+2000/10^{20})r_1$. The difference with respect to the centre of charges is negligible, as 1 part in $10^{17}$ is beyond the accuracy of any experiment. But the opposite is observed: the centre of the H-atom is not determined by charges, it is determined by masses. Charge symmetry is apparently broken by particle mass (the essence of Pauling's remark, 1985) but this leads to question marks about the meaning of the factor $10^{40}$.

This huge factor does not respect two-dimensional scaling either. Only asymptotes or $a_n/R_e$-values can be used for scaling, the main conclusion of our results. The factor $x = 10^{40}$ results from comparing *coefficients* $a_n$ as appearing in (1e) or (3b), i.e.

$$xGm_1m_2 = e^2 \qquad (24)$$

This does not take into account the different scales ($R_e$-values or asymptotes) the two different $a_n$ might refer to. If so, the scale ratio itself might be at the origin of this large scaling factor. Therefore (24) is meaningless, as long as the exact form of the complete potentials leading to (24) is not known. The hypothesis that these different $a_n$ values apply to the same scale is in clear contradiction with experiment as the centre of mass of a simple H-atom clearly shows. A generalised form of (24) is

$$Gm_1m_2/R_x = e^2/r_{Bohr} \qquad (25)$$

How the scaling potential $Gm/R_x$ must be determined is another problem. But the factor $10^{40}$ as it stands now can certainly **not** be used as such to determine the hierarchy of forces (Van Hooydonk, 1999a).

**16. Conclusion**

If the 13 bonds used here are a representative sample for observed PECs, intriguing conclusions can be made. Gauge symmetry and the discrete (symmetry) aspects of a first principle's electrostatic Coulomb scaling power law in real particle systems have not been fully exploited in the past. Solutions for molecular PECs based on this law are in agreement with experiment with a more than reasonable confidence level. The universal function, the *Holy Grail of Spectroscopy*, seems to be very close, if not identical with, a derivative of a scaling Coulomb law, characterised by a dependence on (1-1/m), where m is a number. The shape and scale invariance of PECs and especially the observed left-right asymmetry around the minimum seen in this context indicate that intra-atomic charge inversion must be allowed in the Hamiltonian. Coulomb's law, simple electrostatics, is much more powerful than hitherto believed and, if true, we have been wrong-footed by nature (Van Hooydonk 1999). Cancelling the nucleon-lepton interactions in the Hamiltonian of four-particle systems (bonds), a generic consequence of intra-atomic charge inversion, is confirmed by experiment. A Coulomb function is an *ab initio* scaling model potential and can probably be used as a first principle's guide line for inverting energy levels. One important aspect of quantum behaviour, i.e. the appearance of integral numbers, is not discussed in this report: these solutions are all known and are obtained by standard wave mechanics.

But the shape of a PEC for N-particle systems (N > 2) belongs to the Coulomb domain and some points can now be clarified. The first concerns the H-L theory and its interpretation of chemical bonding by means of 'exchange forces': this exotic exchange process reduces to intra-atomic charge inversion in one of the two atoms and to resonance between two (degenerate) ionic structures. The underlying static central force system is an acceptable working model and is nothing else than classical 'ionic' bonding, applying to covalent bonds as well. Despite the historical evolution, ionic bonding schemes remain of fundamental physical importance in the context of N-particle systems. It is a pity it took so long to show that the contributions of Davy (18$^{th}$ century) and Berzelius (1835, 19$^{th}$ century), almost 200 years old, are essentially correct. With us, only Kossel (1916) and Luck (1957) found enough evidence to defend ionic or electrostatic approximations to bonding in this century, although these contributions are all against the establishment in the post H-L era. The charge inversion technique (Van Hooydonk, 1985), based upon atom chirality can play a role in the future because of its generic character, its simplicity (the magnet metaphor) and the scale- and shape-invariant PECs it leads to. With respect to symmetry and parity breaking, many efforts in SUSY are based on the mathematical harmonic oscillator, a poor physical model.

With Bouchiat and Bouchiat (1997), we hope that a feedback with the Standard Model is possible. Cosmology and super-unification are at stake. The question about anti-matter may even be a false one. Whether physical processes *are continuous or discrete* is still uncertain and this is a long-standing discussion. When we try to measure a *continuous* 1/R-dependence we use a *discrete* measure, light. But a continuous physical 1/R process can be described with a static mass-less Coulomb law and its implicit *discrete* elements (power law, gauge-symmetry, species independent charge) and its perfect scaling ability.

**Acknowledgements**

We are in debt to various colleagues (correspondence, e-mails, pre- and reprints): RL Hall, JR Le Roy, M Molski, WC Stwalley, YP Varhsni and L Von Szentpaly. Coinage metal PEC-data were kindly provided by JR Le Roy. I thank Prof. Voitek for his hospitality at the National Library in Prague in July 1999. The Fund for Scientific Research-Flanders supported part of the spectroscopic aspects of this work, contract G.0073.96.

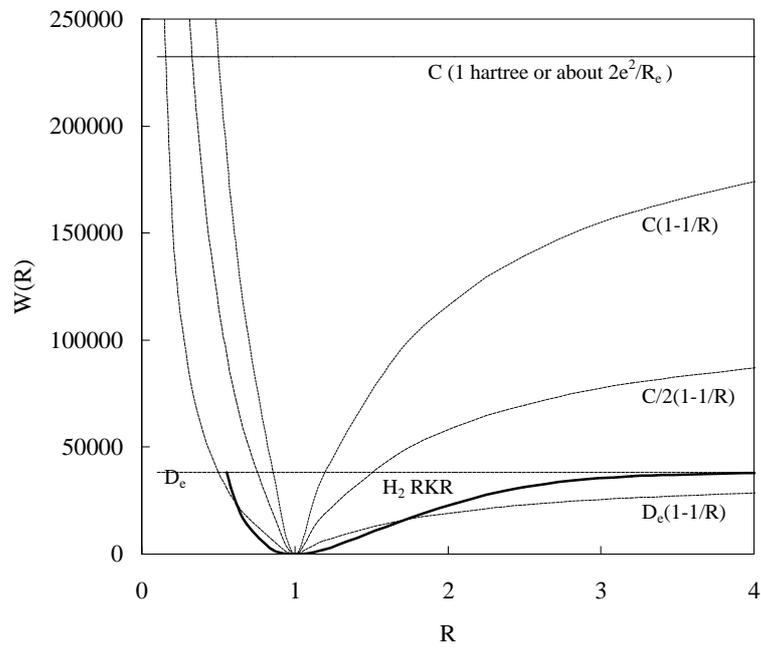

Fig. 1a General effect of Coulomb potential on standard asymptotes and RKR for $H_2$.

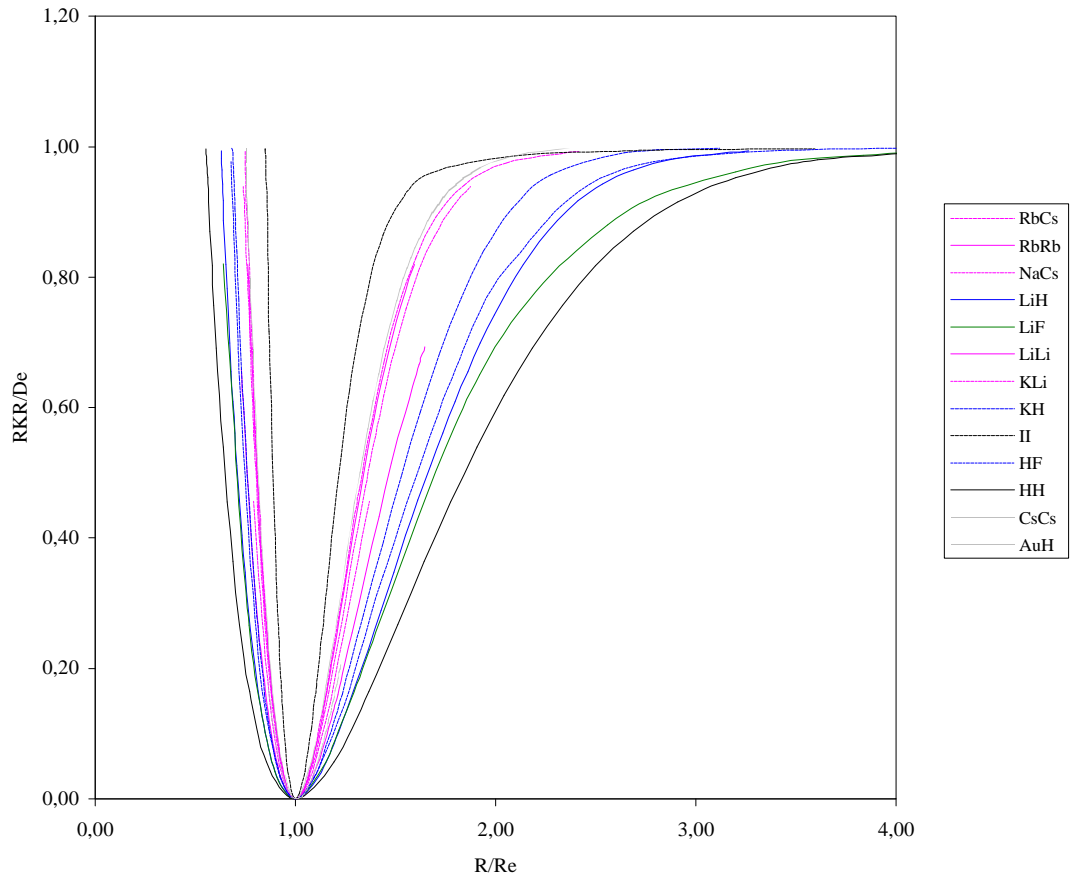

Fig 1b Observed PECs (RKR/De) for 13 bonds versus $R/R_e$

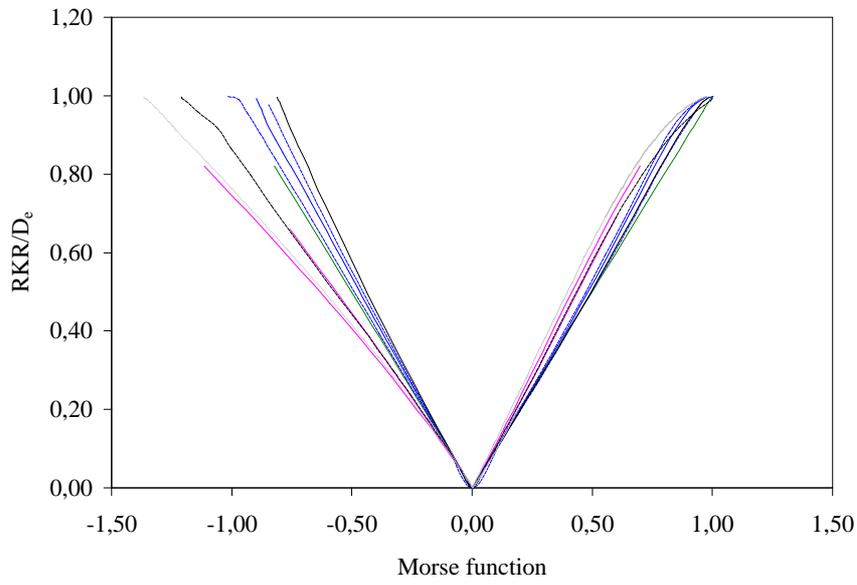

Fig 1c Observed PECs (RKR/$D_e$) for the same 13 bonds (V-shape) versus Morse function

Fig 1d Benchmark: observed PECs (RKR/$D_e$) for the same bonds
(straight line) versus Morse function

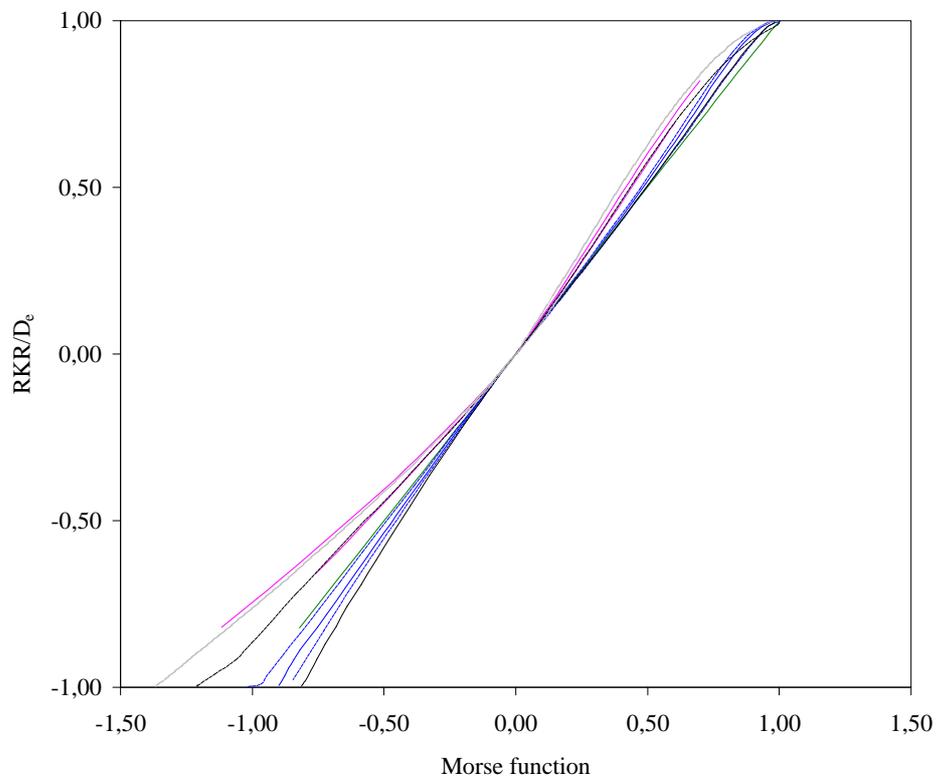

Fig. 2a Potential n = 1

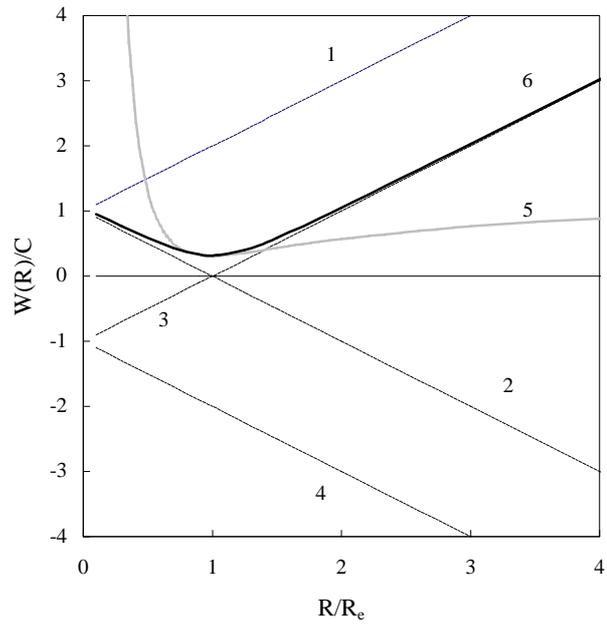

1: +1+R; 2: +1-R; 3: -1+R; 4: -1-R; 5: Kratzer; 6: perturbed Coulomb

Fig. 2b Potential n = 2

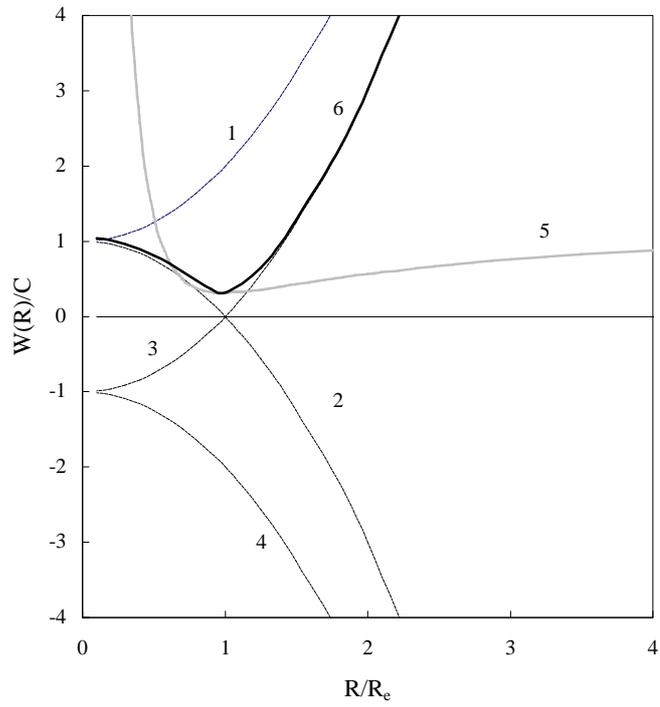

1:$+1+R^2$; 2:$+1-R^2$; 3: $-1+R^2$; 4: $-1-R^2$; 5: Kratzer; 6: perturbed Coulomb

Fig. 3a Potential n = -1

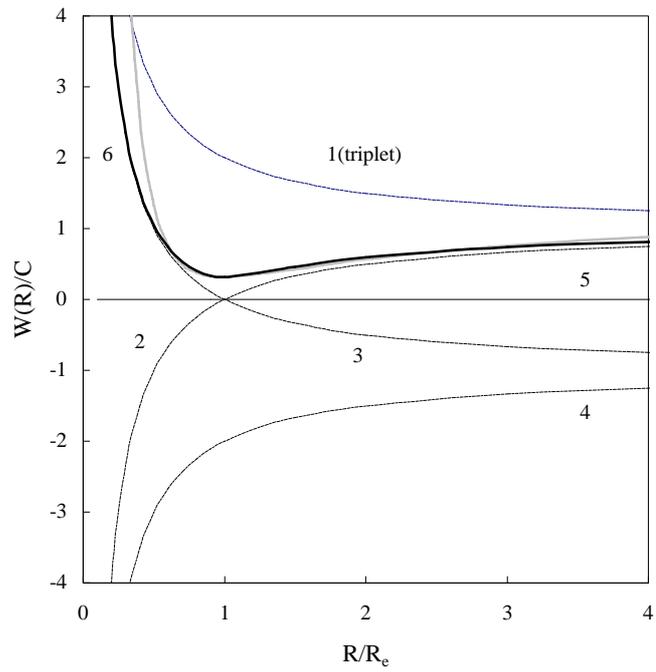

1: +1+1/R; 2: +1-1/R; 3: -1+1/R; 4: -1-1/R; 5: Kratzer; 6:perturbed Coulomb

Fig. 3b Potential n = -2

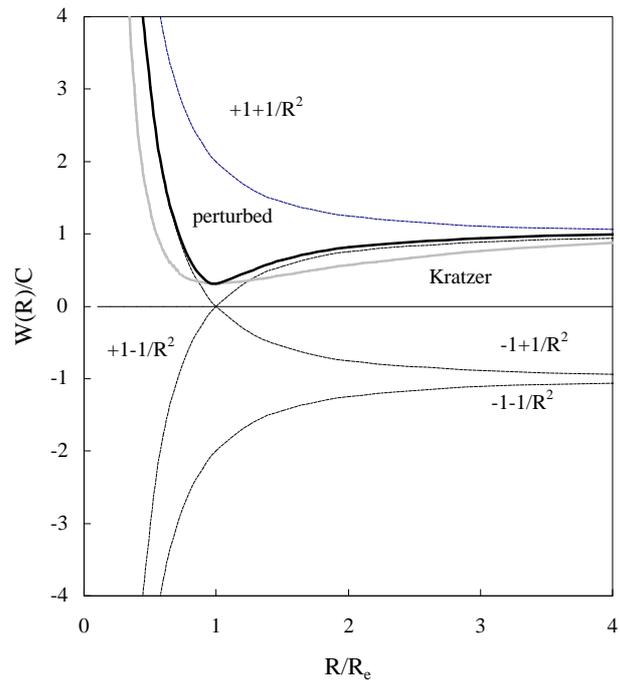

Fig. 4 Coulomb's law

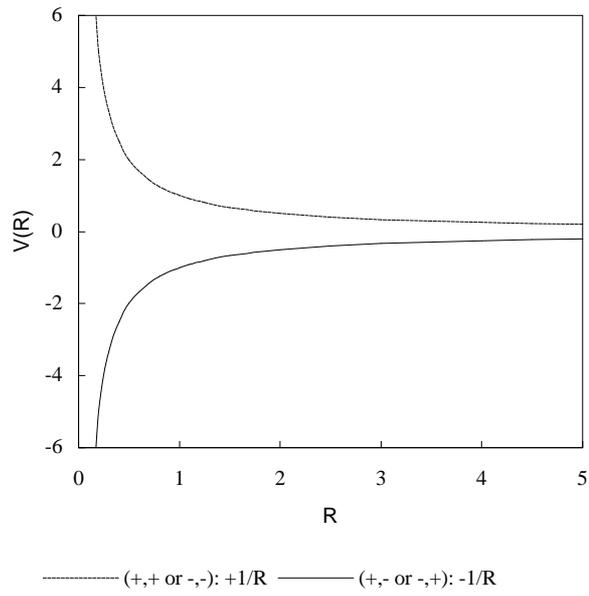

---------- (+,+ or -,-): +1/R  ———— (+,- or -,+): -1/R

Fig. 5 Gauge symmetry and Coulomb's law

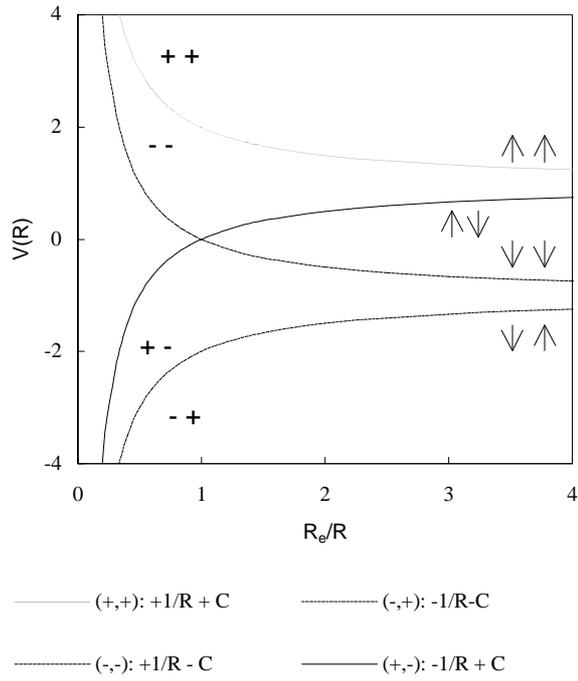

——— (+,+): +1/R + C        --------- (-,+): -1/R-C

--------- (-,-): +1/R - C        ——— (+,-): -1/R + C

Fig. 6a. Coulomb PECs for various asymptotes versus R

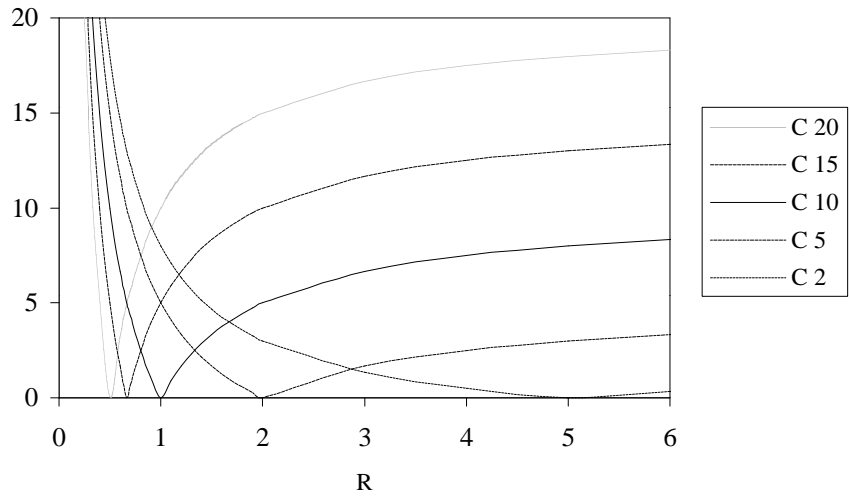

Fig. 6b. Coulomd PECs for several asymptotes versus reduced R

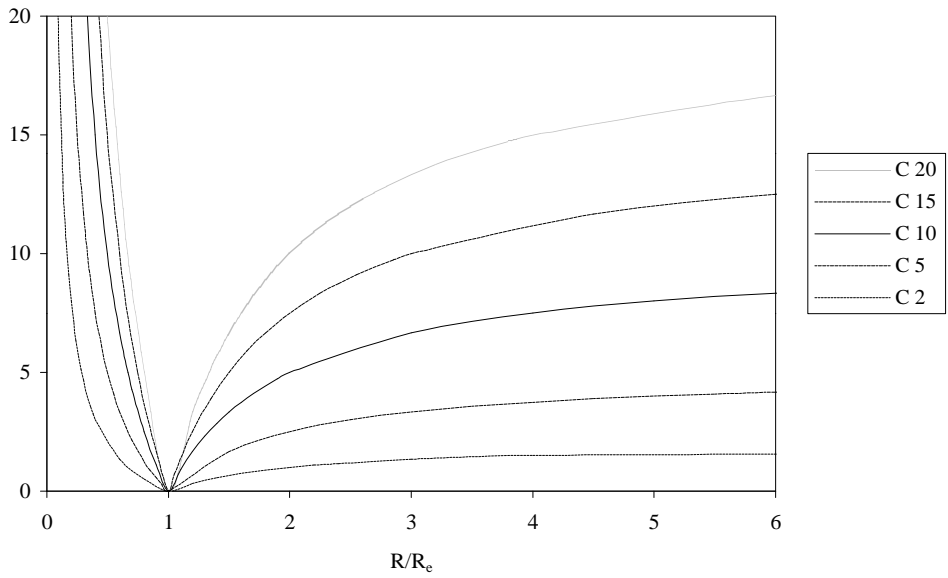

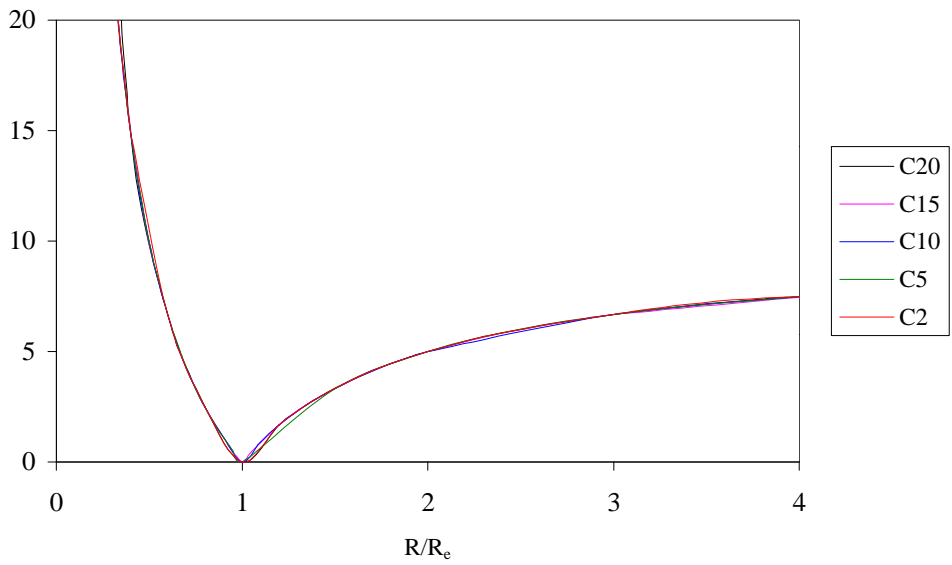

Fig. 6c. Scaled Coulomb asymptotes at asymptote 10 versus reduced R: generic result of universal Coulomb scaling

Fig. 6d Coulomb scheme/gauge symmetry (-) and observed PECs
(o) versus R (13 bonds)(log-scale)

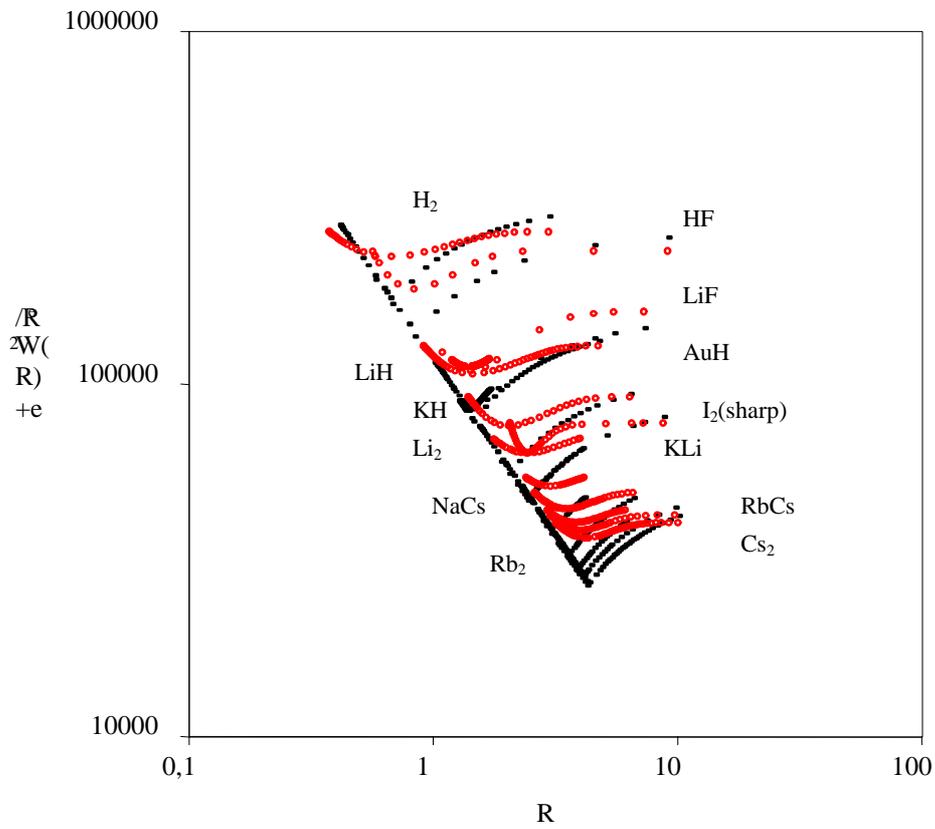

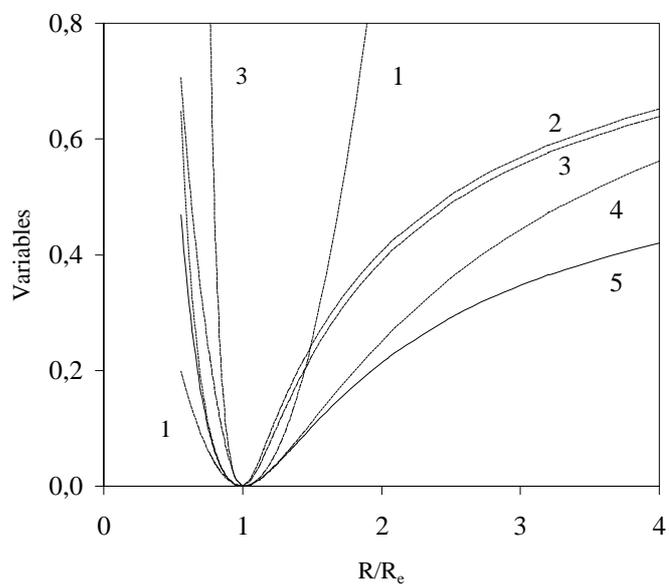

Fig. 7 Dunham, Born-Landé, Kratzer and Generic variables

1: Dunham, 2: Generic (perturbation 0,1), 3: Born-Landé,
4: Kratzer, 5: Generic (perturbation 0,35)

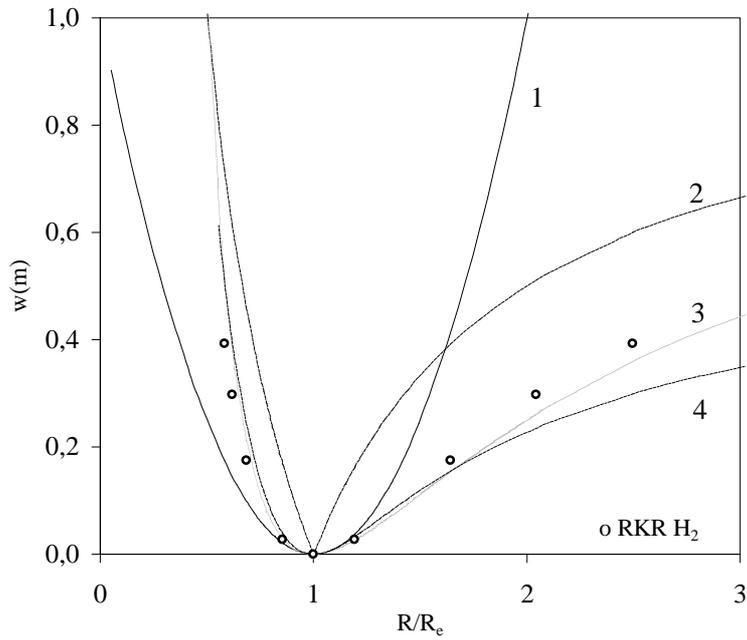

Fig. 8 Comparison of Dunham, Kratzer and generic functions w(m) with experimental data for $H_2$

1: Dunham, 2: Coulomb (unperturbed), 3: Kratzer, 4: Coulomb (perturbation 0,35)

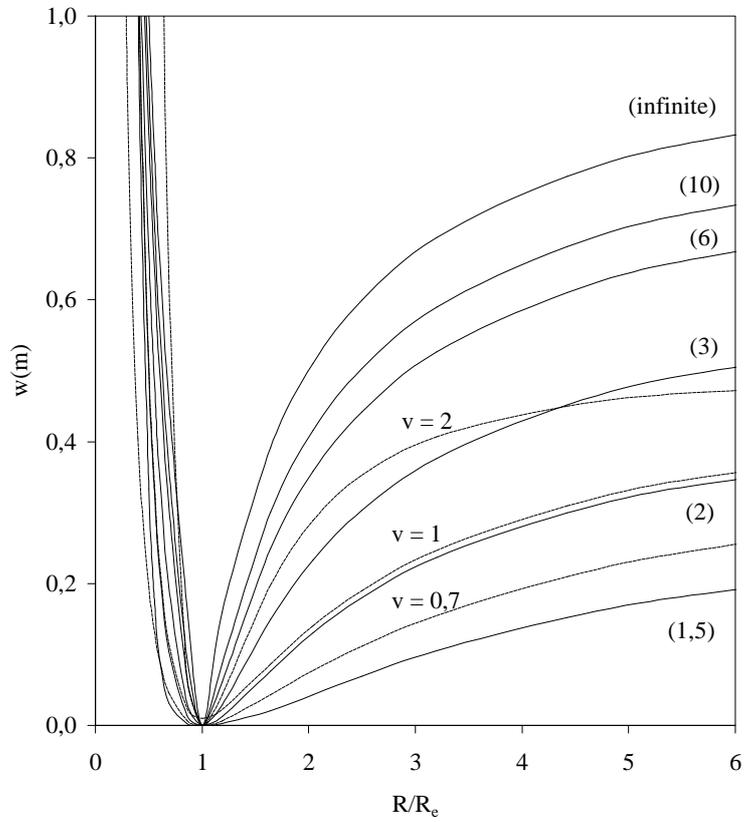

Fig. 9 Perturbed generic and generalised Kratzer-Varshni functions

Generic: full-line, b-values between brackects; Kratzer: dashed lines

Fig. 10a. PEC for $H_2$ from atomic data

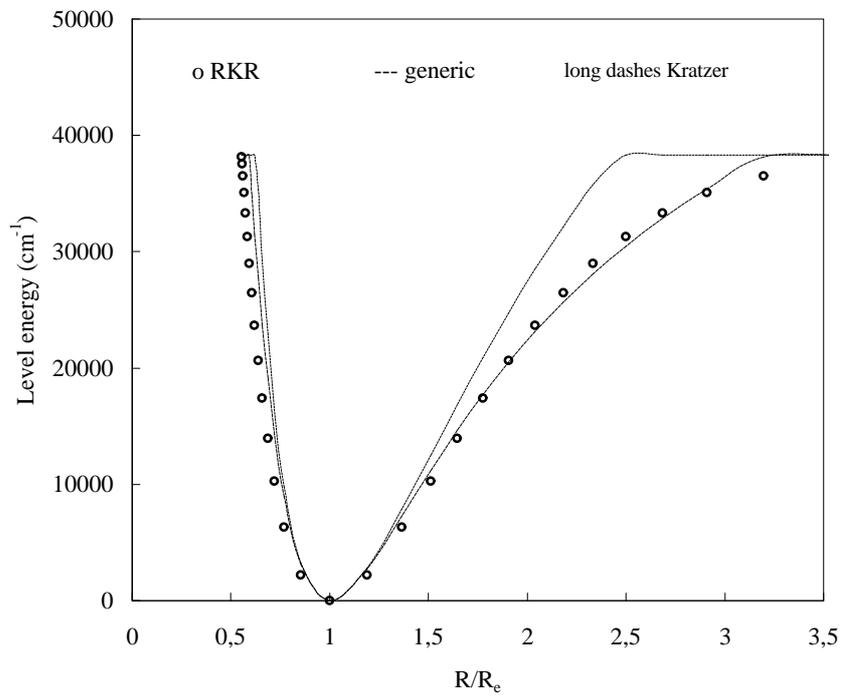

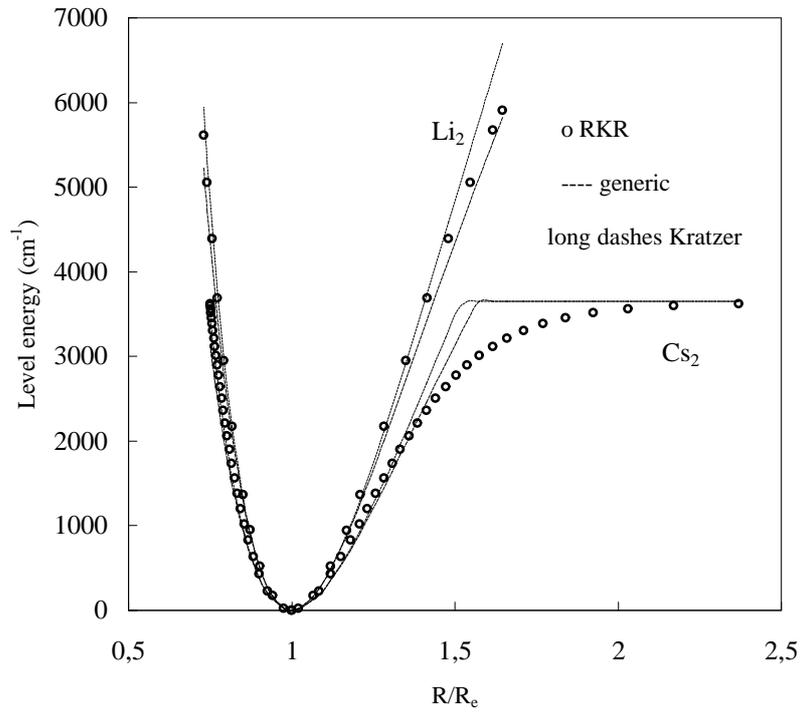

Fig. 10b. PECs for $Li_2$ and $Cs_2$ from atomic data

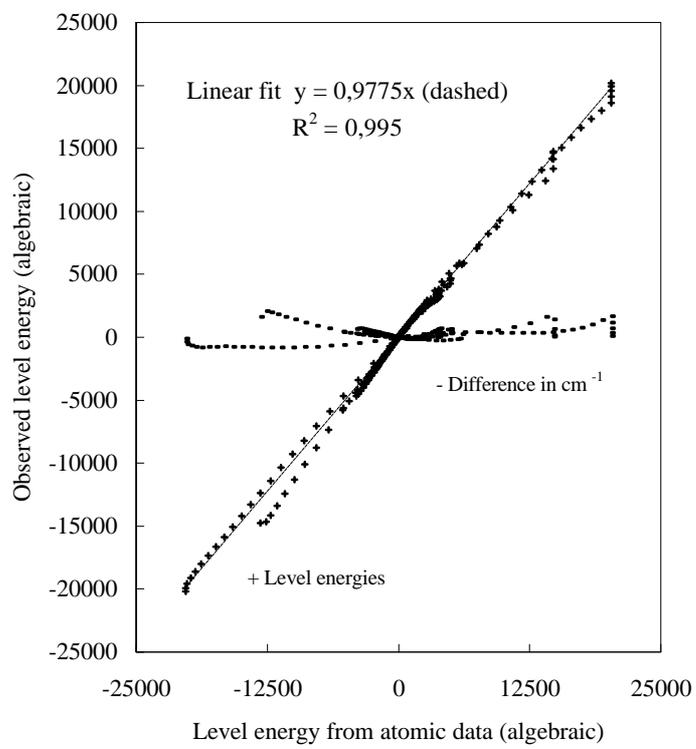

Fig. 10c Observed versus level energies computed from atomic data for 8 bonds (300 data points)

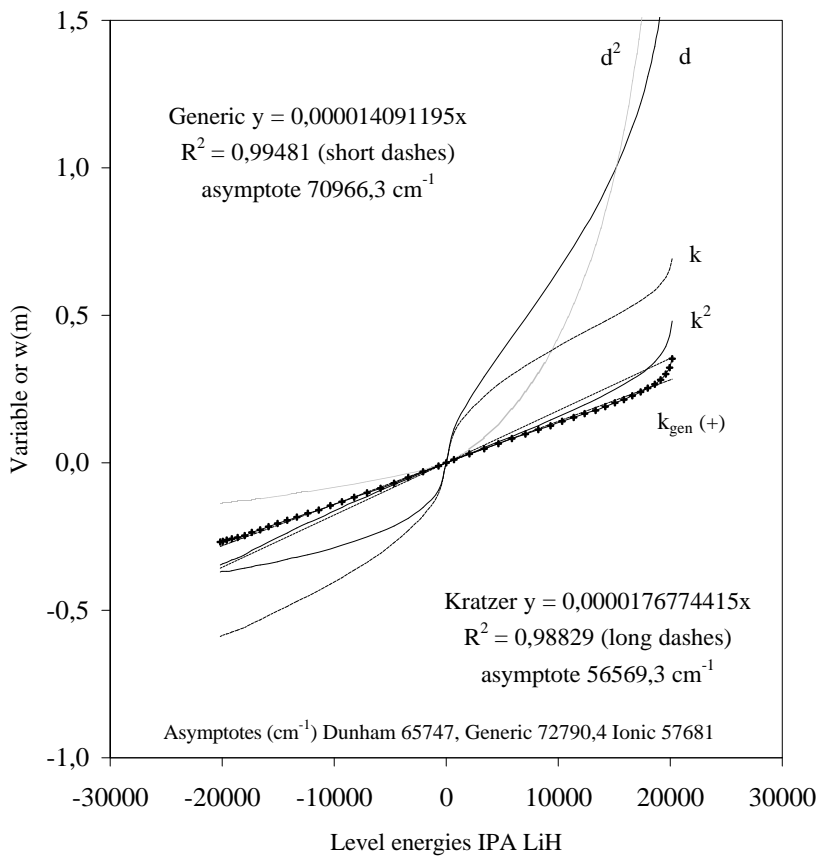

Fig. 11a. Variables versus level energies, IPA for LiH

Fig. 11b Kratzer and Generic functions versus IPA for
LiH less 10 extreme turning points

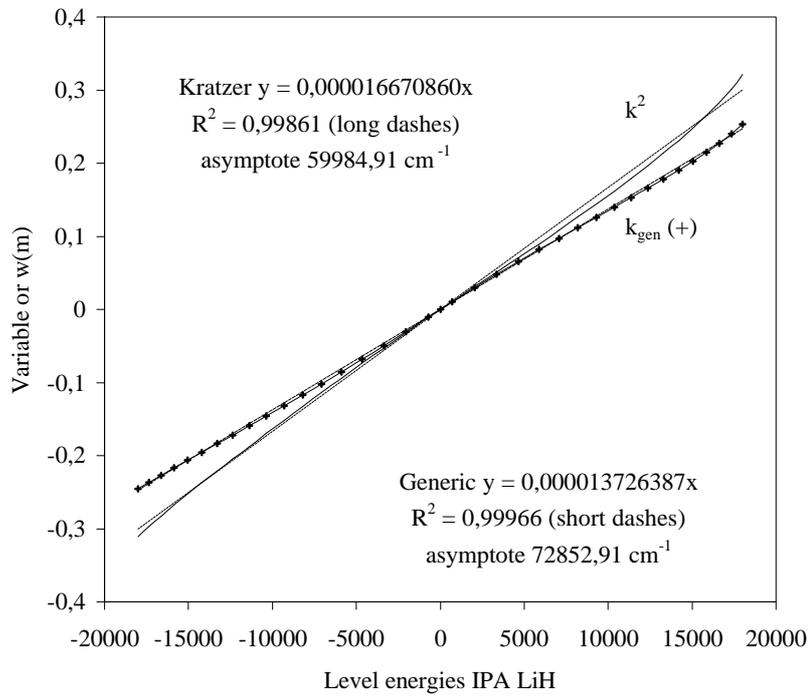

Kratzer y = 0,000016670860x
$R^2$ = 0,99861 (long dashes)
asymptote 59984,91 cm$^{-1}$

$k^2$

$k_{gen}$ (+)

Generic y = 0,000013726387x
$R^2$ = 0,99966 (short dashes)
asymptote 72852,91 cm$^{-1}$

Level energies IPA LiH

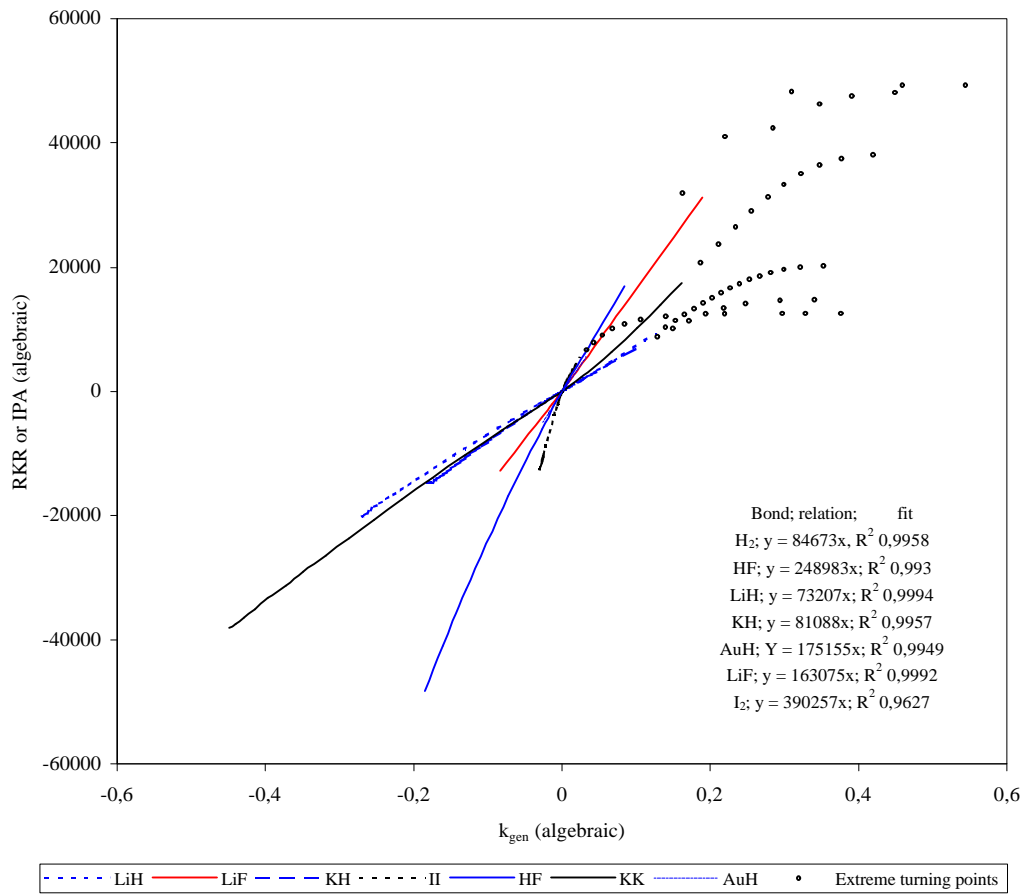

Fig. 12a General fitting procedure for 6 bonds (small $R_e$) and $I_2$ to determine the asymptote

Bond; relation;    fit
$H_2$; y = 84673x, $R^2$ 0,9958
HF; y = 248983x; $R^2$ 0,993
LiH; y = 73207x; $R^2$ 0,9994
KH; y = 81088x; $R^2$ 0,9957
AuH; Y = 175155x; $R^2$ 0,9949
LiF; y = 163075x; $R^2$ 0,9992
$I_2$; y = 390257x; $R^2$ 0,9627

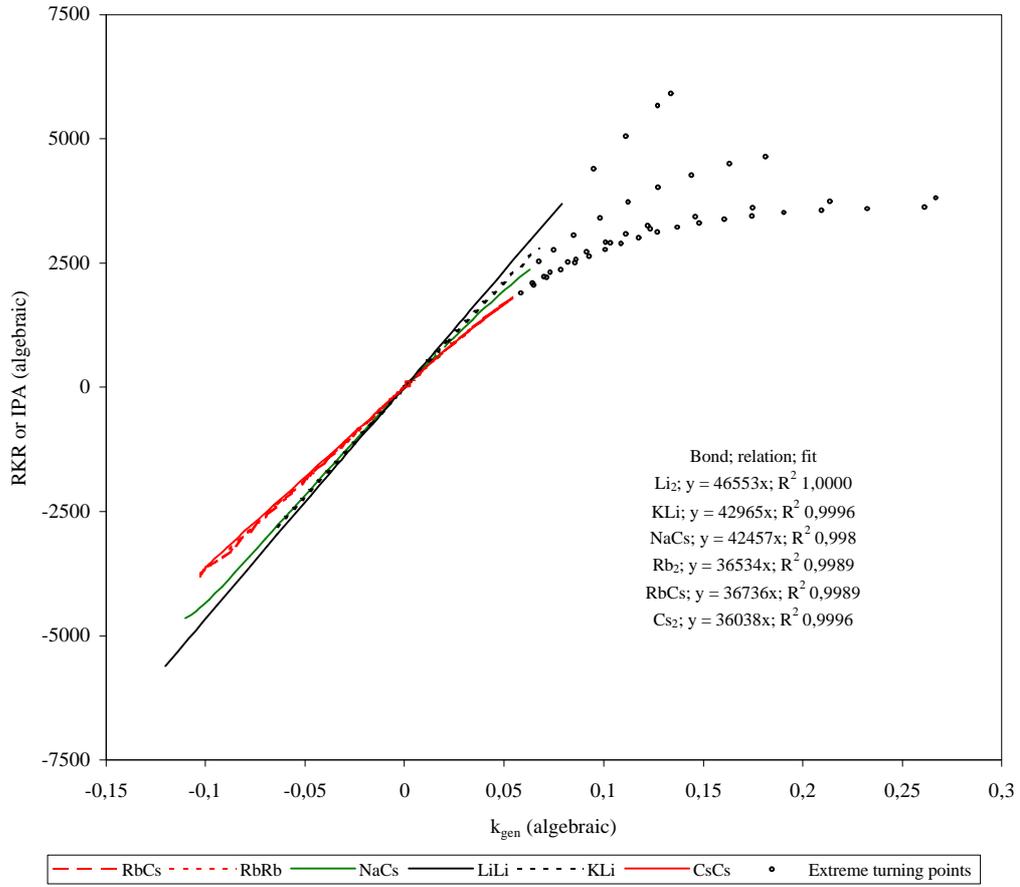

Fig. 12b General fitting procedure for 6 bonds (large $R_e$) to determine the asymptote

Bond; relation; fit
Li$_2$; y = 46553x; $R^2$ 1,0000
KLi; y = 42965x; $R^2$ 0,9996
NaCs; y = 42457x; $R^2$ 0,998
Rb$_2$; y = 36534x; $R^2$ 0,9989
RbCs; y = 36736x; $R^2$ 0,9989
Cs$_2$; y = 36038x; $R^2$ 0,9996

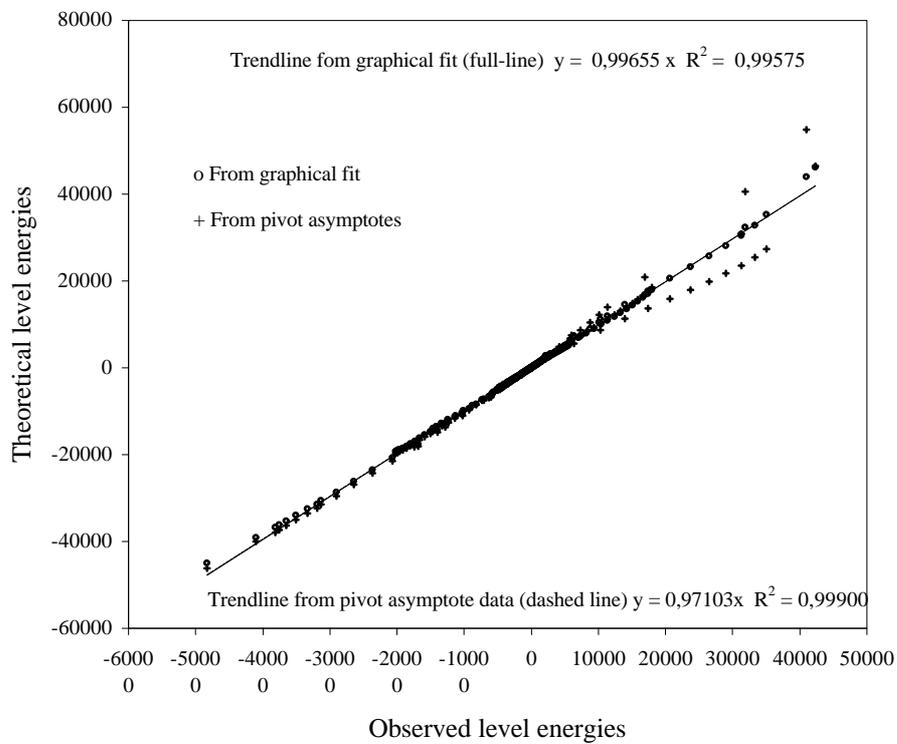

Fig. 13a About 400 theoretical level energies from asymptotes obtained with graphical fitting procedure and pivot table results versus observed level energies

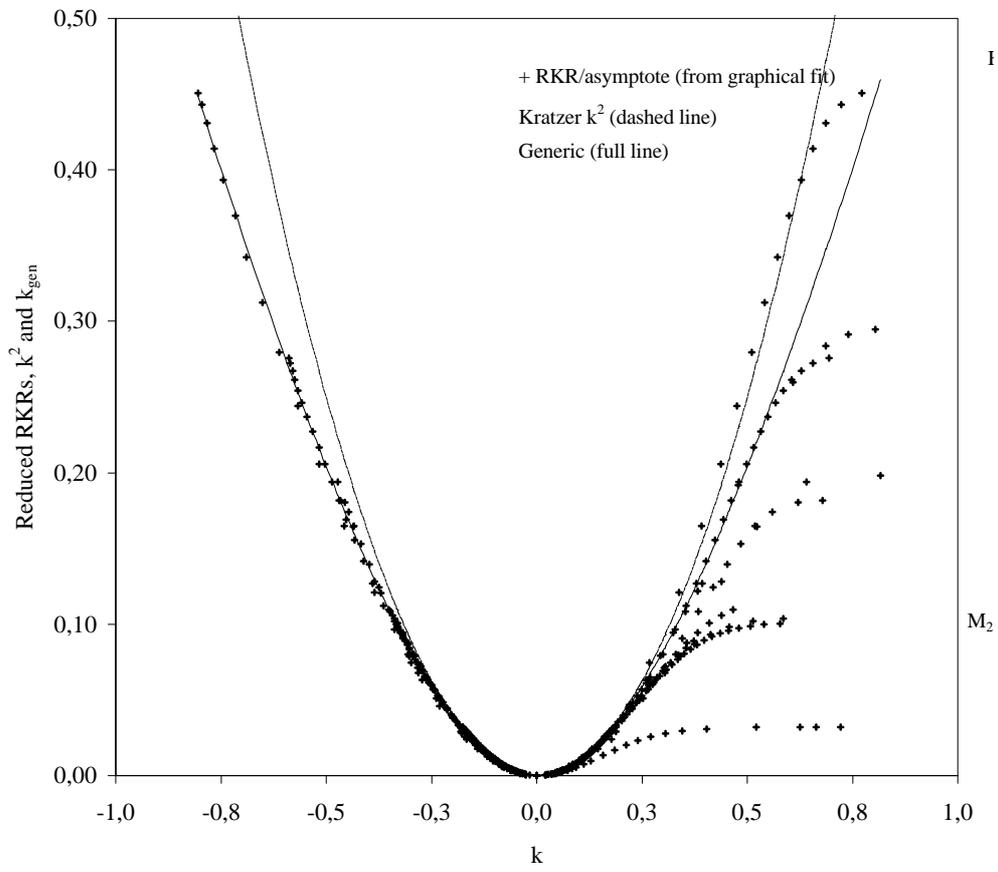

Fig. 13b Reduced RKRs, Kratzer and Generic variables versus k
(all data)

Fig 14a. Reduced RKRs, V-shape

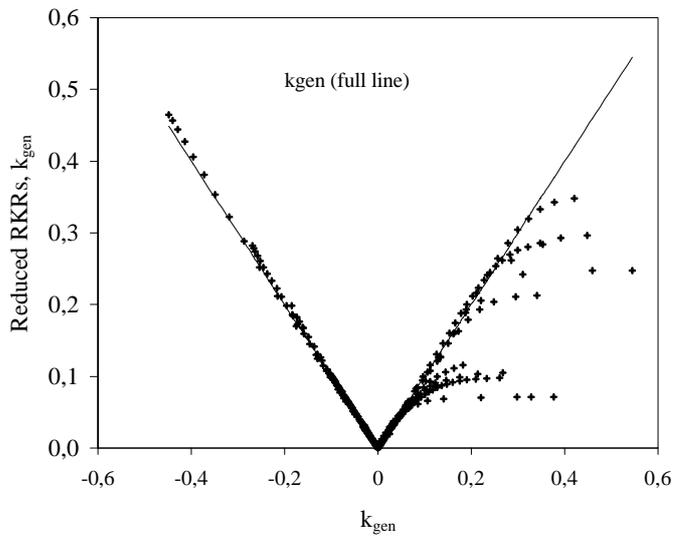

Fig. 14b Rduced RKRs versus k<sub>gen</sub>, linear form

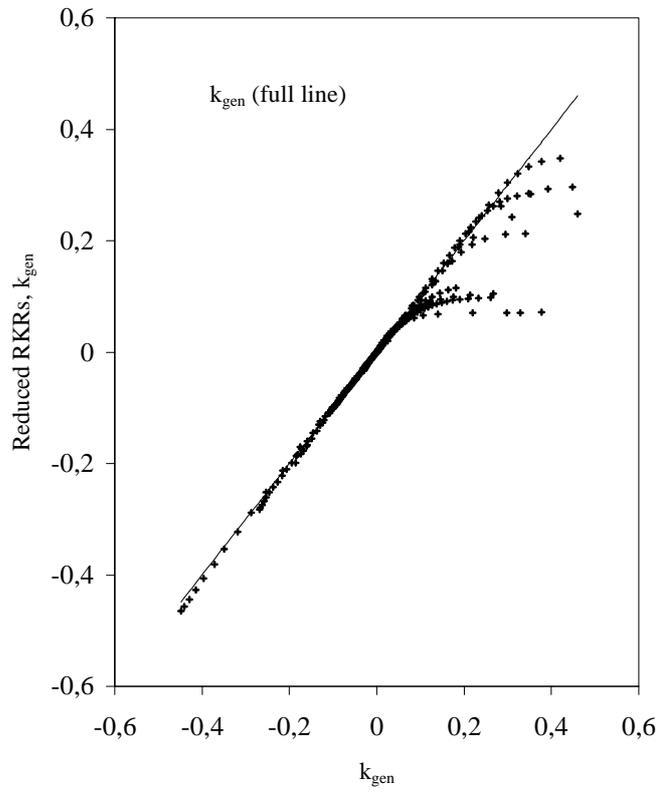

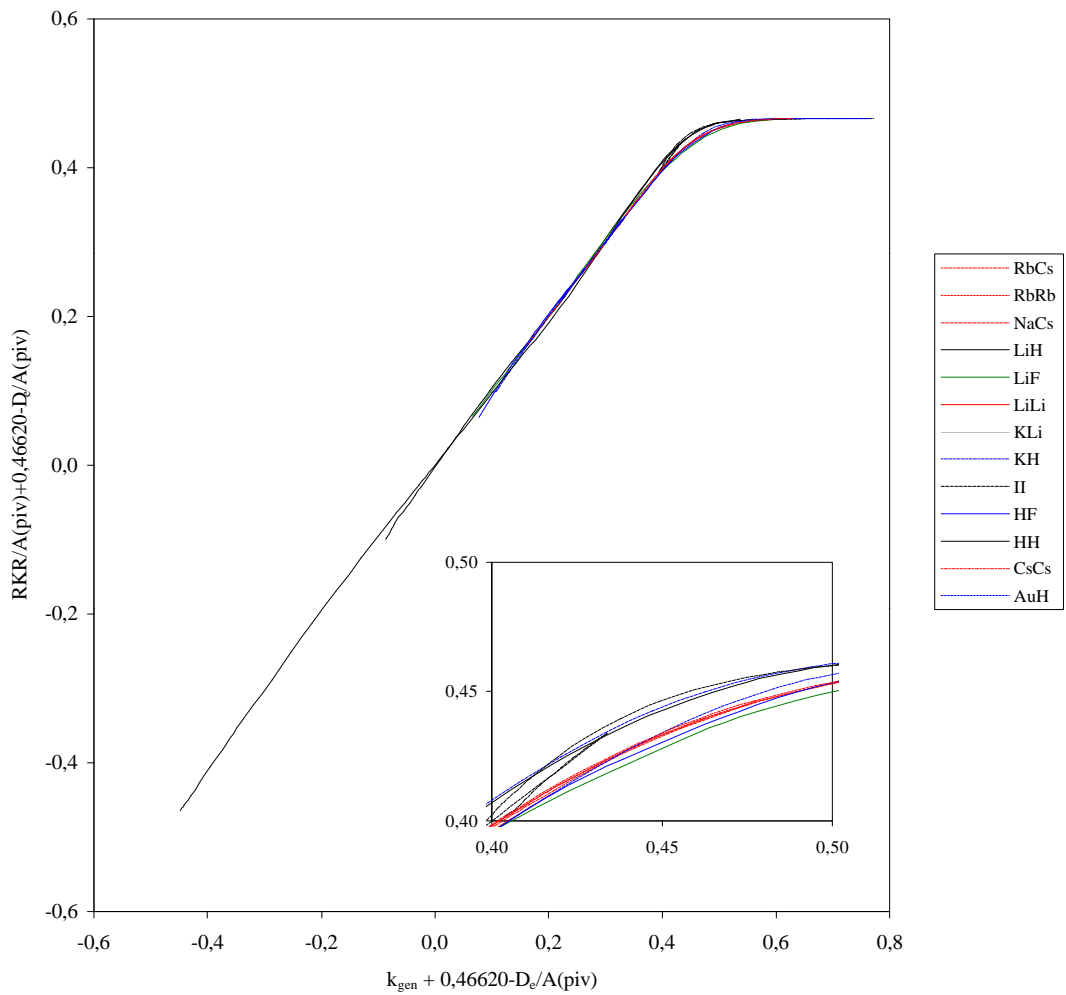

Fig. 15a RKR/A(piv) versus $k_{gen}$ both shifted with $0,46620-D_e/A(piv)$ for 13 bonds

Fig. 15b Observed RKR/$D_e$ versus theoretical $k_{gen}A/D_e$

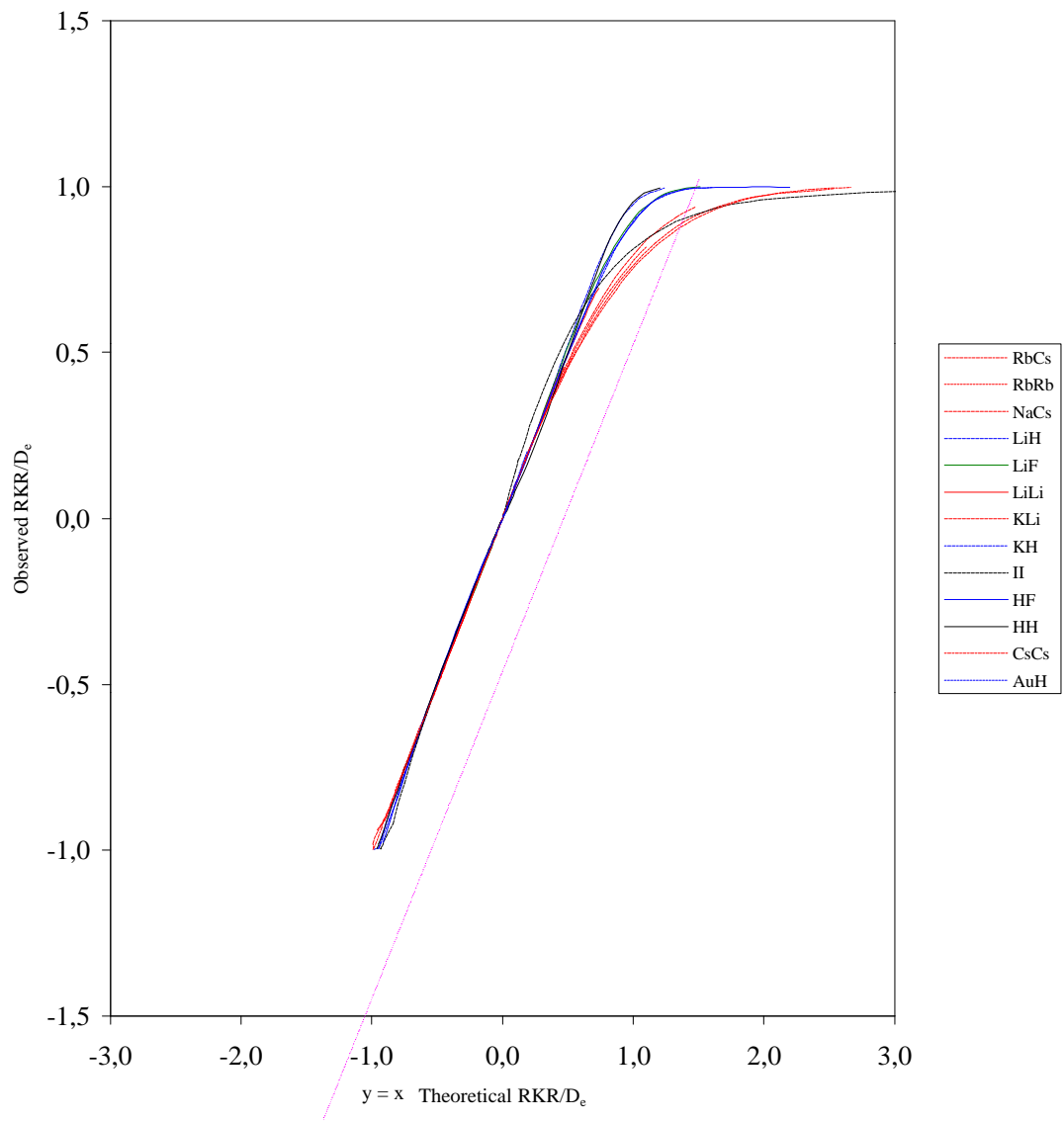

Fig. 15c Observed RKR*$R_e$+75258-$D_e R_e$ versus theoretical

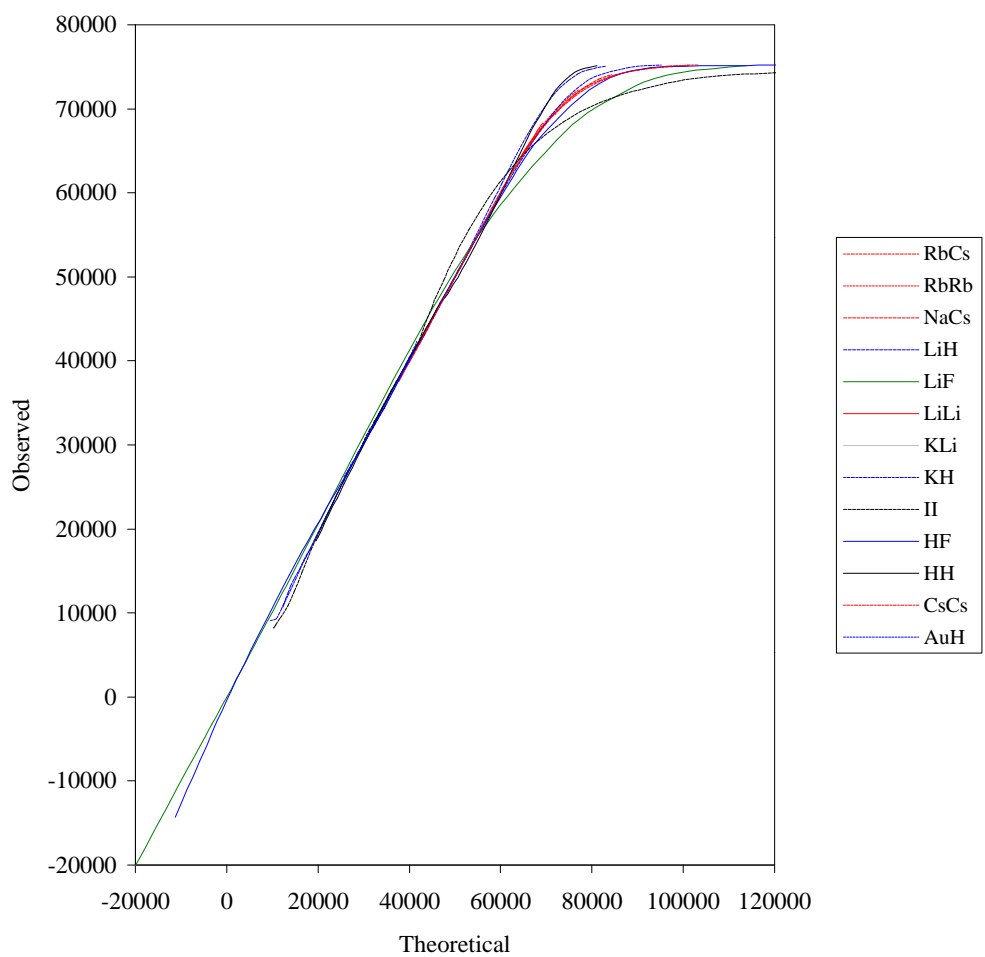

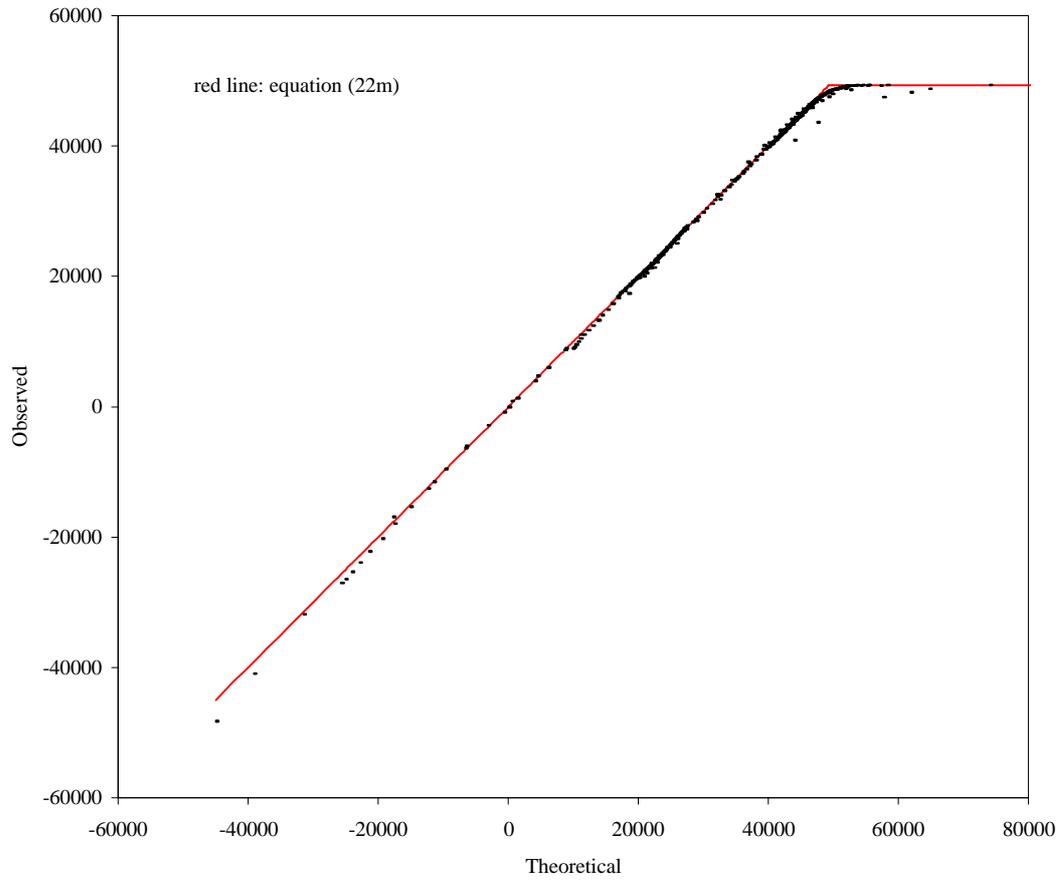

Fig. 15d Observed versus theoretical level energies +49406-$D_e$ for 12 bonds (not $I_2$)

red line: equation (22m)

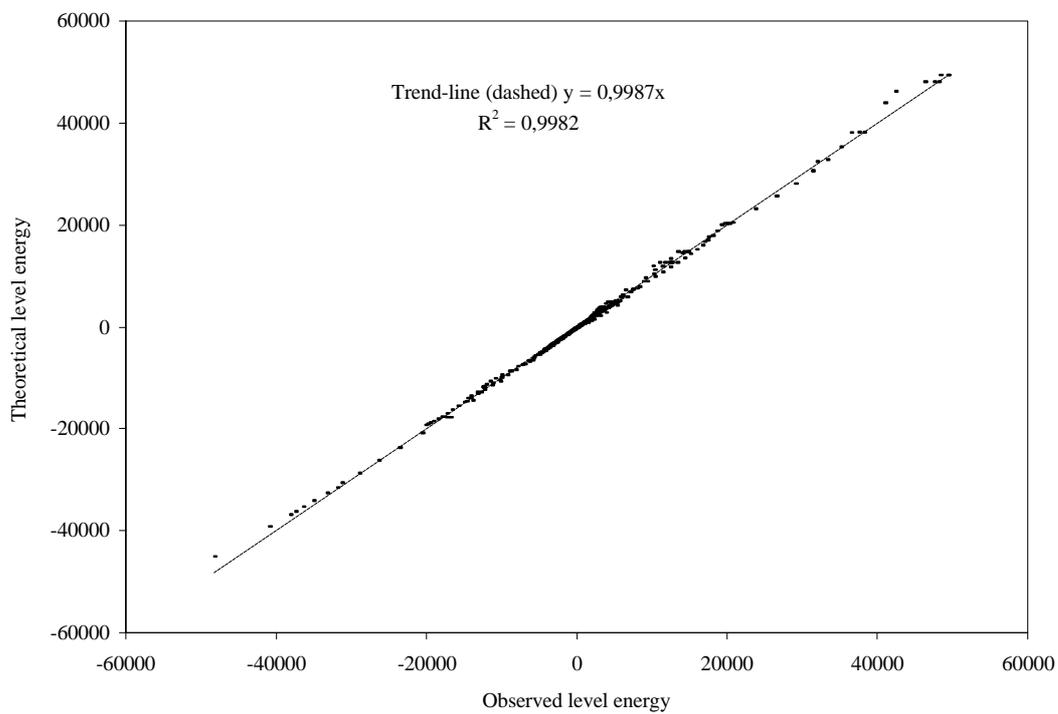

Fig. 16 Theoretical level energies (-) versus observed using (22m) for all 13 bonds, complete range

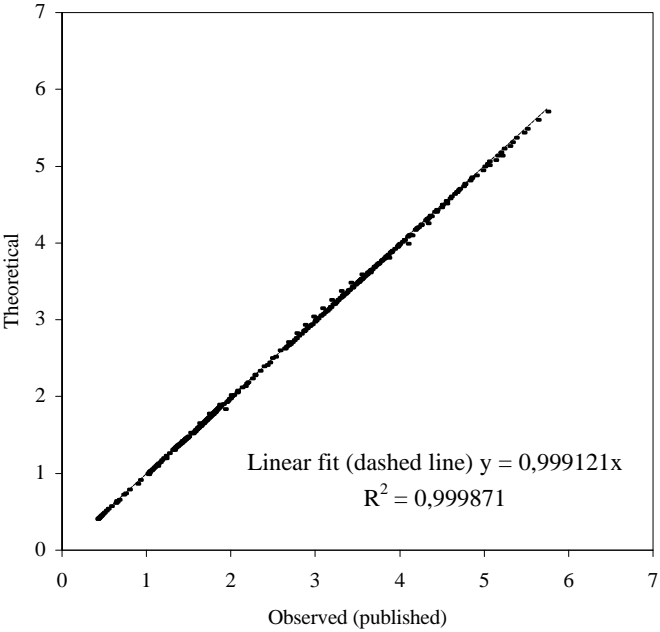

Fig. 17 Theoretical turning points (394 at <50 % of $D_e$) versus observed (published) in Angstrom

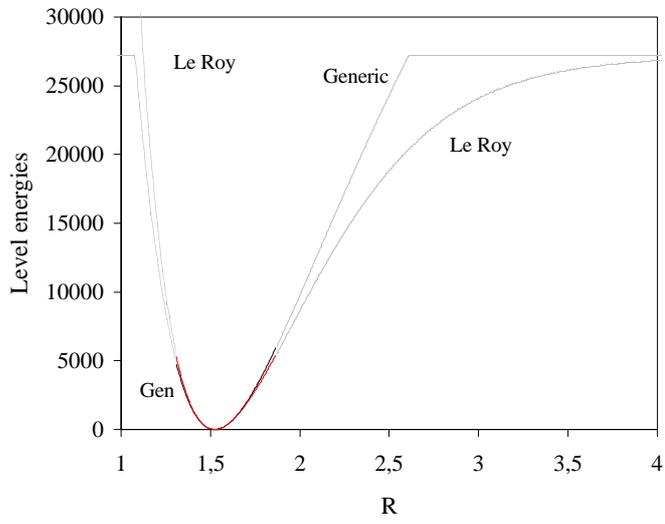

Fig. 18 Complete PEC for AuH: Generic (22m) and Le Roy functions

Fig. 19 Observed and theoretical RKRs for HF

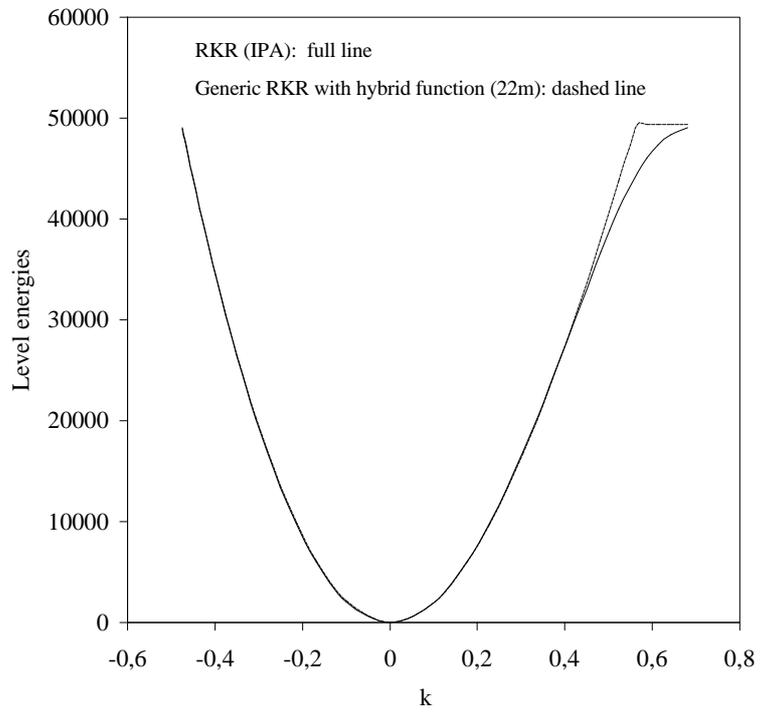

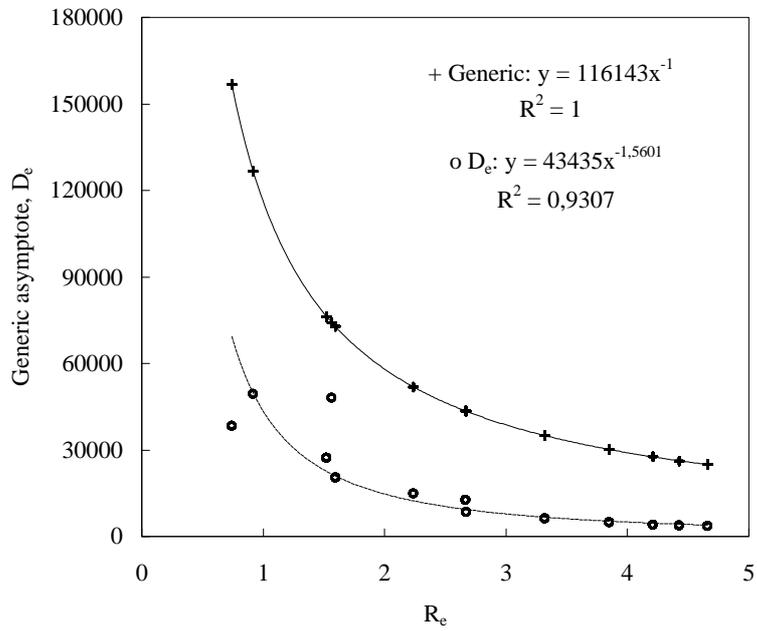

Fig. 20 Power law for asymptotes Coulomb and $D_e$